\documentclass[11pt]{article}
\usepackage[utf8]{inputenc}
\usepackage{amsmath, amssymb, amsthm, bbm, bm, mathtools}
\usepackage[margin=1in]{geometry}
\usepackage{natbib}
\usepackage{color}
\usepackage{graphicx}
\usepackage{qtree}
\usepackage{tikz}
\usepackage{comment}
\usepackage{algorithm, algpseudocode}
\usepackage{multirow}
\usepackage{relsize}
\usepackage{booktabs}
\usepackage{subcaption}

\usepackage{hyperref}
\hypersetup{colorlinks, linkcolor=red, citecolor=blue}

\usepackage{array,epsfig}

\newtheorem{theorem}{Theorem}
\newtheorem{lemma}{Lemma}

\newtheorem{proposition}{Proposition}
\theoremstyle{definition}
\newtheorem{definition}{Definition}

\newtheorem{remark}{Remark}

\newcommand\norm[1]{\left\lVert#1\right\rVert}

\renewcommand{\baselinestretch}{1.2}

\title{Multiscale Tests for Point Processes and Longitudinal Networks}
\author{Youmeng Jiang and Min Xu\thanks{Corresponding author: 	
			Min Xu, Department of Statistics, Rutgers University, New Brunswick, NJ 08854, USA (E-mail: \textit{mx76@stat.rutgers.edu}}
   \\ Statistics Department \\ Rutgers University \\ New Brunswick, NJ, USA}
\date{May, 2024}

\begin{document}

\maketitle

\begin{abstract}
We propose a new testing framework applicable to both the two-sample problem on point processes and the community detection problem on rectangular arrays of point processes, which we refer to as longitudinal networks; the latter problem is useful in situations where we observe interactions among a group of individuals over time. Our framework is based on a multiscale discretization scheme that consider not just the global null but also a collection of nulls local to small regions in the domain; in the two-sample problem, the local rejections tell us where the intensity functions differ and in the longitudinal network problem, the local rejections tell us when the community structure is most salient. We provide theoretical analysis for the two-sample problem and show that our method has minimax optimal power under a Holder continuity condition. We provide extensive simulation and real data analysis demonstrating the practicality of our proposed method. 
\end{abstract}

\section{Introduction}

In many applications involving network data, we observe just not a single static network but rather interactions over time. For example, in business applications, we may observe the timestamp of emails exchanged among employees in a company or transactions over time between people on an e-commerce website. In biology, animal behavioral researchers often use wearable devices to monitor physical interactions among a group of animals to understand their social dynamics \citep{gelardi2020measuring}. 

In this paper, we study testing problems for interactions over time under the framework of \emph{longitudinal networks}, also known as temporal networks. A longitudinal network $A$ is a $n \times n$ array where each entry is an independent realization from a point process; for example, in an animal interaction network, the entry $A_{jk}$ contains all the timestamps of when an interaction event between animal $j$ and $k$ was initiated. Based on the observed longitudinal network $A$, we ask whether the collection of intensity functions, one for each of the $\binom{n}{2}$ (or $n^2$ if network is directed) point processes, contain community structure. For static networks, the problem of testing for community has been extensively studied, including \cite{lei2016goodness} who proposed tests based on random matrix theory and \cite{gao2017testing} and \cite{jin2021optimal} who proposed tests based on subgraph count statistics. The longitudinal setting however introduces a new dimension to the problem: we may be interested in not just \emph{whether} there is a community structure but also \emph{when} the community structure is most apparent. For example, in an animal interaction network, the community structure may only be apparent during a specific time period, such as in the morning when the group is most active.

In this work, we propose a new multiscale testing framework based on discretization. In order to succinctly describe our framework, we first focus on the simpler problem of the two-sample test. Suppose we have two Poisson point processes over the same interval support $\mathcal{X} \subset \mathbb{R}$, with intensity functions $\lambda_a$ and $\lambda_b$ respectively. Our null hypothesis is that the two intensity functions are the same, i.e., $H_0 \,:\, \lambda_a = \lambda_b$. Our testing framework first partitions the ambient space into disjoint bins, which discretizes the Poisson process into a collection of independent Poisson random variables. The partition is chosen hierarchically at different scales to avoid the need to choose a smoothing parameter. In this way, we reduce the problem of testing Poisson processes to a hierarchical collection of tests on Poisson random variables, which we conduct by combining p-values obtained from Binomial exact tests and making the multiple testing adjustments via resampling under the null. 

The advantage of this approach, aside from its computational simplicity, is that it can give granular local information: we can tell not just whether $\lambda_a \neq \lambda_b$ but where in support $\mathcal{X}$ that they differ significantly. We do this by testing not just the global null that $\lambda_a = \lambda_b$ but also a collection of local nulls that $\lambda_a \big|_I = \lambda_b \big|_I$ when restricted to a sub-region $I \subseteq \mathcal{X}$ in our hierarchical partition. To correct for sequential/multiple testing, we apply the adjustment method in \cite{meinshausen2008hierarchical} to control the family-wise error rate. Somewhat surprisingly, the simultaneously valid tests for the local nulls can be done on top of the test for the global null \emph{for free} in the sense that they can be done without any increase in computational complexity and without any decrease in in statistical power. 

This framework may be directly applied to longitudinal networks where we discretize the network into independent Poisson-weighted networks at different scales. This reduces the problem of testing for community in a longitudinal network to a hierarchical collection of tests for community in Poisson networks. To tackle the latter problem, we study tests based on the maximum eigenvalue as well as tests based on subgraph count statistics. We then combine the resulting p-values and make adjustments by resampling under the null. To generate samples under the null for networks, we propose a MCMC procedure based on a sampling algorithm for contingency tables. Although there are existing work on estimation for longitudinal networks and the related multi-layer network setting \citep{zhangefficient, huang2023spectral}, we do not know of prior work focused on community structure testing for longitudinal networks. 

One may ask whether a simple discretization scheme results in too much loss in power compared to existing tests on point processes based on say kernel smoothing \citep{fromont2013two, schrab2021mmd} or wavelets \citep{taleb2021multiresolution}. To that end, we analyze the power of our proposed framework theoretically in the two-sample testing problem and prove, under a Holder continuity condition, that when the dimension of the domain is small, our proposed test has optimal power in the sense that it attains minimax separation rate with respect to the distance $\int_I (\sqrt{\lambda_a} - \sqrt{\lambda_b})^2$ between the two intensity functions $\lambda_a$ and $\lambda_b$. We also perform empirical studies validating that the discretization-based test has competitive power compared to existing approaches.

The remainder of the paper is organized in the following way. In Section~\ref{sec:process_tests}, we define Poisson point process and the two-sample problem; we describe in detail our testing procedure in Section~\ref{sec:two_sample_test_procedure}. In Section~\ref{sec:array_tests}, we define the notion of longitudinal networks and testing for community structure; we describe our tests for three settings: symmetric networks with homogeneous baseline rate (Section~\ref{sec:test_array_homogeneous}), asymmetric networks with homogeneous baseline rate (Section~\ref{sec:asymmetric}), and degree-corrected networks with heterogeneous baseline rates (Section~\ref{sec:dc_array_test}). In Section~\ref{sec:power_analysis}, we provide theorems characterizing the power of our proposed method for the two-sample test. Finally, in Section~\ref{sec:experiments}, we provide both simulation and real data experiments. 

\noindent \textbf{Notation:} Given an integer $K$, we write $[K] := \{1, 2, \ldots, K\}$ and $[K]_0 := \{0, 1, 2, \ldots, K\}$. For a finite set $L$, we write $|L|$ to denote its cardinality. For a matrix $A$, we write $\lambda_1(A)$ to denote its maximum eigenvalue.

\section{Tests on point processes}
\label{sec:process_tests}

We first formally define a Poisson point process. Let the domain $\mathcal{X}$ be a compact subset of $\mathbb{R}^q$ with $\mathcal{B}(\mathcal{X})$ as the corresponding Borel $\sigma$-algebra. We say that $N \,:\, \mathcal{B}(\mathcal{X}) \rightarrow \mathbb{N}$ is a point process realization if it is a counting measure on $\mathcal{I}$ that is finite on every subset $I \in \mathcal{B}(\mathcal{X})$. We write $N(I) \in \mathbb{N}$ as the count of occurrences in $I \subset \mathcal{X}$ and write $N := N(\mathcal{X})$ as the total number of occurrences. We have $N < \infty$ since $\mathcal{X}$ is bounded. We write $X_1, X_2, \ldots, X_N \in \mathcal{I}$ as the locations of the occurrences.

For a finite measure $\Lambda(\cdot)$ on $\mathcal{X}$, we say that a random point process realization $N(\cdot)$ is generated by the inhomogeneous Poisson process $\text{PP}(\Lambda)$ if for all $k \in \mathbb{N}$, all disjoint subsets $A_1, A_2, \ldots A_k \in \mathcal{B}(\mathcal{I})$, and all $m_1, m_2, \ldots, m_k \in \mathbb{N}$, we have
\[
\mathbb{P}( N(A_1) = m_1, \ldots, N(A_k) = m_k ) = \prod_{i=1}^k \mathbb{P}( N(A_i) = m_i) =  \prod_{i=1}^k \frac{e^{- \Lambda(A_i)} \Lambda(A_i)^{m_i} }{m_i !}.
\]
We refer to $\Lambda(\cdot)$ as the intensity measure. In the case where $\Lambda(\cdot)$ has a density $\lambda(\cdot)$ (with respect to the Lebesgue measure), we also write $\text{PP}(\lambda)$ as the same Poisson point process. We refer the readers to  \cite{diggle2013statistical} and \cite{kallenberg2017random} for additional details. 

\subsection{Two-sample test}
\label{sec:two_sample_test}

Before considering the longitudinal network setting where we observe an array of point process realizations, we first study the two-sample setting where we observe two realizations $N_a(\cdot) \sim \text{PP}(\Lambda_a)$ and $N_b(\cdot) \sim \text{PP}(\Lambda_b)$. Our goal is to test whether they have the same intensity measure, that is, we consider the null hypothesis
\begin{align}
\label{eq:two-sample-test}
H_0 \,:\, \Lambda_a = \Lambda_b. \qquad \text{(Two-sample test)}
\end{align}




The null defined in \eqref{eq:two-sample-test} requires that $\Lambda_a = \Lambda_b$ everywhere on $\mathcal{X}$ so we refer to it as the \emph{global null}. We can consider local tests where we ask whether $\Lambda_a = \Lambda_b$ when restricted to a sub-region. To formalize this, we define the notion of a hierarchical partitioning of $\mathcal{X}$.

\begin{definition}
\label{defn:dyadic_partition}
Let $R \in \mathbb{N}$ be a resolution level. We say that $\bm{I} = \bigl\{ \bm{I}^{(0)}, \bm{I}^{(1)}, \bm{I}^{(2)}, \ldots, \bm{I}^{(R)} \bigr\}$ is a hierarchical dyadic partition of $\mathcal{X}$ if $\bm{I}^{(0)} = \{ I_1^{(0)} \}$ with $I_1^{(0)} = \mathcal{X}$ and 
\begin{enumerate}
\item when $r=1$, we let $\bm{I}^{(1)} = \{ I_1^{(1)}, I_2^{(1)} \}$ be a partition of $\mathcal{X}$,
\item and for each $r > 1$, for each $\ell \in [2^r]$, let $I^{(r)}_{\ell}, I^{(r)}_{\ell + 1}$ be a partition of $I^{(r-1)}_{(\ell + 1)/2}$.
\end{enumerate}
\end{definition}

\begin{figure}[h!]
\centering
  \includegraphics[width=0.6\textwidth]{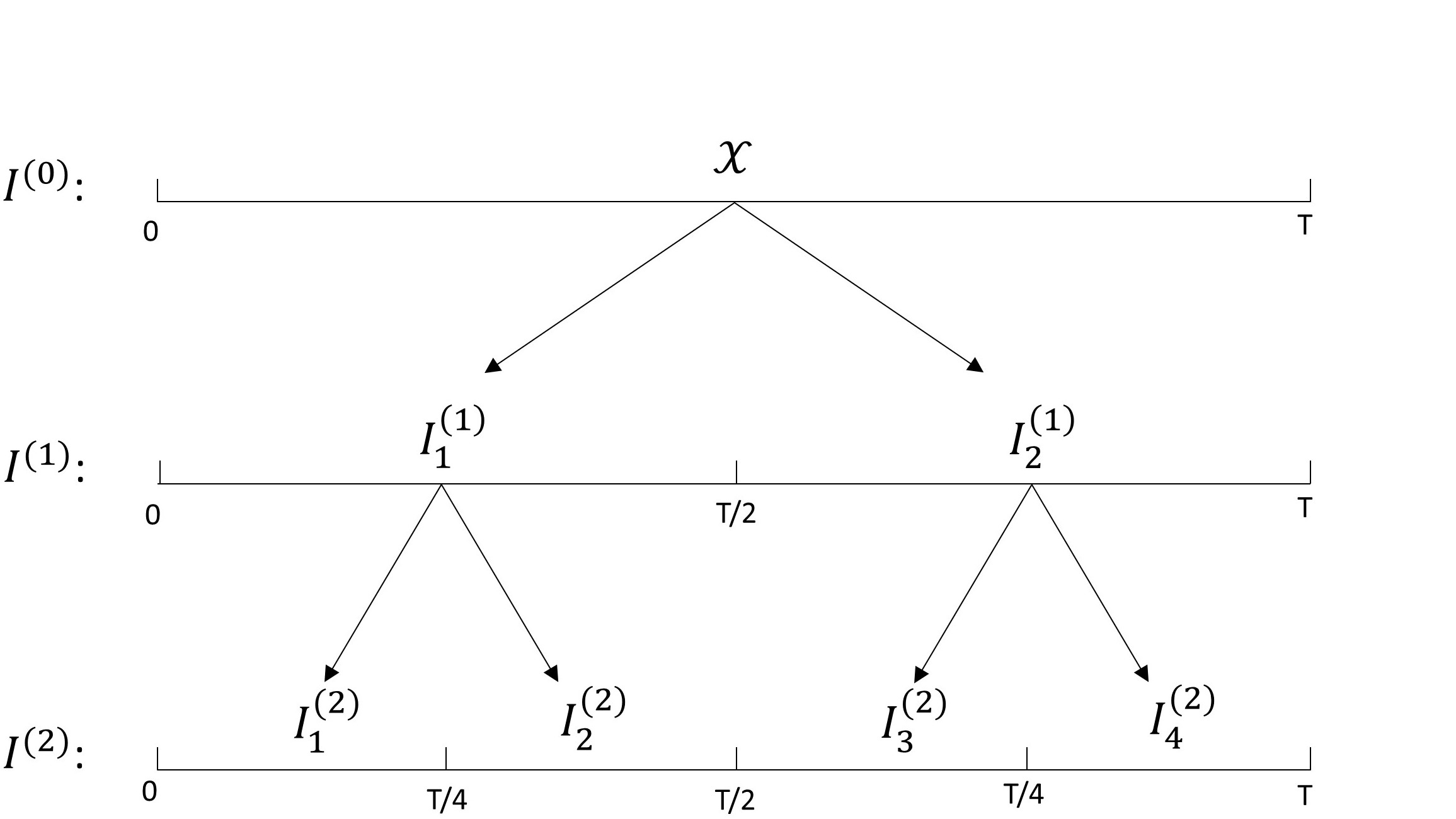}
  \caption{}
  \label{fig:partition}
\end{figure}

For each resolution level $r \in [R]_0$, the collection of intervals $\bm{I}^{(r)} = \bigl\{ I_{\ell}^{(r)} \bigr\}_{\ell \in [2^r]}$ is a partition of $\mathcal{X}$. If we fix an interval $I^{(s)}_j$ where $s \in [R]_0$ and $j \in [2^s]$, then, defining
\begin{align*}
L(s, j, r) &:= \{ 2^{r-s} (j - 1) + k \,:\, k = 1, 2, \ldots, 2^{r-s} \}, \,\, \text{with } r > s,
\end{align*}
we see that $\{I^{(s)}_j\}_{j \in L(s, j, r)}$ is a partition of $I_j^{(s)}$ at resolution level $r$. For example, we have $L(s, j, s+1) = \{2j - 1, 2j \}$ so that $I^{(s+1)}_{2j - 1}, I^{(s+1)}_{2j}$ is a dyadic partition of $I^{(s)}_j$ at one higher resolution level. See Figure~\ref{fig:partition}.

To simplify notation, if $N(\cdot)$ is a point process realization on $\mathcal{X}$, we write $N^{(s,j)} := N( I_{j}^{(s)})$ 
as the number of occurrences in region $I_j^{(s)}$. When $\mathcal{X}$ is a one-dimensional interval, we can form the hierarchical partition $\mathbf{I}$ by recursively dividing each interval in halves. If $\mathcal{X}$ has dimension two or above, we can take any partitioning method that in some sense "evenly" divides each region. We also discuss how to construct $\bm{I}$ in Remark~\ref{rem:random_partition}. For now, we assume that such a partition $\bm{I}$ is given and does not depend on the random realizations. Moreover, all of our discussions generalize to $k$-yadic partition in a straightforward way but we will work with the dyadic version for simplicity of presentation.

For a given $\mathbf{I}$, we may then define the notion of a local null for the interval $I^{(s)}_j$:
\begin{align}
H_0^{(s, j)}  \,:\, \Lambda_a(\cdot) = \Lambda_b(\cdot) \text{ on $I^{(s)}_j$}. \qquad\qquad \text{(Local null)} \label{eq:localnull}
\end{align}

We note that, since $I^{(0)}_1 = \mathcal{X}$, the null $H_0^{(0,1)}$ is exactly the global null. For reasons that will become clear, we also define a related notion of the local null which we refer to as the \emph{discretized local null}:
\begin{align}
\bar{H}_0^{(s, j)}  \,:\, \Lambda_a(I_j^{(s)}) = \Lambda_b(I_j^{(s)}). \qquad \qquad \text{(Discretized local null)} \label{eq:discrete_localnull}
\end{align}

We note that $H_0^{(s, j)}$ implies $\bar{H}_0^{(s, j)}$ but the two are generally not equivalent. They may be similar if $\Lambda_a(\cdot), \Lambda_b(\cdot)$ have a smooth density and if the region $I_j^{(s)}$ has a small diameter.

An important observation that underpins our testing procedure is the fact that the collection of local nulls $H_0^{(s,j)}$'s has a logical tree structure:
\begin{align}
H_0^{(s,j)} \Rightarrow H_0^{(r,\ell)} \text{ for any $r \in \{s, s+1, \ldots, R\}$ and any $\ell \in L(s, j, r)$}. \label{eq:logicalstructure}
\end{align}
This holds because $\Lambda_a = \Lambda_b$ on the region $I^{(s)}_j$ implies that $\Lambda_a = \Lambda_b$ on every sub-region $I^{(r, \ell)} \subseteq I^{(s)}_j$. Because of this logical structure, if we do not reject $H_0^{(s,j)}$, then we should not reject $H_0^{(r,\ell)}$ for any sub-region $I^{(r, \ell)} \subseteq I^{(s,j)}$. 

\begin{remark}
It is clear that $H_0^{(s, j)} \Rightarrow H_0^{(s+1, 2j-1)} \cap H_0^{(s+1, 2j)}$ (each local null implies its children). For the two sample test, we in fact also have the reverse 
\begin{align}
H_0^{s+1, 2j-1} \cap H_0^{s+1, 2j} \Rightarrow H_0^{(s, j)}, \label{eq:reverse_logic}
\end{align}
that is, $H_0^{(s,j)}$ must be true if its two direct children are true. This allows us to obtain some small improvement in power when performing the multiple testing adjustment (see Section~\ref{sec:two_sample_step4}). 
We note that \eqref{eq:reverse_logic} does not hold in the longitudinal network setting (c.f. Remark~\ref{rem:no_reverse_logic}).
\end{remark}

\subsubsection*{Related work on testing for point processes}

For the two-sample problem, \cite{fromont2013two} proposed tests for the global null using U-statistics based on kernel functions and proved their optimality; similar methods appear in \cite{gretton2012kernel}. Methods based on scan statistics have been studied in \cite{kulldorff2009scan}, \cite{walther2010optimal}, and \cite{picard2018continuous}. In contrast to these work, our focus is on simultaneous testing of both the global and local nulls as well as on having a framework that easily extends to the longitudinal networks. There are also work on testing homogeneity \citep{fromont2011adaptive} and for testing whether the proportion of two intensity functions is a constant or increasing \citep{bovett1980comparing, deshpande1999testing}.

\subsection{Testing procedure for two-sample test}
\label{sec:two_sample_test_procedure}

Our test procedure will produce simultaneously valid p-values for $H_0$ and the entire family $H^{(s, j)}$ simultaneously in the following sense: we produce a collection of p-values $p^{(s,j)}$ such that if we reject
\begin{align}
\mathcal{R}_\alpha = \{ H_0^{(s,j)} \,:\, p^{(s,j)} \leq \alpha \text{ and } p^{(s^*, j^*)} \leq \alpha \text{ for all $I^{(s^*, j^*)}$ such that $I^{(s,j)} \subseteq I^{(s^*, j^*)}$} \},
\label{eq:rejection_rule}
\end{align}
then we control family-wise error rate at level $\alpha$, that is,
$\mathcal{R}_\alpha$ contains no false positives with probability at least $\geq 1 - \alpha$.

On a high level, our testing strategy is to approximate the local null $H_0^{(s,j)}$ by an intersection of discretized local nulls:
\begin{align}
\check{H}_0^{(s, j)} := \bigcap_{r = s}^R \bigcap_{\ell \in L(s, j, r)} \bar{H}_0^{(r, \ell)}. \qquad \text{(Approximate local null)}
\end{align}

For a given $r \in \{s, s+1, \ldots, R\}$, we see that the intersection $\bigcap_{\ell \in L(s, j, r)} \bar{H}_0^{(r,\ell)}$ approximates $H_0^{(s, j)}$ at discretization/resolution level $r$. The additional intersection over $r$ accounts for all the resolution levels. In the case where $s = 0$ and $j = 1$, we get an approximation of the global null for the two sample test \eqref{eq:two-sample-test}:
\begin{align}
\check{H}_0 \equiv \tilde{H}_0^{(0, 1)} = \bigcap_{r=0}^R \bigcap_{\ell=1}^{2^r} \bar{H}_0^{(r, \ell)}.
\end{align}

Our testing procedure proceeds in four steps: (i) construct p-values $\bar{p}^{(r, \ell)}$'s for $\bar{H}_0^{(r, \ell)}$'s with exact tests, (ii) combine $p^{(r,\ell)}$ across $\ell$ at the same resolution level, (iii) combine p-value across the different resolution levels and use resampling to obtain individually valid p-values $\check{p}_F^{(s, j)}$ for each $H_0^{(s, j)}$, and (iv), apply sequential testing adjustment to obtain simultaneously valid p-values $p_F^{(s,j)}$. We explain each of the steps in detail below and give a concise description of the whole procedure in Algorithm~\ref{alg:twosample}. We also illustrate how the procedure works in an in-depth numerical example in Section~\ref{sec:example}.  


\subsubsection{Step 1: compute p-value for each discretized local null.} 
\label{sec:two_sample_step1}

In the first step, we compute a p-value for each of the discretized local nulls, that is, for each $\bar{H}_0^{(r, \ell)}$ where $r \in [R], \ell \in [2^r]$. To test $\bar{H}_0^{(r,\ell)}$, we observe that the random counts $N_a^{(r, \ell)} := N_a(I_\ell^{(r)})$ and $N_b^{(r, \ell)} := N_b(I_\ell^{(r)})$ are Poisson random variables with means $\Lambda_a( I_\ell^{(r)}), \Lambda_b( I_\ell^{(r)}) \geq 0$ respectively. Define 
\[
N(\cdot) := N_a(\cdot) + N_b(\cdot)
\]
as the aggregated realization. We then have that, under the null hypothesis and conditional on $N^{(r, \ell)}$, the random counts $N_a^{(r, \ell)}$ and $N_b^{(r, \ell)}$ have the binomial $\text{Bin}( \frac{1}{2}, N^{(r, \ell)} )$ distribution. 

We take $N_a^{(r, \ell)}$ as the test statistic. Let $S_{\text{Bin}( \frac{1}{2}, m)}(\cdot) := \mathbb{P}(  |\text{Bin}( \frac{1}{2}, m ) - \frac{m}{2}| \geq \cdot)$ be the two-sided tail probability function, and write 
$
\tilde{p}^{(r, \ell)} \equiv \tilde{p}^{(r,\ell)}\bigl( N_a(I_{\ell}^{(r)}), N^{(r, \ell)} \bigr) := S_{\text{Bin}(\frac{1}{2}, N^{(r, \ell)})}\biggl( \biggl| N_a^{(r, \ell)} - \frac{N^{(r, \ell)}}{2} \biggr| \biggr)
$
as the p-value. We may then reject the local null $\bar{H}_0^{(r,\ell)}$ at level $\alpha \in (0,1)$ if $\tilde{p}^{(r, \ell)} \leq \alpha$. However, under the null and conditional on $N^{(r, \ell)}$, $\tilde{p}^{(r, \ell)}$ has a discrete distribution. We may thus gain additional power by randomizing $\tilde{p}^{(r, \ell)}$ so that its distribution is continuous and uniform under the null. To that end, we generate an independent $U \sim \text{Unif}[0,1]$, define $\tilde{S} := S_{\text{Bin}(\frac{1}{2}, N^{(r,\ell)})}\bigl( \bigl| N_a^{(r,\ell)} - \frac{N^{(r,\ell)}}{2} \bigr| + 1 \bigr)$, and define
$
\bar{p}^{(r,\ell)} = U \tilde{p}^{(r,\ell)} + (1 - U) \tilde{S}.
$
The randomized p-value $\bar{p}^{(r,\ell)}$ has the $\text{Unif}[0,1]$ distribution under $\bar{H}_0^{(r,\ell)}$. We also have that $\bar{p}^{(r,\ell)} \leq \tilde{p}^{(r,\ell)}$ so there is no loss in power (we give a proof in Proposition~\ref{prop:pval_transform} in the appendix for completeness).  

\subsubsection{Step 2: combining p-values of the same resolution level.}
\label{sec:step2_two_sample}

In the second step, for each $(s, j)$, we will consider each $r \in \{s, s+1, \ldots, R\}$ and combine the p-values
$
\{ \bar{p}^{(r, \ell)} \,:\, \ell \in L(s, j, r) \}.
$
In the case where $s = 0$ and $j = 1$ so that $I^{(s)}_j = \mathcal{X}$, this amounts to combining the p-values $\{ \bar{p}^{(r, \ell)} \}_{ \ell \in [2^r] }$ for each $r \in [R]$. To simplify exposition, we describe the p-value combination method for when $s = 0$ and $j = 1$; the same method applies immediately for any $(s,j)$. 

We combine the p-values $\{ \bar{p}^{(r, \ell)} \}_{\ell \in [2^r]}$ by specifying a function $f \,:\, [0,1]^{2^r} \rightarrow [0,1]$ and taking 
$
f\bigl( \bar{p}^{(r, 1)}, \ldots, \bar{p}^{(r, 2^r)} \bigr).
$
By choosing the combining function $f$ carefully and using the fact that $\{ \bar{p}^{(r,1)}, \ldots, \bar{p}^{(r, 2^r)}\}$ are independent random variables uniform on $[0,1]$, we can guarantee that $f\bigl( \bar{p}^{(r, 1)}, \ldots, \bar{p}^{(r, 2^r)} \bigr)$ has the uniform distribution under the null. There are a number of reasonable choices for $f(\cdot)$ but we focus on two:
\begin{align}
p_{\text{F}}^{(r)} &:= S_{\chi^2_{2^r}}\biggl( - 2 \sum_{\ell \in [2^r]} \log \bar{p}^{(r, \ell)} \biggr), \qquad \text{(Fisher combination)} \label{eq:fisher_combination} \\
p_{\text{M}}^{(r)} &:= S_{\beta_{2^r}}\biggl( \min_{\ell \in [2^r]} \bar{p}^{(r, \ell)} \biggr), \qquad \text{(Minimum combination)} \label{eq:min_combination}
\end{align}
where $S_{\chi^2_{2^r}}(\cdot) := \mathbb{P}( \chi^2_{2^r} \geq \cdot)$ is the right tail probability function for the $\chi^2$ distribution with $2^r$ degree of freedom and $S_{\beta_{2^r}}(\cdot) := \mathbb{P}( \text{Beta}_{1, 2^r} \leq \cdot)$ is the left tail probability of the Beta distribution with parameter $(1, 2^r)$. The fact that $p_F^{(r)}$ and $p_M^{(r)}$ have the uniform distribution under the null follows from the fact that the negative sum of logarithm of independent $\text{Unif}[0,1]$ random variables has the $\chi^2$ distribution and that the minimum of independent $\text{Unif}[0,1]$ random variables has the Beta distribution. For the remainder of this paper, we use Fisher combination by default but it would be trivial to use minimum combination instead.

We can follow the same procedure to compute, for any $(s, j)$ and $r \in \{s, s+1, \ldots, R\}$, the p-value $p_F^{(s, j, r)}$ which combines $\{ \bar{p}^{(r, \ell)} \}_{\ell \in L(s, j, r)}$. Moreover, we derive a dynamic program that, in the process of computing $p_F^{(1)},\ldots, p_F^{(R)}$ for the global null case ($s=0, j = 1$), can simultaneously and without any additional computational burden, compute the whole collection 
$
\bigl\{ \, \{ p_{F}^{(s, j, r)} \}_{r = s}^R \, \bigr\}_{s \in [R], j \in [2^r]}.
$
The dynamic program uses an iterative bottom-up approach and runs in time $O(N)$. We give the details in Algorithm~\ref{alg:all_local_null}. 

\begin{remark}
The question of which combination method has more power depends on what the alternative is. We show through our theoretical analysis in Section~\ref{sec:power_analysis} that when the integrated difference $\int (\frac{\lambda_a - \lambda_b}{\lambda})^2$ is large, then Fisher combination has higher power. On the other hand, if $|\lambda_a - \lambda_b|$ is large only on a small region and 0 elsewhere, then the minimum combination method has higher power. 
\end{remark}

\subsubsection{Step 3: combining across different resolution levels}
\label{sec:two_sample_step3}

Finally, to obtain the p-value for $H_0^{(s,j)}$, we combine $\{ p_F^{(s, j, r)} \}_{r=s}^R$ across resolution levels $\{s, s+1, \ldots, R\}$. There are again a number of choices, but we propose
\[
\tilde{p}^{(s, j)}_F := \min \bigl\{ p^{(s, j, r)}_F \,:\, r \in \{s, s+1, \ldots, R\} \bigr\}.
\]
Since the random variables $\{ p_F^{(s, j, r)} \}_{r=s}^R$ are not independent, the distribution of $\tilde{p}^{(s, j)}_F$ under $H_0$ is difficult to characterize exactly. Instead, we make adjustments to $\tilde{p}^{(s,j)}_F$. One straightforward way is to make the Bonferroni adjustment, where we let
\begin{align}
\check{p}_F^{(s, j)} = (R - s + 1) \cdot \tilde{p}_F.
\end{align}

We may also adjust $\tilde{p}_F$ by resampling. We note that the realizations $N_a(\cdot)$ and $N_b(\cdot)$ can be equivalently characterized by two sequences of random variables $X_1, X_2, \ldots X_N$ taking value on $\mathcal{X}$ and $M_1, M_2, \ldots, M_N$ taking value on $\{-1, 1\}$ where $N = N_a + N_b$ is the random length of the sequence. The occurrences of $N_a(\cdot)$ comprise of all $X_i$ where $M_i = -1$ and the occurrences of $N_b(\cdot)$ comprise of those points for which $M_i = 1$. 

Under $H_0$, $M_1,\ldots, M_N$ would be Rademacher random variables, that is, $\mathbb{P}_0(M_i = 1) = 1/2$, and independent of $X_1, \ldots, X_N$. Hence, to resample $B$ samples from $H_0$, we do the following, for $b^* = 1, 2, \ldots, B$:
\begin{enumerate}
\item Generate $M^{(b^*)}_1, \ldots, M^{(b^*)}_N \sim \text{Rademacher}$ independently.
\item Take $N^{(b^*)}_a(\cdot) = \{X_i \,:\, M^{(b^*)}_i = -1 \}$ and $N^{(b^*)}_b(\cdot) = \{X_i \,:\, M^{(b^*)}_i = 1\}$.
\end{enumerate}
On each sample $N^{(b^*)}_a(\cdot), N^{(b^*)}_b(\cdot)$, we then repeat steps 1, 2, and the first part of step 3 to compute $\tilde{p}^{(s, j)}_{F,\, b^*}$ for each $s \in [R]$ and $j \in [2^s]$. We may then define
\begin{align}
\check{p}^{(s, j)}_F = \frac{1}{B} \sum_{b^* = 1}^B \mathbbm{1}\bigl( \tilde{p}^{(s, j)}_{F, b^*} \leq \tilde{p}^{(s, j)}_{F}  \bigr).
\label{eq:p_check}
\end{align}

\subsubsection{Step 4: sequential/multiple testing adjustment}
\label{sec:two_sample_step4}

The p-values $\check{p}_F^{(s,j)}$'s produced from step 3 are individually valid in that under $H_0^{(s,j)}$ we have that $\check{p}_F^{(s,j)} \leq \alpha$ with probability at most $\alpha$. To account for sequential/multiple testing, we use the adjustment method proposed by \cite{meinshausen2008hierarchical}. For each $(s,j)$ where $s \in [R]$ and $j \in [2^s]$, define
\begin{align*}
\mathcal{L}(s, j) = 
\begin{cases}
\#\{\text{terminal nodes emanating from $(s,j)$ in $\bm{I}$}\} & \text{ if $(s,j)$ is not a terminal node} \\
2 & \text{ if $(s,j)$ is a terminal node}
\end{cases}
\end{align*}
where a terminal node in $\bm{I}$ is a region $I^{(r)}_\ell$ with no sub-region. If $\bm{I}$ is a full binary tree with $R$ resolution levels, then $\mathcal{L}(s,j) = 2^{R - s}$ if $s < R$ and $\mathcal{L}(R, j) = 2$. The total number of terminal regions is $\mathcal{L}(0,1)$ which, in the case of a full binary tree, is $2^R$. We then define the final adjusted p-value:
\begin{align}
p^{(s,j)}_F = \check{p}_F^{(s,j)} \cdot \frac{\mathcal{L}(0,1)}{\mathcal{L}(s,j)} = \check{p}_F^{(s,j)} \cdot 2^{s \wedge (R-1)}
\label{eq:final_pval}
\end{align}

We note in particular that the p-value $p_F^{(0, 1)}$ for the global null does not receive any adjustment so that any rejections we make of the local nulls $H_0^{(s, j)}$ comes "for free" on top of our test for the global null. In other words, conducting the tests for the local nulls does not decrease our power for the global null. 

Using the fact that under $H_0^{(s,j)}$, we have $\mathbb{P}(\check{p}^{(s,j)} \leq \alpha) \leq \alpha$, and Theorem 2 in \cite{meinshausen2008hierarchical}, the following FWER guarantee immediately follows:

\begin{theorem}
The rejection set $\mathcal{R}_\alpha$ formed via~\eqref{eq:rejection_rule} with p-values $\{ p^{(s,j)}_F \}_{s \in [R], j \in [2^s]}$, as defined in~\eqref{eq:final_pval}, has family-wise error rate (FWER) at most $\alpha$.
\end{theorem}

{\color{black} The overall procedure (Algorithm~\ref{alg:twosample}) is computationally efficient. The whole collection $\{ p_F^{(s, j, r)} \}$ can be computed in $O(N)$ time using Algorithm~\ref{alg:all_local_null} so that the overall procedure has runtime complexity $O(N B)$ where $B$ is the number of resampling repetitions.}

\begin{remark}
\label{rem:random_partition}
Recall that we can equivalently describe the two realizations $N_a(\cdot)$ and $N_b(\cdot)$ through a sequence (of random length) of positions $X_1, \ldots, X_N \in \mathcal{X}$ of the union of $N_a(\cdot)$ and $N_b(\cdot)$ and a set of markers $M_1, \ldots, M_N \in \{-1, +1\}$, where $M_i = -1$ implies that $X_i$ belong to $N_a(\cdot)$. We can then see that the type I error guarantee holds conditional on the aggregate positions $X_1, \ldots, X_N$ of the union of the two realizations. This is because the p-values $\bar{p}^{(r, \ell)}$ produced in step 1 for the discretized local nulls are valid conditional on the positions; we only use the fact that the random markers $M_1, \ldots, M_N$ are, conditionally on $N$, independent Rademacher random variables. An important implication of this fact is that the hierarchical partition $\bm{I}$ can depend on the aggregate positions $\{X_1, \ldots, X_N\}$ so long as it does not depend on $\{M_1, \ldots, M_n\}$; in particular, we can split each region such that each sub-region has equal number of "unmarked" points. {\color{black} In practice, we recommend choosing the hierarchical partition in this way and setting the maximum resolution level $R = O(\log n)$ such that each bin at level $R$ contains a constant number of points, say 10 or 20}. 
\end{remark}

\begin{remark}
The adjustment method in \cite{meinshausen2008hierarchical} can in fact be improved by looking at the test sequentially and removing any previously rejected hypotheses from the set of terminal nodes under consideration. This is analogous to how Holm's method improves upon Bonferroni method. We refer the readers to the excellent paper by  \cite{goeman2010sequential} for more detail. 
\end{remark}

{
\renewcommand{\baselinestretch}{1}
\begin{algorithm}[htp]
\caption{Computing simultaneously valid p-values $p_F^{(s,j)}$ for all $H_0^{(s,j)}$.}
\label{alg:twosample}
\begin{flushleft}
\textbf{INPUT:} Poisson process realizations $N_a(\cdot)$ and $N_b(\cdot)$ and a hierarchical partitioning $\bm{I} = \{ I^{(r)}_\ell \}_{r \in [R], \ell \in [2^r]}$ of the domain. \\
\textbf{OUTPUT:} Simultaneously valid p-values $p^{(s,j)}_F$ for each $H_0^{(s,j)}$.
\begin{algorithmic}[1]
\For{ each $r \in [R]$ }
  \For{ each $\ell \in [2^r]$} 
    \State Set $\tilde{p}^{(r, \ell)} = S_{\text{Bin}(\frac{1}{2}, N^{(r, \ell)})}\bigl( | N_a^{(r, \ell)} - N^{(r, \ell)}/2 | \bigr)$
    \State Use randomization described in Section~\ref{sec:two_sample_step1} to obtain $\bar{p}^{(r, \ell)}$.
  \EndFor
\EndFor
\State Apply Algorithm~\ref{alg:all_local_null} on $\{ \bar{p}^{(r, \ell)} \}_{r \in [R], \ell \in [2^r]}$ to obtain $\{ \{ p_F^{(s, j, r)} \}_{r=s}^R \}_{s \in [R], j \in 2^s}$.
\State Compute $\tilde{p}^{(s,j)}_F := \min \{ p_F^{(s, j, r)} \,:\, r \in \{s, s+1, \ldots, R\} \}$. 
\For { $b^* \in \{1,2,\ldots, B\}$ }:
  \State Generate $M^{(b^*)}_1, \ldots, M^{(b^*)}_N \sim \text{Rademacher}$ independently.
 \State Take $N^{(b^*)}_a(\cdot) = \{X_i \,:\, M^{(b^*)}_i = -1 \}$ and $N^{(b^*)}_b(\cdot) = \{X_i \,:\, M^{(b^*)}_i = 1\}$.
 \State Repeat lines 1 to 8 on $N_a^{(b^*)}(\cdot)$ and $N_b^{(b^*)}(\cdot)$ to obtain $\tilde{p}_{F, b^*}^{(s,j)}$.
 \EndFor
\State Compute the raw p-values $\check{p}^{(s,j)}_F := \frac{1}{B} \sum_{b^* = 1}^B \mathbbm{1}\{ \tilde{p}_{F, b^*}^{(s, j)} \leq \tilde{p}^{(s,j)}_F \} $.
\State Compute the adjusted p-values $p^{(s,j)}_F = \check{p}_F^{(s,j)} 2^{s \wedge (R-1)}$.
\end{algorithmic}
\end{flushleft}
\end{algorithm}

\begin{algorithm}[htp]
\caption{Dynamic program for computing the collection of p-values $\bigl\{ \, \{ p_{F}^{(s, j, r)} \}_{r = s}^R \, \bigr\}_{s \in [R], j \in [2^r]}$.}
\label{alg:all_local_null}
\begin{flushleft}
\textbf{INPUT:} a collection $\{ \bar{p}^{(r, \ell)} \}_{r \in [R], \ell \in [2^r]}$. \\
\textbf{OUTPUT:} The collection $\bigl\{ \, \{ p_{F}^{(s, j, r)} \}_{r = s}^R \, \bigr\}_{s \in [R], j \in [2^r]}$
\begin{algorithmic}[1]
\State For every $r \in [R], \ell \in [2^r]$, set $m_0^{r, \ell} = -2\log \bar{p}^{(r, \ell)}$ 
\For {$k \in \{1,2, \ldots, R-1\}$}:
    \State For every $r \in [R-k]$ and $\ell \in [2^r]$, set $m_k^{(r, \ell)} = m_{k-1}^{r+1, 2\ell-1} + m_{k-1}^{r+1, 2\ell}$.
\EndFor
\For{$s \in [R], j \in [2^s], r \in \{s, s+1, \ldots, R\}$}:
 \State Set $p_F^{(s, j, r)} = S_{\chi^2_{2^{r-s}}}(m_{r-s}^{r, j})$.
\EndFor
\end{algorithmic}
\end{flushleft}
\end{algorithm}
}

\section{Tests on longitudinal networks}
\label{sec:array_tests}

In this section, we can consider interaction processes among a group of $n$ individuals. For each pair of individuals $u, v \in [n]$, we write $N_{uv}(\cdot)$ as the realization that captures the interaction events between $u$ and $v$ over time. The collection $\{ N_{uv}(\cdot) \,:\, u,v \in [n] \}$ is therefore an array of point process realizations which we refer to as a longitudinal network. Here, we take the network to be symmetric/undirected in that $N_{uv}(\cdot) = N_{vu}(\cdot)$; we study the directed/asymmetric setting in Section~\ref{sec:asymmetric}. Suppose $N_{uv}(\cdot) \sim PP(\Lambda_{uv})$ for $\binom{n}{2}$ intensity measures $\{ \Lambda_{uv} \}$. We aim to test whether the intensity measures $\Lambda_{uv}$'s are all identical. When $n$ is large however, the space of alternative hypothesis is enormous so that it is important for us to designate a plausible alternative with which to test against. 

We therefore assume that there is an underlying block/community structure. More precisely, suppose each individual $u$ belongs to one of $K$ communities and write $\bm{\sigma}(u) \in [K]$ as the community membership of $u$, where $\bm{\sigma} \,:\, [n] \rightarrow [K]$ is the community assignment function. We assume that the probability distribution of the interactions between $u,v$ depends only on the community memberships of $u$ and $v$. More precisely, for each pairs of communities $s,t \in [K]$, let $\Gamma_{st}(\cdot)$ be an intensity measure and suppose 
\[
\Lambda_{uv} = \Gamma_{\bm{\sigma}(u)\bm{\sigma}(v)} \quad \text{for individuals $u,v \in [n]$.}
\]
We then define the null hypothesis to be $H_0 : K=1$ and the alternative to be $H_1 : K > 1$. More precisely, we define the null hypothesis
\begin{align}
\label{eq:symm-array-test}
H_0 \,:\, N_{uv}(\cdot) \sim \text{PP}(\Gamma), \text{ for some $\Gamma$, for all $u,v \in [n]$}. \qquad \text{(Symmetric array test)}
\end{align}

This is the generalization of the two-sample test to the array case. In many applications however, individuals may have different baseline rates of interactions. To capture potential rate heterogeneity, we propose to augment the block model with a vector $\bm{\theta} \in [0, \infty)^n$ of non-negative scalars and let
$
N_{uv}(\cdot) \sim \text{PP}( \Gamma_{\bm{\sigma}(u) \bm{\sigma}(v)} \theta_u \theta_v ).
$

We may then consider the same test of whether there exists a community structure in the interactions. An equivalent formulation is to define
\begin{align}
\label{eq:dc_array_test}
H_0 \,:\, N_{uv}(\cdot) \sim \text{PP}(\Gamma \theta_u \theta_v),  \text{ for some $\Gamma$ and $\bm{\theta}$, for all $u,v \in [n]$}.
\end{align}
We refer to~\eqref{eq:dc_array_test} as the \emph{degree-corrected array test}. In the next section, we focus on the symmetric array test and consider the degree-corrected array test in Section~\ref{sec:dc_array_test}.

\subsubsection*{Related work on longitudinal networks}

Recently, there has been increased attention on the modeling of dynamic networks \citep{holme2012temporal}. For example, \cite{xu2013dynamic} employ a state space model to describe temporal changes at the level of the connectivity pattern, \cite{dubois2013stochastic} introduced a family of relational event models that captures the heterogeneity in underlying interaction dynamics of network data over time. Modelling temporal interaction between two nodes by Poisson processes is also considered in \cite{corneli2016block} and \cite{matias2018semiparametric}, where they provide likelihood-based algorithms for membership estimation. \cite{zhangefficient} studies longitudinal networks from a tensor factorization perspective; they discretize the time into bins and propose an adaptive merging method to ensure that the discretized network is not too sparse. We refer the readers to the introduction in \cite{zhangefficient} for a more extensive review of estimation methods for longitudinal networks. Unlike estimation, testing for longitudinal networks has not received much attention. This is where our work enters the picture. 

\subsection{Testing procedure for longitudinal networks}
\label{sec:test_array_homogeneous}

We now consider the test of interaction processes among a group of individuals, defined in~\eqref{eq:symm-array-test}. As with the two-sample test, we construct our testing procedure based on a hierarchical partitioning $\bm{I} = \{ I^{(r)}_\ell\}$ of the support $\mathcal{X}$ of the interactions, as described in Definition~\ref{defn:dyadic_partition}. The test follows the same steps as that described in Section~\ref{sec:two_sample_test}. The only differences are in step 1, where we specify different test statistics for the discretized local null, and in step 4, where we specify different resampling algorithms.

Following the two-sample test described in Section~\ref{sec:two_sample_test}, we define, for a resolution level $r \in [R]$ and $\ell \in [2^r]$, the local null
\begin{align*}
H_0^{(r,\ell)} \,:\, \Lambda_{uv} = \Gamma, \text{ on $I^{(r)}_\ell$ for some common $\Gamma$, for all $u,v \in [n]$}.
\end{align*}

We also define discretized local null
\begin{align*}
\bar{H}_0^{(r, \ell)} \,:\, \Lambda_{uv}(I^{(r)}_\ell) = \gamma, \text{ for some common $\gamma \geq 0$, for all $u,v \in [n]$.}
\end{align*}


Our testing procedure follows the same steps as the two-sample test. 

\subsubsection{Step 1: compute p-values for each discretized local null.} 
\label{sec:array_homo_symm_step1}

In the first step, we test the discretized local null $\bar{H}_0^{(r,\ell)}$ for some $r \in [R]$ and $\ell \in [2^r]$. Define $N^{(r, \ell)}_{uv} := N_{uv}( I^{(r)}_\ell)$ and observe that $N^{(r, \ell)}_{uv} \sim \text{Poisson}( \Lambda_{uv}(I^{(r)}_\ell) )$. To motivate our test, define an integer matrix $A^{(r, \ell)} \in \mathbb{N}^{n \times n}$ where 
\begin{align}
A^{(r,\ell)}_{uv} = \begin{cases}
N^{(r,\ell)}_{uv}, & u \ne v \\
0, & u=v
\end{cases}. \label{eq:ad_matrix_local}
\end{align}

We view $A^{(r, \ell)}$ as the adjacency matrix of a weighted network. If the intensity measures $\{\Lambda_{uv} \}$ has a block structure in that $\Lambda_{uv} = \Gamma_{\bm{\sigma}(u) \bm{\sigma}(v)}$ where $\bm{\sigma}(u), \bm{\sigma}(v) \in [K]$ are the cluster membership of $u$ and $v$ (c.f. Section~\ref{sec:array_tests}), then $A^{(r, \ell)}$ is a random matrix that follows a Poisson Stochastic Block Model (SBM). To be precise, define a matrix $\bm{\gamma} \in \mathbb{R}^{K \times K}$ where, for $s,t \in [K]$, we have $\bm{\gamma}_{st} = \Gamma_{st}(I^{(r)}_\ell)$. For $u \neq v$, we then have
$
A^{(r, \ell)}_{uv} \sim \text{Poisson}( \bm{\gamma}_{\bm{\sigma}(u) \bm{\sigma}(v)}).
$

Without loss of generality, we may assume that no clusters are empty and that the rows of $\bm{\gamma}$ are distinct. Now let $\bm{P}$ be a $n \times n$ matrix given by 
\begin{align}
\label{eq:nodes_prob_matrix}
    \bm{P}_{uv} = \bm{\gamma}_{\sigma(u)\sigma(v)}
\end{align}
We can then see that $\mathbb{E}_{\bm{\gamma}, \sigma} \bigl[ A \bigr] =\bm{P}-\text{diag}(\bm{P})$ and now for each $r \in [R]$ and $\ell \in [2^r]$, we can restate the \emph{discretized local null} as
\begin{align}
\label{eq:local_homogeneous_SBM_test}
\bar{H}^{(r,\ell)}_0 : \bm{P} =  \gamma^{(r,\ell)} \mathbf{1}_n \mathbf{1}_n^\mathsf{T} \quad \text{for some constant $\gamma^{(r,\ell)} > 0$}
\end{align}
where $\mathbf{1}$ a vector of all ones of length $n$. 

Given an observed adjacency matrix $A^{(r,\ell)}$, an intuitive idea for the goodness-of-fit test is to remove the signal using an estimate of the true mean $\gamma^{(r,\ell)}$ and test whether the residual matrix is a noise matrix. Let $\hat{\gamma}^{(r,\ell)} = \frac{2}{n^2-n}\sum_{u<v}A^{(r,\ell)}_{uv}$ be an estimator of the true Poisson mean, we denote the empirically centered and re-scaled adjacency matrix by $\tilde{A}^{(r,\ell)}$
\begin{align}
\label{eq:empirical_ad}
\tilde{A}^{(r,\ell)}_{uv}:=\begin{cases} 
\frac{A^{(r,\ell)}_{uv}-\hat{\gamma}^{(r,\ell)}}{\sqrt{(n-1)\hat{\gamma}^{(r,\ell)}}}, & u\neq v, \\
0, & u=v.
\end{cases}
\end{align}
The asymptotic distribution of the extreme eigenvalues of the empirically centered and re-scaled adjacency matrix has been studied in \cite{bickel2013hypothesis} and \cite{lei2016goodness} while the entries are Bernoulli. We extend their result to the case with Poisson distributed edges.
\begin{theorem}
\label{thm:tracy-widom}
For each $r \in [R], \ell \in [2^r]$, Let $A^{(r,\ell)}$ be the adjacency matrix generated from a Poisson Stochastric Block Model and $\tilde{A}^{(r,\ell)}$ be defined as in~\eqref{eq:empirical_ad}, then under the local null hypotheses~\eqref{eq:local_homogeneous_SBM_test} and as $n \rightarrow \infty$, we have
$
n^{2/3}(\lambda_{1}(\tilde{A}^{(r,\ell)}) - 2) \stackrel{d}{\longrightarrow} \text{TW}_1,
$
where $\text{TW}_1$ is the Tracy--Widom law with $\beta = 1$. 
\end{theorem}

We relegate the proof of Theorem~\ref{thm:tracy-widom} to Section~\ref{sec:tracy-widom-proof} of the Appendix. In Section~\ref{sec:eigen_dist_alt}, we also characterize the behavior of $\lambda_1(\tilde{A}^{(r,\ell)})$ under the alternative setting where $K > 1$, showing that $\lambda_1(\tilde{A}^{(r,\ell)})$ diverges as $n$ increases. 

Theorem~\ref{thm:tracy-widom} shows that, if we take $\lambda_1(A^{(r,\ell)})$ to be the test statistics for the local null $\bar{H}_0^{(r,\ell)}$, we may obtain the asymptotically valid p-value using the Tracy-Widom distribution. Denote $F_{\scalebox{1}{$\scriptscriptstyle {\text{TW1}}$}}(\cdot)$ as the CDF of the Tracy-Widom law with $\beta = 1$ and define the local p-value
\[
\bar{p}^{(r, \ell)} \equiv \bar{p}^{(r,\ell)}\bigl(A^{(r,\ell)}\bigr) := 2\text{min}\bigg(F_{\scalebox{1}{$\scriptscriptstyle \text{TW1}$}}\Big(n^{2/3}\big(\lambda_1(A^{(r,\ell)})) - 2\big)\Big), 1-F_{\scalebox{1}{$\scriptscriptstyle \text{TW1}$}}\Big(n^{2/3}\big(\lambda_1(A^{(r,\ell)})) - 2\big)\Big)\bigg)
\]
In the same manner, we could then reject the discretized local null $\bar{H}_0^{(r,\ell)}$ at level $\alpha\in(0,1)$ if $p_0^{(r, \ell)}\leq \alpha$. The Tracy--Widom law is asymptotic, but we will perform resampling in step 3 so that the final p-values that we produce are still valid for any finite $n$. 

{\color{black}
We note that when the resolution level is very high, the discretized networks may be very sparse and thus any tests on them may have low power. This presents no intrinsic difficulty since we aggregate results from multiple resolution levels. In practice, we recommend setting $R$ so that all the discretized networks are connected. 
}

\subsubsection{Step 2: combining p-values of the same resolution level}

We follow exactly the same procedure described in Section~\ref{sec:step2_two_sample} to obtain $p_F^{(s, j, r)}$ (or $p_M^{(s, j, r)}$) for every $s \in [R]$, $j \in [2^s]$, and $r \in \{s, s+1, \ldots, R\}$.

\subsubsection{Step 3: combining across different resolution levels}
\label{sec:symmetric_resample_method}
We again define $\tilde{p}^{(s,j)}_F = \min \{ p_F^{(s, j, r)} \,:\, r \in \{s, s+1, \ldots, R\} \}$. We can make a Bonferroni adjustment just as before but we propose to adjust with resampling.

Unlike the two-sample test setting where we only have two realizations, here, we have ${n \choose 2}$ realizations, denoted as $\{N_{uv}(\cdot): u,v\in[n], u < v\}$. They could also be equivalently characterized by two sequences of random variables $X_1, X_2, \dots, X_N \in \mathcal{X}$ and random tuples $M_1, M_2, \dots, M_N$ taking value on $ \binom{[n]}{2} := \{(u,v): u,v\in[n], u<v\}$, where $N = \sum_{u < v} N_{uv}$ is the random length of the sequence. 
Under the global null $H_0$, $M_1, M_2, \dots, M_N$ are multinomial random tuples distributed uniformly over the set $\binom{[n]}{2}$, that is, $\mathbb{P}_0 \{M_i=(u,v)\} = \frac{2}{n(n-1)}$ for each $i \in [N]$ and $(u, v) \in \binom{[n]}{2}$. Thus we could resample B samples, $b^* = 1,2,\dots, B$, from $H_0$ in the following way:
\begin{enumerate}
	\item Generate $M^{(b^*)}_1, \ldots, M^{(b^*)}_N \sim \text{Uniform}\binom{[n]}{2}$ independently.
	\item Take $N_{uv}^{(b^*)}(\cdot) = \{X_i \,:\, M_i^{(b^*)} = (u,v)\}$ for each pair of $(u,v) \in \binom{[n]}{2}$.
\end{enumerate}
On each sample collection $\{N^{(b^*)}_{uv}(\cdot): u,v\in[n], u < v\}$, we could then compute the simulated unadjusted p-value $\tilde{p}^{(s, j)}_{F, b^*}$ as in Section~\ref{sec:two_sample_step3} and output $\check{p}^{(s,j)}_F := \frac{1}{B} \sum_{b^* = 1}^B \mathbbm{1}\bigl( \tilde{p}^{(s, j)}_{F, b^*} \leq \tilde{p}^{(s,j)}_F \bigr)$ as the individually valid p-value for $H_0^{(s,j)}$. 

\subsubsection{Step 4: sequential/multiple testing adjustment}
\label{sec:symmetric_array_step4}

Following Section~\ref{sec:two_sample_step4}, we define $p^{(s,j)}_F = \check{p}^{(s,j)}_F \cdot 2^{s \wedge (R-1)}$.

\subsection{Degree-corrected symmetric array test}
\label{sec:dc_array_test}

In this setting, we define the local null at $r \in [R]$ and $\ell \in [2^r]$ as
\begin{align*}
H^{(r, \ell)}_0 \,:\, \Lambda_{uv} = \Gamma \theta_u \theta_v, \, \text{ on $I^{(r)}_\ell$ for some $\Gamma, \bm{\theta}$, for all $u,v \in [n]$},
\end{align*}
where $\bm{\theta} \in [0, \infty)^n$ is a vector of baseline rate for each of the $n$ individuals. We define the discretized local null as
$\bar{H}^{(r, \ell)}_0 \,:\, \Lambda_{uv}(I^{(r)}_\ell) = \gamma \theta_u \theta_v$, for some $\gamma, \bm{\theta}$, for all $u,v \in [n]$.

\begin{remark}
\label{rem:no_reverse_logic}

Note that in our definition, we do not require two local nulls $H_0^{(r, \ell)}$ and $H_0^{(r', \ell')}$ to have the same degree correction parameter $\bm{\theta}$. This means that, although we have the logical implication that $H_0^{(r, \ell)} \Rightarrow H_0^{(r+1, 2\ell-1)} \cap H_0^{(r+1, 2\ell)}$ (each local null implies its children), we do not have the reverse: if $H_0^{(r, \ell)}$ is false, it may still be that both $H_0^{(r+1, 2\ell-1)}$ and $H_0^{(r+1, 2\ell)}$ are true. This implies that we may have no power against certain alternatives. This issue is difficult to overcome completely but it can be ameliorated by performing the test with different choices of the hierarchical partition. 
\end{remark}

\subsubsection{Step 1: compute p-value for each discretized local nulls}

For $r \in [R]$ and $\ell \in [2^r]$, define $N^{(r, \ell)}_{uv} := N_{uv}( I^{(r)}_\ell)$ as with Section~\ref{sec:test_array_homogeneous} and define the adjacency matrix $A^{(r, \ell)}_{uv} = N^{(r, \ell)}_{uv}$ for $u \neq v $ and $A^{(r,\ell)}_{uv} = 0$ if $u = v$. 

Suppose the intensity measures $\{ \Lambda_{uv} \}$ has a block structure so that $\Lambda_{uv} = \Gamma_{\sigma(u) \sigma(v)}$ where $\sigma(u), \sigma(v) \in [K]$ are the cluster memberships of individual $u$ and $v$.
In the setting where each individual $u \in [n]$ has its own baseline rate of interactions $\theta_u > 0$, we have that
\begin{align*}
A^{(r, \ell)}_{uv} \sim \text{Poisson}( \bm{\gamma}_{\sigma(u) \sigma(v)} \theta_u \theta_v),
\end{align*}
where $\bm{\gamma}_{st} := \Gamma_{st}(I^{(r)}_\ell)$. This model is similar to the so-called degree-corrected stochastic block model (DCSBM) where the edges are Bernoulli distributed binary random variables instead of Poisson random integers as we have in our setting.

To test each discretized local null, we apply the Signed Triangle (SgnT) and the Signed Quadrilateral (SgnQ) statistics introduced and analyzed by \cite{jin2021optimal}. To define the SgnT and SgnQ statistic, first
define a vector $\hat{\eta}^{(r, \ell)}$  and a scalar $V^{(r, \ell)}$ as
\begin{align}
\hat{\eta}^{(r, \ell)} = \Big(1/V^{(r, \ell)}\Big)^{\frac{1}{2}} A^{(r, \ell)} \mathbf{1}_n, \qquad \text{where} \ V^{(r, \ell)} = \mathbf{1}_n^{'}A^{(r, \ell)} \mathbf{1}_n.
\end{align}

The Signed-Polygon(SgnT) statistic $T_n$ is then defined as
\begin{equation}
\label{equation:sgnt_defn}
  \begin{aligned}
T^{(r, \ell)} \equiv T(I_\ell^{(r)}) &= \sum_{\substack{u_1, u_2, u_3 \in [n] \\ u_1 \ne u_2 \ne u_3}}\big(A_{u_1 u_2}^{(r, \ell)} - \hat{\eta}_{u_1}^{(r, \ell)} \hat{\eta}_{u_2}^{(r, \ell)} \big) \cdot \big(A_{u_2 u_3}^{(r, \ell)} - \hat{\eta}_{u_2}^{(r, \ell)} \hat{\eta}_{u_3}^{(r, \ell)} \big)\\
&\quad \cdot \big(A_{u_3 u_1}^{(r, \ell)} - \hat{\eta}_{u_3}^{(r, \ell)} \hat{\eta}_{u_1}^{(r, \ell)} \big) 
  \end{aligned}  
\end{equation}
In a similar manner, we define the Signed-Quadrilateral(SgnQ) statistic as
\begin{equation}
\label{equation:sgnq_defn}
  \begin{aligned}
Q^{(r, \ell)} \equiv Q(I^{(r)}_{\ell}) &= \sum_{\substack{u_1, u_2, u_3, u4 \in [n] \\ u_1 \ne u_2 \ne u_3 \ne u_4}}
\big(A_{u_1 u_2}^{(r, \ell)} - \hat{\eta}_{u_1}^{(r, \ell)} \hat{\eta}_{u_2}^{(r, \ell)} \big)
\big(A_{u_2 u_3}^{(r, \ell)} - \hat{\eta}_{u_2}^{(r, \ell)} \hat{\eta}_{u_3}^{(r, \ell)} \big) \\
&\qquad \cdot \big(A_{u_3 u_4}^{(r, \ell)} - \hat{\eta}_{u_3}^{(r, \ell)} \hat{\eta}_{u_4}^{(r, \ell)} \big)
\big(A_{u_4 u_1}^{(r, \ell)} - \hat{\eta}_{u_4}^{(r, \ell)} \hat{\eta}_{u_1}^{(r, \ell)} \bigr)
  \end{aligned}  
\end{equation}

The intuition behind the SgnT and SgnQ test statistics is that a network with community structure tend to have more triangles and quadrilaterals than a network with similar number of edges but without community structure. We refer the readers to \cite{gao2017testing, jin2018network, jin2021optimal} for a more detailed discussion. 

Theorem 2.1 and 2.2 in \cite{jin2021optimal} prove asymptotic normality for $T^{(r, \ell)}$ and $Q^{(r, \ell)}$ for degree corrected stochastic block model where the edges are Bernoulli (they actually prove it for the more general mixed membership model). However, a careful examination of their proof shows that their result applies, without modification, to the setting where the edges are Poisson. We restate their result below; see Section~\ref{sec:dc_array_simulation} for experimental validation.

\begin{proposition} (follows from Theorem 2.1 and 2.2 in \cite{jin2021optimal})
\label{prop:signed_polygon} 
Suppose $\| \theta \|_2 \rightarrow \infty, \|\theta\|_\infty \rightarrow 0, \frac{ \| \theta\|_2^2 }{\|\theta \|_1} \sqrt{ \log \| \theta\|_1 } \rightarrow 0$ as $n \rightarrow \infty$, then
\begin{align}
\frac{T^{(r, \ell)}}{\sqrt{6} (\| \hat{\eta}^{(r, \ell)}\|^2 - 1 )^{3/2} } \stackrel{d}{\longrightarrow} N(0,1)  \\
\text{and } \,\, \frac{Q^{(r, \ell)} - 2 (\| \hat{\eta}^{(r, \ell)} \|^2 - 1 )^2}{\sqrt{8} (\|\hat{\eta}^{(r, \ell)}\|^2 - 1)^2} \stackrel{d}{\longrightarrow} N(0,1)
\end{align}
\end{proposition}

Based on Proposition~\ref{prop:signed_polygon}, we could take $T^{(r, \ell)}$ or $Q^{(r,\ell)}$ as our test statistic for the discretized local null $\bar{H}_0^{(r,l)}$, define the corresponding local p-value as
\begin{align*}
\bar{p}^{(r, \ell)} \equiv \bar{p}^{(r,\ell)}\bigl(T^{(r, \ell)} \bigr) &:= 2\biggl[ 1-\Phi\biggl( \biggl| \frac{T^{(r, \ell)}}{\sqrt{6} (\| \hat{\eta}^{(r, \ell)}\|^2 - 1)^{3/2} }
\biggr| \biggr) \biggr] \\
\bar{p}^{(r, \ell)} \equiv \bar{p}^{(r,\ell)}\bigl( Q^{(r,\ell)}\bigr) 
&:= 2 \biggl[ 1 - \Phi \biggl( \biggl| 
\frac{Q^{(r, \ell)} - 2 (\| \hat{\eta}^{(r, \ell)} \|^2 - 1 )^2}{\sqrt{8} (\|\hat{\eta}^{(r, \ell)}\|^2 - 1)^2} 
\biggr| \biggr) \biggr].
\end{align*}

\subsubsection{Step 2: computing p-values of the same resolution level}

We follow exactly the same procedure described in Section~\ref{sec:step2_two_sample} to obtain $p_F^{(s, j, r)}$ (or $p_M^{(s, j, r)}$) for every $s \in [R]$, $j \in [2^s]$, and $r \in \{s, s+1, \ldots, R\}$.

\subsubsection{Step 3: combine across different resolution levels}
\label{sec:sampling_dc_network}

We again define $\tilde{p}^{(s,j)}_F = \min \{ p_F^{(s, j, r)} \,:\, r \in \{s, s+1, \ldots, R\} \}$. We can make a Bonferroni adjustment just as before but we propose to adjust with resampling. The added challenge here however is that we do not observe the degree correction parameter vector $\bm{\theta}$. 

To describe the resampling procedure, we again characterize the $\binom{n}{2}$ realizations $\{ N_{uv}(\cdot) \,:\, u, v \in [n], u < v\}$ as a random sequence $X_1, \ldots, X_N \in \mathcal{X}$ and $M_1, \ldots, M_N$ taking value on $\binom{[n]}{2}$. Then, under $H_0$, we have
\begin{align}
\mathbb{P}_0 \{ M_i=(u,v) \}=\frac{\theta_{u}\theta_{v}}{\sum_{u'<v'}\theta_{u'}\theta_{v'}} 
\label{eq:M_prob1}
\end{align}
We cannot directly use~\eqref{eq:M_prob1} to resample $M_1, \ldots, M_N$ since we do not observe $\bm{\theta}$. To overcome this problem, we condition on the degree of all the individuals, which is a sufficient statistic for $\bm{\theta}$. To be precise, We write $A^{(0)}_{uv} := N_{uv}(\mathcal{X})$ as the total number of interactions between $u$ and $v$ and define
$
D_u \equiv D_u(A^{(0)}) := \sum_{v \neq u} A^{(0)}_{uv}
$
as the degree of individual $u$. Equivalently, we may express $D_u$ as a function of $M_1,\ldots, M_N$ via the equation $D_u(M) = \sum_{i=1}^N \mathbbm{1}\{ u \in M_i\}$. Write $\bm{m}:=\{(u_1,v_1),\ldots,(u_N,v_N)\}$ as a possible outcome for $M_1, \ldots, M_N$ and write $\bm{D}(\bm{m}) := \{ D_u(\bm{m}) \}_{u \in [n]}$ as the corresponding vector of all the degrees, then
\begin{align*}
\mathbb{P}_0\bigl(\{M_1,\ldots,M_N\}=\bm{m}\bigr)=\frac{\prod_{i=1}^{N}\theta_{u_i}\theta_{v_i}}{(\sum_{u'<v'}\theta_{u'}\theta_{v'})^N}
=\frac{\prod_{u\in[n]}(\theta_u)^{D_u(\bm{m})}}{(\sum_{u'<v'}\theta_{u'}\theta_{v'})^N}.
\end{align*}

For a vector $\bm{d} \in \mathbb{N}^N$, define $\mathcal{M}_{\bm{d}} := \bigl\{ \bm{m} = \bigl((u_1, v_1), \ldots, (u_N, v_N) \bigr) \,:\, \bm{D}(\bm{m}) = \bm{d} \bigr\}$ as the set of all possible outcomes of $M_1,\ldots,M_N$ that result in degree vector $\bm{d}$. Then, we have that
\begin{align*}
\mathbb{P}_0\bigl(\{M_1,\ldots,M_N\}=\bm{m}\,|\, \bm{D}(\bm{m}) = \bm{d}\bigr)= \frac{\mathbb{P}_0\bigl(\{M_1, \ldots, M_N\} = \bm{m}\bigr)}
{\sum_{\bm{m'}\in\mathcal{M}_{\bm{d}}} \mathbb{P}_0\bigl(\{M_1, \ldots, M_N\} = \bm{m}\bigr) }
=\frac{1}{|\mathcal{M}_{\bm{d}}|}.
\end{align*}
Importantly, conditional on the degree, the distribution of $M_1, \ldots, M_N$ does not depend on $\bm{\theta}$. Therefore, writing $\bm{D}^{\text{obs}}$ as the observed degree vector, we propose to generate $b^* = 1, \ldots, B$ Monte Carlo samples $M_1^{(b*)},\ldots, M_N^{(b*)}$ from the conditional distribution 
\begin{align}
\mathbb{P}_0\bigl( \{M_1,\ldots,M_N\} = \cdot \,|\, \bm{D}(\cdot) = \bm{D}^{\text{obs}} \bigr). \label{eq:cond_distr_deg}
\end{align}

To generate from~\eqref{eq:cond_distr_deg}, we use a Metropolis--Hastings algorithm. Given current state $\bm{m} = \{ (u_1, v_1), \ldots, (u_N, v_N) \}$, we generate a proposal $\bm{m}'$ by choosing a pair $i, j \in [n]$ where $i < j$. Denote $m_i = (u_i, v_i)$ and $m_j = (u_j, v_j)$. We then generate $m_i', m_j'$ by drawing uniformly from each of the following five outcomes:
\begin{align}
\begin{psmallmatrix} m'_i \\ m'_j \end{psmallmatrix} \sim 
\text{Unif} \left\{ 
\begin{psmallmatrix} u_j & v_j \\ u_i & v_i \end{psmallmatrix},\, 
\begin{psmallmatrix} u_i & v_j \\ u_j & v_i \end{psmallmatrix},\, 
\begin{psmallmatrix} u_j & v_i \\ u_i & v_j \end{psmallmatrix},\, 
\begin{psmallmatrix} u_i & u_j \\ v_i & v_j \end{psmallmatrix},\, 
\begin{psmallmatrix} v_i & v_j \\ u_i & u_j \end{psmallmatrix}
\right\}.  \label{eq:MH_proposal}
\end{align}
We complete the proposal by letting $m'_k = m_k$ for every $k \neq i, j$. If $m'_i$ or $m'_j$ contain multi-edge, that is, both end-points of $m'_i$ or $m'_j$ refer to the same node, then reject $m'$. Otherwise, it is straightforward to verify that the Metropolis--Hastings ratio is exactly 1 and we accept the proposal $m'$. 

\begin{proposition}
\label{prop:mcmc_ergodicity}
The Markov Chain specified via~\eqref{eq:MH_proposal} and the acceptance rule given below~\eqref{eq:MH_proposal} is ergodic and has stationary distribution that is uniform on $\mathcal{M}_{\bm{d}}$. 
\end{proposition}

The simpler version of Proposition~\ref{prop:mcmc_ergodicity} for contingency tables is well known (see e.g. \cite{diaconis1998algebraic}). We give a proof for the longitudinal network setting in Section~\ref{sec:mcmc_ergodicity_proof} of the appendix.

We may thus generate $M_1^{(b*)}, \ldots, M_N^{(b*)}$ by taking some steps of the Metropolis--Hastings algorithm and obtain our resampled point process realizations $\{ N_{uv}^{(b^*)}(\cdot)\}_{u,v}$. For each $b^*$, we may then obtain $\tilde{p}^{(s,j)}_{F, b^*}$ and construct the final p-value as
$
\check{p}^{(s,j)}_F = \frac{1}{B} \sum_{b^* = 1}^B \mathbbm{1}\bigl\{ \tilde{p}^{(s,j)}_{F, b^*} \leq \tilde{p}^{(s,j)}_F \bigr \}.
$

\noindent \textbf{Step 4: sequential/multiple testing adjustment.}

We make the sequential/multiple testing adjustment on $\check{p}_F^{(s, j)}$ to obtain simultaneously valid $p_F^{(s,j)}$ just as in Section~\ref{sec:two_sample_step4}, with one key difference because we do not have the logical implication that $H_0^{(s+1, 2j-1)} \cap H_0^{(s+1, 2j)} \Rightarrow H_0^{(s, j)}$, that is, if a local null is false, it could still be that the two children are true. As a consequence, we must redefine $\mathcal{L}(s, j)$ as
\begin{align*}
\mathcal{L}(s, j) = 
\begin{cases}
\#\{\text{terminal nodes emanating from $(s,j)$ in $\bm{I}$}\} & \text{ if $(s,j)$ is not a terminal node} \\
1 & \text{ if $(s,j)$ is a terminal node.}
\end{cases}
\end{align*}
In the case where the hierarchical partitions $\bm{I}$ is a full binary tree, we have that $\mathcal{L}(s, j) = 2^{R - s}$ and $\mathcal{L}(0, 1) = 2^R$. We then define the simultaneously valid p-value as
$
p_F^{(s,j)} = \check{p}_F^{(s,j)} \frac{\mathcal{L}(0,1)}{\mathcal{L}(s,j)} = 2^s \check{p}_F^{(s,j)}.
$
If we form our rejections via~\eqref{eq:rejection_rule} using $\{p_F^{(s,j)} \}$, then our FWER is controlled at $\alpha$. We succinctly summarize all the steps in Algorithm~\ref{alg:dc_graph}.

\section{Theoretical analysis}
\label{sec:power_analysis}

In this section, we analyze the power of our proposed test under the two-sample test setting where we have two intensities functions $\lambda_a$ and $\lambda_b$; we write $\lambda = \lambda_a + \lambda_b$. Under the global null, we suppose that $\lambda_a = \lambda_b = \frac{\lambda}{2}$ on the support $\mathcal{X}$. We state our results in terms of power against the global null but they are applicable to local nulls as well. Throughout this section, we let $\nu$ be the base measure with respect to which the $\lambda_a, \lambda_b$ are defined; one can assume $\nu$ is the Lebesgue measure for simplicity. We also take the partitions $\bm{I}$ to be fixed and state our results in terms of deterministic conditions on $\bm{I}$.

Our results are of the following form: under an alternative hypothesis where $\lambda_a$ and $\lambda_b$ are sufficiently separated according to some notion of distance, our proposed tests at level $\alpha$ will have power at least $\beta$. More precisely, writing $p$ as the p-value that we output, we have that 
$
\mathbb{P}_0(p \leq \alpha) \leq \alpha \text{  and  } \mathbb{P}(p \leq \alpha) \geq 1 - \beta.
$
Our first two results consider the Fisher combination test. For a given hierarchical partition $\bm{I}$ and a resolution level $r$, it measures the separation between $\lambda_a$ and $\lambda_b$ in terms of a discretized $L_2$ divergence. 

\begin{theorem}
\label{thm:fisher_power}
Let $\alpha, \beta \in (0,1)$ and let $C$ be a universal constant greater than $1$ whose value is specified in the proof. Assume $\int_{I_\ell^{(R)}} \lambda d\nu \geq 2$ for all $\ell \in [2^R]$. Assume there exists $r \in [R]$ such that 
\begin{align}
\frac{1}{4} \sum_{\ell=1}^{2^r} \biggl( \frac{ \int_{I^{(r)}_\ell} \lambda_a - \lambda_b \, d\nu }{ \int_{I^{(r)}_\ell} \lambda \, d\nu } \biggr)^2 \int_{I^{(r)}_\ell} \lambda d\nu \geq 2^{r/2} \biggl( \frac{C^{1/2}}{\beta} + 2 \log^{1/2} \frac{R}{\alpha} \biggr) +2 \log \frac{R}{\alpha}. \label{eq:special_r}
\end{align}

Then, we have that 
$
\mathbb{P}( p^{(0,1)}_F \leq \alpha) \geq 1 - 2\beta.
$
\end{theorem}

We prove Theorem~\ref{thm:fisher_power} in Section~\ref{sec:fisher_power_proof} of the Appendix. To better understand the implications of Theorem~\ref{thm:fisher_power}, we next take the support $\mathcal{X}$ to be a compact subset of $\mathbb{R}^q$ and take the hierarchical partition $\bm{I}$ to be any partition such that each split divides a region into two sub-regions whose volume is halved and whose diameter is reduced by a factor of $O(2^{-\frac{1}{q}})$. Since there is no fixed sample size for Poisson processes, we write $n := \int_{\mathcal{X}} \lambda d \nu$ so that it is the "expected sample size"; note then that $\frac{\lambda}{n}$ is a probability measure.

\begin{theorem}
\label{thm:fisher_power_smooth}
Let $\mathcal{X} \subset \mathbb{R}^q$ and let $\lambda = \lambda_a + \lambda_b$. Suppose $0 < c_{\min} := \inf_{x \in \mathcal{X}} \lambda(x) \leq \sup_{x \in \mathcal{X}} \lambda(x) =: c_{\max} < \infty$. Let $n := \int_{\mathcal{X}} \lambda \, d \nu$ and suppose $\frac{ \lambda_a - \lambda_b }{\lambda}$ is $\gamma$-Holder continuous:
\[
\biggl| \frac{\lambda_a(x) - \lambda_b(x)}{\lambda(x)} - \frac{\lambda_a(y) - \lambda_b(y)}{\lambda(y)} \biggr| \leq C_H \| x - y\|^\gamma_2 \qquad \text{ for all $x, y \in \mathcal{X}$}.
\]
Let $R = \lfloor \log_2 \frac{n}{2} - \log_2 \bigl( \frac{c_{\max} }{ c_{\min} } \bigr) \rfloor$ and $\{ I_l^{(r)} \}_{r \in [R], l \in [2^r] }$ be a dyadic partition of $\mathcal{X}$ such that for all $r \in [R]$ and $l \in [2^r]$, $\nu( I_l^{(r)} ) = \frac{\nu(I)}{2^r}$ and $\text{diam}(I_l^{(r)}) \leq C_{d} 2^{-r/q}$. Let $\alpha, \beta \in (0,1)$ and suppose
\begin{align}
\int_{\mathcal{X}} \biggl( \frac{\lambda_a - \lambda_b}{\lambda} \biggr)^2 \frac{\lambda}{n} \, d \nu \geq \biggl\{
 \begin{array}{cc}
 C_1 n^{- \frac{4\gamma}{q + 4\gamma}} \bigl( \beta^{-1} + \log \frac{\log n}{\alpha} \bigr) & \text{ if $\gamma/q \geq 1/4$} \\
 C_1 n^{- \frac{2\gamma}{q} } \bigl( \beta^{-1} + \log \frac{\log n}{\alpha} \bigr) & \text{ if $\gamma/q \leq 1/4$ } 
 \end{array} 
 \biggr. , \label{eq:strong_signal} 
\end{align}  
where $C_1 > 0$ depends only on $c_{\max}/c_{\min}$, $C_H$, and $C_d$. Then, 
$
\mathbb{P}( p_F^{(0,1)} \leq \alpha) \geq 1 - 2\beta.
$
\end{theorem}

We prove Theorem~\ref{thm:fisher_power_smooth} in Section~\ref{sec:fisher_power_smooth_proof} of the Appendix. It is important to note that Theorem~\ref{thm:fisher_power_smooth} only requires the difference $\frac{\lambda_a - \lambda_b}{\lambda}$ to be Holder continuous; no smooth assumptions are made on the individual intensity functions themselves. {\color{black} Moreover, because we use a hierarchical partition instead of a fixed resolution level, our test is able to adapt to the unknown smoothness $\gamma$ of the underlying function $\frac{\lambda_a - \lambda_b}{\lambda}$, that is, it attains the separation rate \eqref{eq:strong_signal} without knowledge of $\gamma$. }

\begin{remark}
As a direct consequence of Theorem~\ref{thm:fisher_power_smooth}, we see that when $\mathcal{X} \subseteq \mathbb{R}$ (so that $q=1$) and when $\frac{\lambda_a - \lambda_b }{\lambda}$ is Lipschitz (so that $\gamma = 1$), then our test has nontrivial power when the separation $\int ( \frac{\lambda_a - \lambda_b}{\lambda} )^2 \frac{\lambda}{n} d\nu$ (which is the squared $L_2(\lambda/n)$ distance between $\lambda_a/\lambda$ and $\lambda_b/\lambda$) is of order $n^{-\frac{4}{5}}$. This matches, up to log-factor, the lower bound on the minimax separation rate given by \cite{fromont2011adaptive} (see Section 2 within), showing that our test has minimax optimal power up to log-factors in this setting. We conjecture that when $\frac{\lambda_a - \lambda_b }{\lambda}$ is Lipschitz, our test is minimax optimal when $q \leq 4$ and suboptimal when $q > 4$.  
\end{remark}

Next, we consider the testing procedure that combines all the p-values of the same resolution level by taking the minimum instead of using the Fisher combination function. The next result is the analog of Theorem~\ref{thm:fisher_power}. However, the separation strength between $\lambda_a$ and $\lambda_b$ is measured by taking the maximum among the regions rather than taking the sum. 

\begin{theorem}
\label{thm:min_power}
Let $\alpha, \beta \in (0,1)$ and let $C$ be a universal constant whose value is specified in the proof. Assume that $\int_{I^{(R)}_l} \lambda d\nu \geq 2$ for all $l \in [2^R]$. Assume there exists $r \in [R]$ such that
\[
\frac{1}{4} \max_{l \in [2^r]} \biggl( 
\frac{ \int_{I_l^{(r)}} \lambda_a - \lambda_b d\nu }{ \int_{I_l^{(r)}} \lambda d \nu } \biggr)^2 \int_{I_l^{(r)}} \lambda d \nu \geq 2r + \frac{C^{1/2}}{\beta} + 2 \log \frac{R}{\alpha}.
\]
Then, we have that 
$
\mathbb{P}( p_M \leq \alpha) \geq 1 - 2\beta.
$
\end{theorem}

We give the proof of Theorem~\ref{thm:min_power} in Section~\ref{sec:min_power_proof} of the Appendix. To more clearly see the implication of Theorem~\ref{thm:min_power}, we consider a setting where $|\lambda_a - \lambda_b|$ is non-zero possibly only on a small region $S$. We show that, so long as the hierarchical partition is chosen so that each split divides the volume equally, the minimum combination method will have non-trivial power. The following result is an analog of Theorem~\ref{thm:fisher_power_smooth}.

\begin{theorem}
\label{thm:min_power_smooth}
Let $\mathcal{X} \subset \mathbb{R}$ be an interval and let $S \subset \mathcal{X}$ be a sub-interval. Let $\{I_l^{(r)} \}_{r \in [R]}$ be a dyadic partition of $\mathcal{X}$ such that $R = \lfloor \log_2 \frac{n}{2} - \log_2 \frac{\lambda_{max}}{\lambda_{min}} \rfloor$ and that for all $r \in [R]$ and $l \in [2^r]$, $\nu(I_l^{(r)}) = \nu(\mathcal{X}) 2^{-r}$. Suppose that $0 < c_{\min} \leq \lambda \leq c_{\max} < \infty$ and that $\int_I \lambda d\nu = n$. Suppose that $\bigl| \frac{\lambda_a(x) - \lambda_b(x)}{2 \lambda(x)} \bigr| \geq \delta_S > 0$ for all $x \in S$ and also that $\frac{\nu(S)}{\nu(\mathcal{X})} \geq \frac{c_{\max}}{c_{\min}} \frac{8}{n}$.

For any $\alpha, \beta \in (0,1)$, if
\begin{align}
\delta_S^2 \frac{\nu(S)}{\nu(\mathcal{X})} \geq  \frac{C_2}{n}\biggl( \log n + \beta^{-1} + \log \frac{1}{\alpha} \biggr) 
\label{eq:delta_S_condition}
\end{align}
for $C_2 > 0$ dependent only on $c_{\max}/c_{\min}$, then we have that 
$
\mathbb{P}( p_M \leq \alpha ) \geq 1 - 2\beta.
$
\end{theorem}

We give the proof of Theorem~\ref{thm:min_power_smooth} in Section~\ref{sec:min_power_smooth_proof} of the Appendix. We note that Theorem~\ref{thm:min_power_smooth} does not require any smoothness on the difference $\lambda_a - \lambda_b$; it only requires that the difference is at least $\delta_S$ in magnitude on the region $S$. 

{\color{black}
\begin{remark}
Although our theoretical results are developed for the two-sample test, the high level proof strategy may be applicable for analyzing the power of longitudinal network tests as well. The main difficulty however is that, when analyzing all the $l \in [2^r]$ bins at high resolution level $r$, many of the bins may have very small signal strength but nevertheless cannot be ignored because on aggregate, they contribute to the overall separation from the null hypothesis. In Theorem~\ref{thm:fisher_power}, we handle the low signal bins through a careful analysis of the tail probability of the binomial distribution (see Lemma~\ref{lem:variance_bound} and~\ref{lem:expectation_bound} in Section~\ref{sec:aux_result} of the appendix). Extending this step to the longitudinal network setting is a nontrivial problem which we leave to a future work. 
\end{remark}
}

\section{Experimental studies}
\label{sec:experiments}

\subsection{An illustrative example}
\label{sec:example}

\begin{figure}[htp]
\begin{center}
\begin{subfigure}{0.49\linewidth}
\includegraphics[scale=.3]{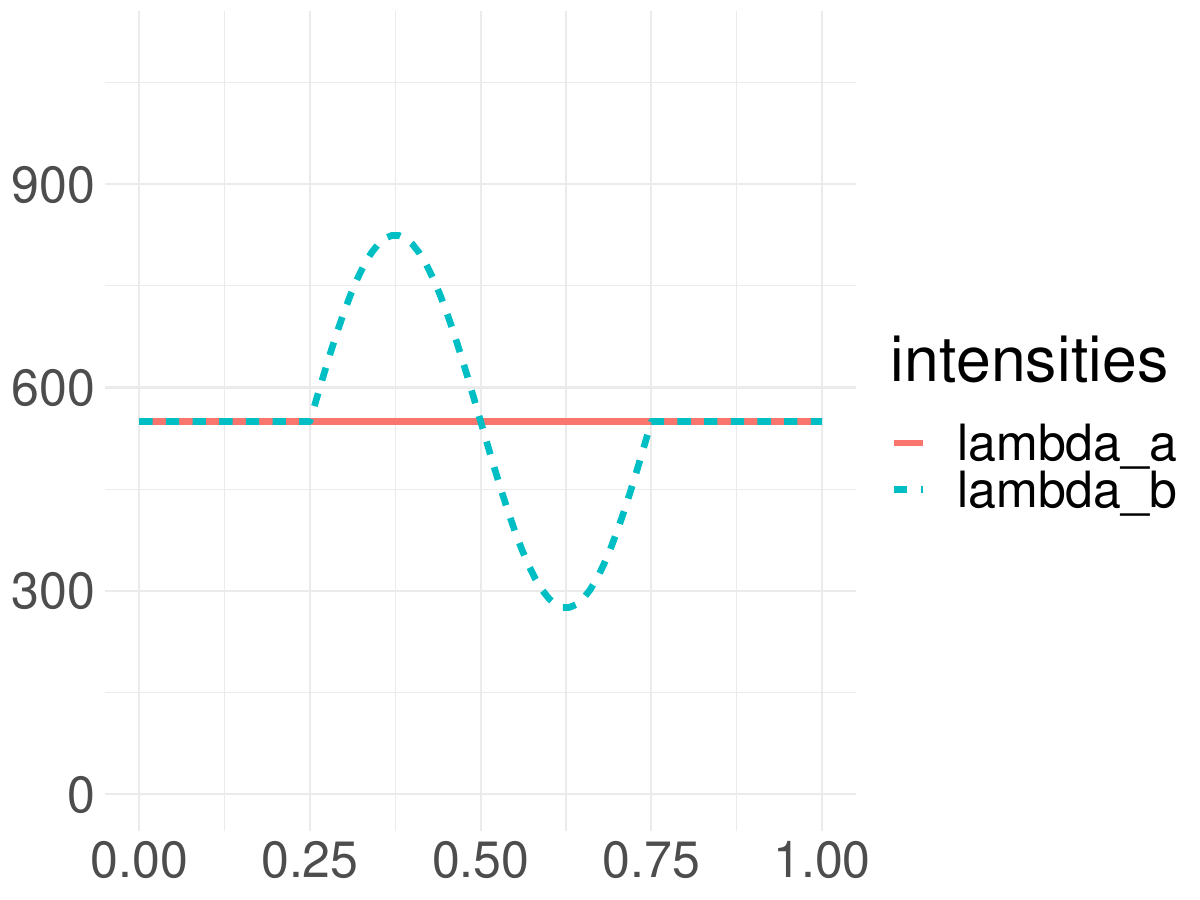}
\caption{Intensity functions $\lambda_a, \lambda_b$}
\label{fig:example_fn}
\end{subfigure}
\begin{subfigure}{0.49\linewidth}
\includegraphics[scale=.3]{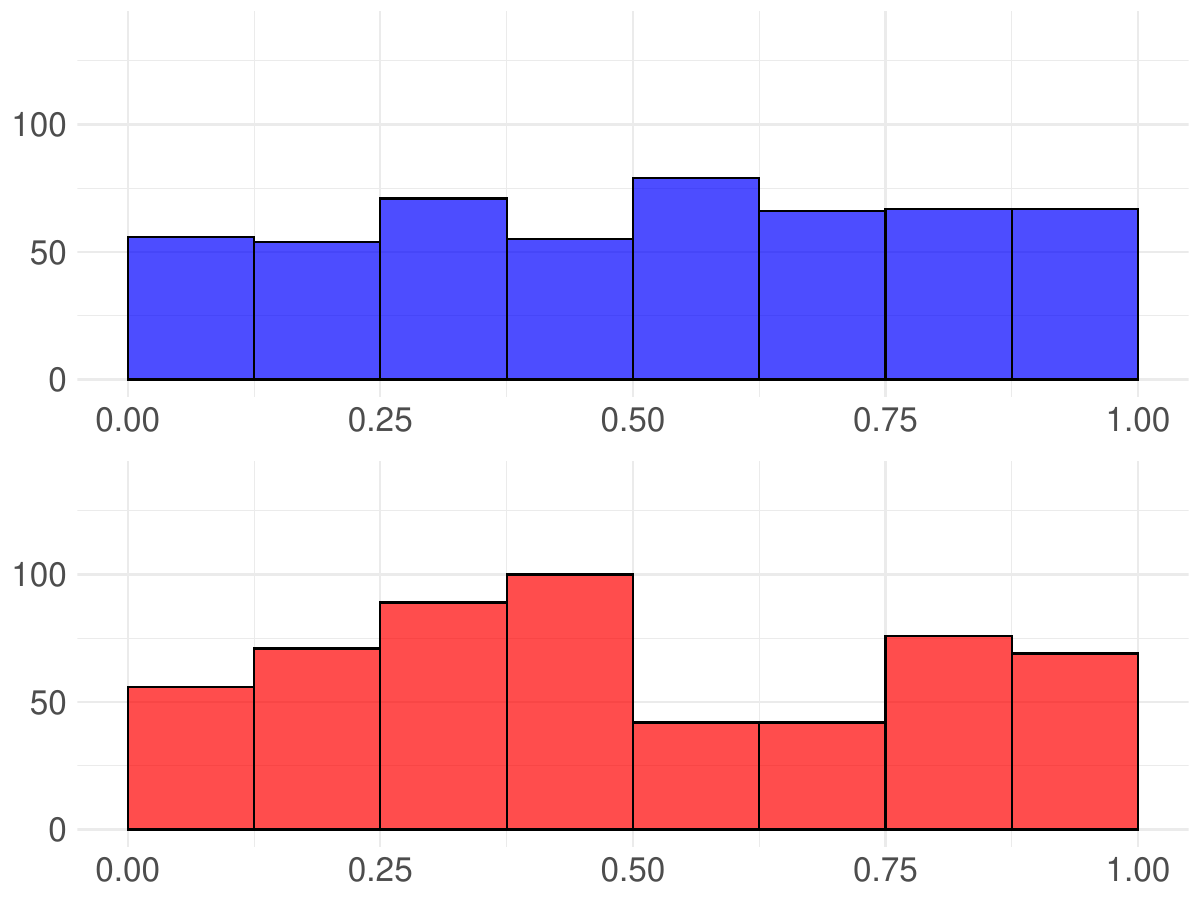}
\caption{Realizations $N_a(\cdot), N_b(\cdot)$}
\label{fig:example_data}
\end{subfigure}
\end{center}
\vspace{-10pt}
\caption{Example of a two-sample test setting}
\label{fig:allfigures}
\end{figure}

We begin with a single example for the two-sample test to give a concrete illustration of our testing procedure. We let the two intensity functions be $\lambda_a(x) = 550 \mathbbm{1}_{[0, 1]}$ and $\lambda_b(x) = 550 + 275 \sin(4 \pi (x - 1/4)) \mathbbm{1}_{[1/4, 3/4]}$; see Figure~\ref{fig:example_fn}. We take our hierarchical partition $\bm{I}$ to have $R=3$ resolution level by dyadic splitting so that $I^{(1)}_1 = [0, 1/2]$ and $I^{(1)}_2 = [1/2, 1]$. We see then that $H_0^{(2, 1)}$ and $H_0^{(2, 4)}$ are true since $\lambda_a = \lambda_b$ on the regions $[0, 1/4]$ and $[3/4, 1]$. We generate two point process realizations and show them in Figure~\ref{fig:example_data}.

\begin{figure}
\begin{center}
\begin{tikzpicture}[level distance=10mm, 
                    every node/.style={},
                    level 1/.style={sibling distance=60mm},
                    level 2/.style={sibling distance=30mm},
                    level 3/.style={sibling distance=15mm}]
\node[label=above:$p_F^{(0,1)}$] {\bf 0.000} 
    child {node[label=above:$p_F^{(1,1)}$] {\bf 0.003} 
        child {node[label=above:$p_F^{(2,1)}$] {1} 
            child {node {1}}
            child {node {0.44}}
        }
        child {node[label=above:$p_F^{(2,2)}$] {\bf 0.004}
            child {node {0.664}}
            child {node {\bf 0.000}}
        }
    }
    child {node[label=above:$p_F^{(1,2)}$] {\bf 0.001}
        child {node[label=above:$p_F^{(2,3)}$] {\bf 0.000}
            child {node {\bf 0.002}}
            child {node {0.078}}
        }
        child {node[label=above:$p_F^{(2,4)}$] {1}
            child {node {1}}
            child {node {1}}
        }
    };
  \end{tikzpicture}
  \end{center}
  \vspace{-20pt}
\caption{Outputted simultaneously valid p-values. Bold indicates rejection at $\alpha = 0.05$.}
\label{fig:example_pvals}
\end{figure}
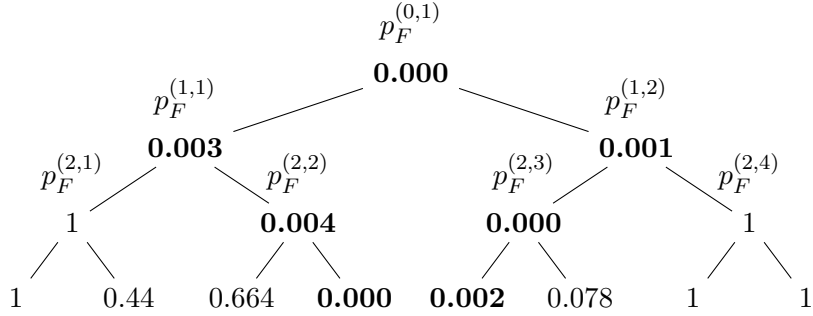

We then perform our testing procedure with $B=1000$ resampling draws and display the final simultaneously valid p-values $\{ p_F^{(s,j)} \}$ in Figure~\ref{fig:example_pvals}. We see that, at level $\alpha = 0.05$, we correctly reject the global null $H_0^{(0,1)}$. Moreover, we correctly reject the local nulls $H_0^{(1,0)}, H_0^{(1,1)}$ at resolution level $r=1$ and the the local nulls $H_0^{(2,2)}, H_0^{(2,3)}$ at resolution level $r = 2$. At resolution level $r=3$, we reject $H_0^{(3,4)}, H_0^{(3,5)}$ but missed $H_0^{(3,3)}, H_0^{(3,6)}$. This is not unexpected since the multiple testing adjustment makes it rather difficult to reject hypotheses at the most granular resolution level. 


\subsection{Simulation for the two-sample test}

Next we provide simulations results of our proposed test for the two-sample problem. We Let $N_a(\cdot)$ and $N_b(\cdot)$ be two Poisson point processes on $\mathcal{X} = [0,1]$, with intensity functions $\lambda_a(\cdot)$ and $\lambda_b(\cdot)$ respectively. In Section~\ref{sec:type1exp} of the appendix, we verify that our tests have the desired type I error. Here, we study the power of our proposed test. For comparison with our proposed test, we also present simulations results of other two-sample testing procedures. The first is the kernel-based test proposed in \cite{gretton2012kernel} and \cite{fromont2013two}. Recall that we can characterize $N_a = \{X_i: M_i = 1\} $ and $N_b = \{X_i = -1\}$, then for any symmetric kernel function $K: \mathcal{X} \times \mathcal{X} \longrightarrow \mathbb{R}$, the test statistic of the kernel-based test is given by
$
T_\text{kernel} = \sum_{i\ne j\in[N]}K(X_i,X_j)M_i M_j.
$
There are many choices of the Kernel but in this simulation we use the Gaussian kernel $
	K(X_i, X_j): = \text{exp}\Bigl\{-\frac{(X_i-X_j)}{2\sigma^2}\Bigr\} 
$ which is shown to have good performance in practice by \cite{fromont2013two}. We also consider the conditional Kolmogorov-Smirnov test for performance comparison, which is commonly used for two-sample problems. We again apply Monte-Carlo method to approximate the exact p-value of these two tests. 
In total, we consider 5 different tests denoted by MF, MM, $\text{KN}_1$, $\text{KN}_2$, KS, where MF and MM represent our \emph{multi-scale binning} test with Fisher combination and minimum combination respectively. The tests $\text{KN}_1$, $\text{KN}_2$ represent the Gaussian kernel test with parameter $\sigma = 0.5$ and $\sigma=0.1$ of the kernel function. The test KS represents the conditional Kolmogorov-Smirnov test.

\begin{figure}[h]
    \centering
    \includegraphics[width=14cm]{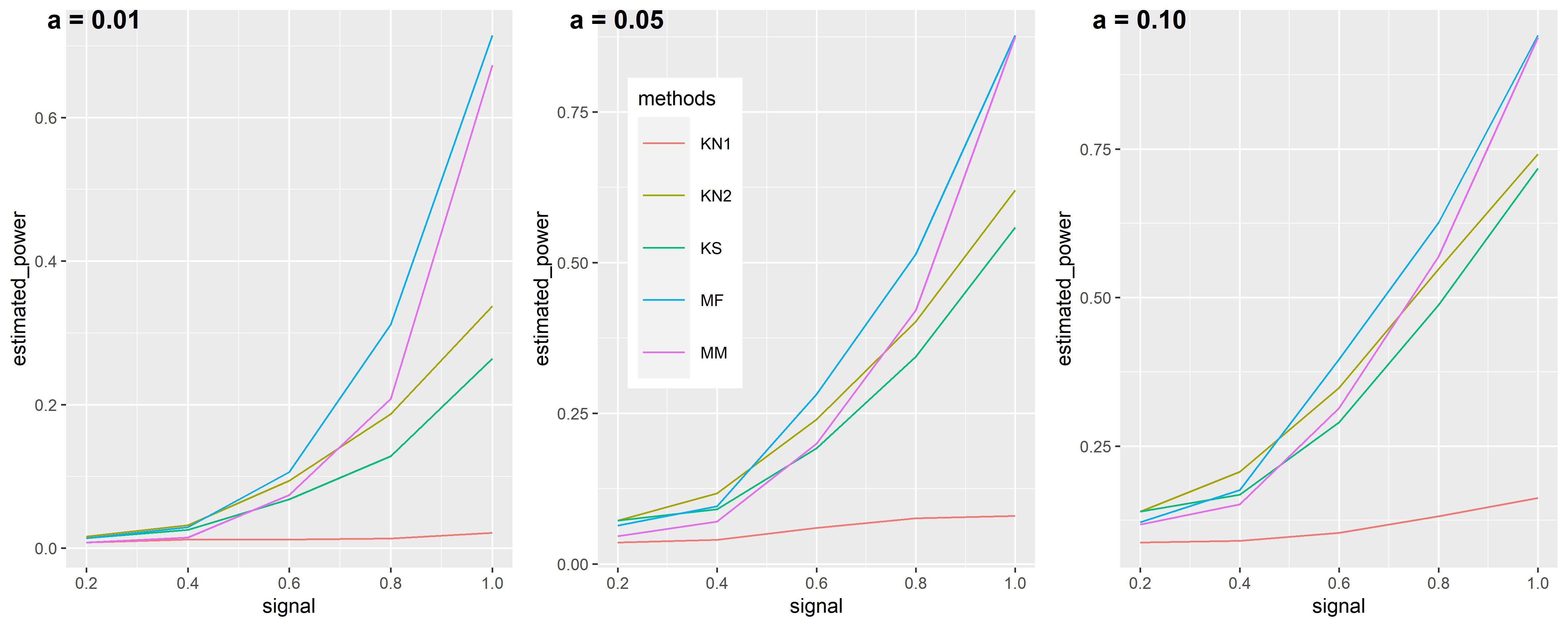}
    \caption{The proportion of rejections of the five tests out of 1000 simulated samples under different signal strength. Left: level $\alpha=0.01$. Center: level $\alpha=0.05$. Right: level $\alpha= 0.10$ .}
            \label{figure:two_sample_signal}
\end{figure}

We let $\lambda_a(\cdot)$ be a constant function and $\lambda_b(\cdot)$ be a piecewise constant function given below
\begin{align*}
\lambda_a(x) = 50\cdot\bm{1}_{[0,1]}(x),\quad
 \lambda_b(x) = 50\bigl((1-p)\cdot\mathbf{1}_{[1,\frac{1}{4}]}(x) + (1+p)\cdot\mathbf{1}_{(\frac{1}{4}, \frac{1}{2}]}(x) + \mathbf{1}_{(\frac{1}{2},1]}(x)\bigr).
\end{align*}
For each $p\in \{0.2, 0.4, 0.6, 0.8, 1.0\}$, we perform 1000 repetitions and use 500 resampling draws in each repetition. In Figure~\ref{figure:two_sample_signal}, we plot the empirical power, i.e., the proportion of rejections out of the 1000 simulations for each of the five tests in previous experiments, at three different levels $\alpha = 0.01, 0.05, 0.10$. 

At all three levels, We can see that the kernel based test with large bandwidth has very bad performance which is not surprising since the large bandwidth smooths the difference between the two Poisson realizations. The other four tests have similar number of rejections when the signal is relatively weak while as the value of $p$ becomes larger, the MF and MM test, i.e., our \emph{multi-scale binning} test with Fisher and Minimum combination dominates the K-S test and kernel based tests. 

\subsection{Simulation study of array test}

Next we study tests for degree-corrected longitudinal networks described in Section~\ref{sec:dc_array_test}. We provide the results on the non-degree-corrected setting in Section~\ref{sec:homogeneous_simulation} of the appendix.

\subsubsection{Testing degree-corrected longitudinal networks}
\label{sec:dc_array_simulation}

We generate a degree corrected longitudinal network with $K = 2$ communities and $n$ nodes; we generate community membership $\bm{\sigma}$ so that the first $n/2$ nodes are in community 1 and the second half are in community 2. To generate the degree correction parameter $\bm{\theta}$, we first create unnormalized $\tilde{\bm{\theta}} \in [0, \infty)^n$ and then set $\theta_u = s \cdot \frac{\tilde{\theta}_u}{\| \tilde{\bm{\theta}} \|}$ for a fixed sparsity parameter $s$ so that $ \| \bm{\theta} \| = s$.  We generate the interactions between $u,v$ as
\begin{align*}
N_{uv}(\cdot) \sim \begin{cases}
   \text{PP}( \lambda_{a} \cdot \theta_u \theta_v), & \sigma_u = \sigma_v, \\
   \text{PP}( \lambda_{b} \cdot \theta_u \theta_v), & \sigma_u \neq \sigma_v,
   \end{cases}
\end{align*}
where $\lambda_a$ is the within-community intensity function and $\lambda_b$ is the between-community intensity function.

\begin{figure}[h]
 \centering
    \caption{Histogram of Signed Polygon test statistics with different values of $p$.}
    \includegraphics[width=13cm]{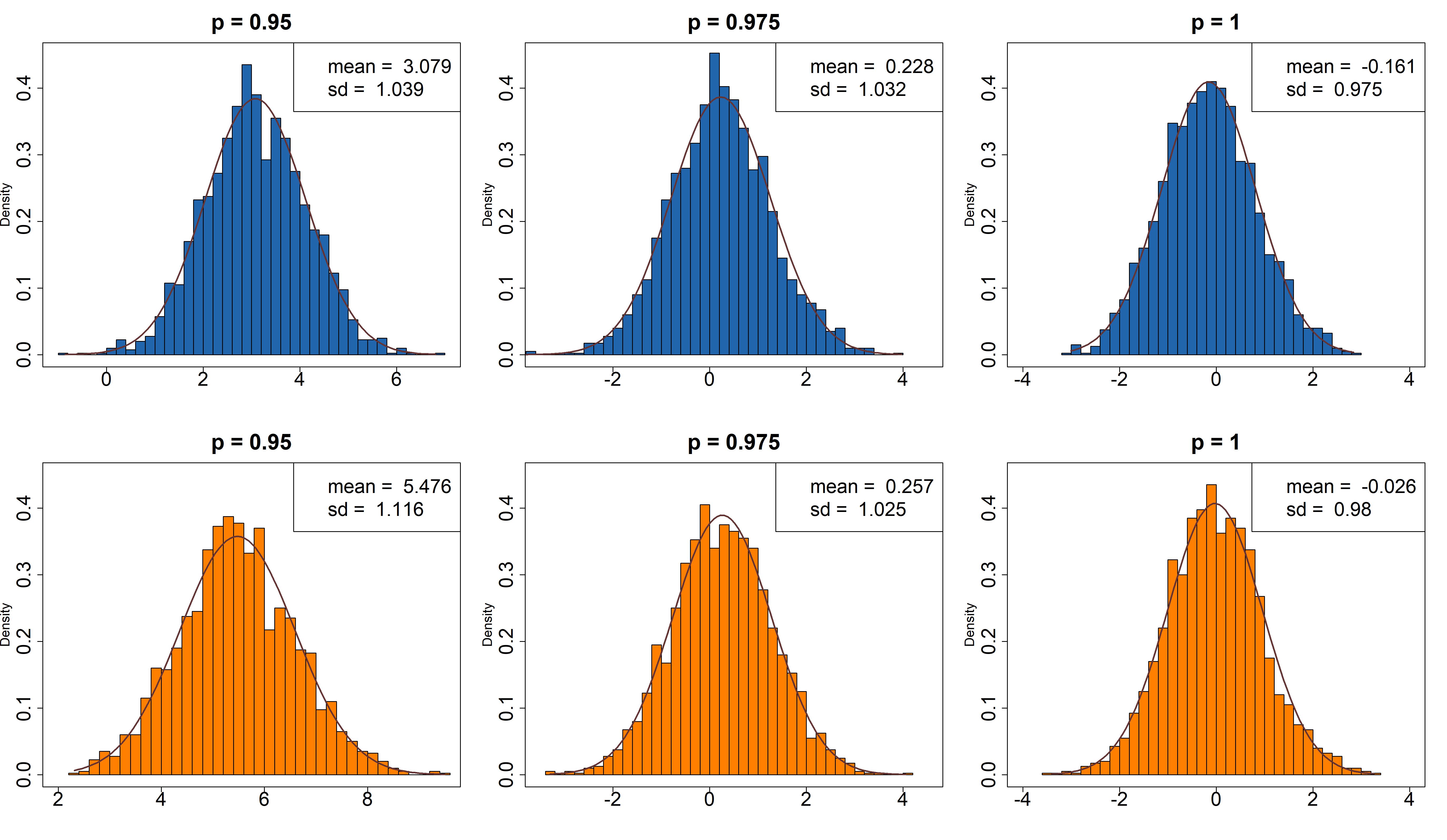}
    \label{figure:static_sgn_hist}
\end{figure}

We first empirically verify the null distribution of $T^{(r,\ell)}$ and $Q^{(r,\ell)}$ -- the SgnT \eqref{equation:sgnt_defn} and SgnQ \eqref{equation:sgnq_defn} test statistics for each discretized local null. We set $n=1000$ number of nodes, $s=100$ sparsity level, and let $\tilde{\theta}_u = u$ for each $u\in[n]$. We set $\lambda_a = \bm{1}_{[0,1]}$ and $\lambda_b = p \bm{1}_{[0,1]}$ where we let $p\in\{0.95, 0.975, 1\}$ so that the null is either true or close to being true. For each value of $p$, we generate 2000 longitudinal networks and reduce each to a single static weighted network $A$ such that $A_{uv} = N_{uv}([0, 1])$. We then compute the centered, scaled SgnT and SgnQ test statistics for $A$ and plot their empirical distribution respectively in Figure \ref{figure:static_sgn_hist}.

Next we study the power of our proposed test. We set $n = 100$, $s = 12$, and consider two levels of degree heterogeneity: (1) \textbf{moderate degree heterogeneity} with \emph{iid} $\tilde{\theta}_u \sim \text{Unif}(2,3)$ and (2) \textbf{severe degree heterogeneity} with fixed $\tilde{\theta}_u = \sqrt{u}$. We let
\begin{align*}
   \lambda_{a}(x) = s\frac{x(1-x)^4}{\int_{0}^{1}x(1-x)^4 dx}\bm{1}_{[0,1]}(x), \quad \lambda_{b}(x) = p\cdot s\frac{x(1-x)^4}{\int_{0}^{1}x(1-x)^4 dx}\bm{1}_{[0,1]}(x).
\end{align*}
We note that $\int_0^1 \lambda_a = s$ and $\int_0^1 \lambda_b = p s \leq s$ so the expected total number of interaction events in the network is at most $s \| \bm{\theta} \|^2 = s^3 = 1728$, which implies that the network is relatively sparse. To study the power of our proposed test at different signal strength, we let $p$ vary within the set $\{0.6, 0.7, 0.8, 0.9, 1.0\}$. We let the number of resolution levels $R$ be equal to 4 and let $I^{(r,\ell)}: =[\frac{\ell-1}{2^r}, \frac{\ell}{2^r})\subset \mathcal{X}$ be the discretized intervals for $l\in [2^r], r\in \{1,2,3,4\}$. For each experiment, we create $B=400$ resampling draws using the Metropolis-Hastings algorithm described in Section \ref{sec:sampling_dc_network} to derive the adjusted p-values. We perform the proposed multiscale binning test for the global null using both SgnT and SgnQ test statistics and summarize the empirical proportion of rejections out of the 200 experiments under the two degree heterogeneity levels in Table \ref{table:dc_array_sims_mod} and Table \ref{table:dc_array_sims_sev} respectively.

{\color{black}
We observe that tests based on either the SgnT or the SgnQ test statistic have the desired type I error control. The SgnQ test however has higher power in all the settings that we have studied. We also see that the severity of degree heterogeneity negatively affects the power of our testing procedure. 
}

{
\renewcommand{\baselinestretch}{1}
\begin{table}[H] \centering
    \begin{tabular}{crcccccrccccc}\toprule
        && \multicolumn{5}{c}{SgnT} && \multicolumn{5}{c}{SgnQ} \\
        \cmidrule{3-7} \cmidrule{9-13}
        $p$ && $0.6$ & $0.7$& $0.8$ & $0.9$ & $1.0$ && $0.6$ & $0.7$& $0.8$ & $0.9$ & $1.0$\\
        \midrule
        $\alpha=0.01$ && 0.98 & 0.15 & 0.02 & 0 & 0 && 0.84 & 0.77 & 0.11 & 0.02 & 0.01\\
        $\alpha=0.05$ && 0.99 & 0.41 & 0.07 & 0.05 & 0.03 && 1 & 0.95 & 0.15 & 0.07 & 0.04\\
        $\alpha=0.10$ && 0.99 & 0.59 & 0.11 & 0.10 & 0.06 && 1 & 0.97 & 0.22 & 0.09 & 0.05\\
        $\alpha=0.25$ && 0.99 & 0.84 & 0.27 & 0.23 & 0.22 && 1 & 0.99 & 0.35 & 0.19 & 0.23\\
        \bottomrule
    \end{tabular}
    \caption{The proportion of rejections of the proposed array test under moderate degree heterogeneity level, where $\Tilde{\theta}_u \sim \text{Unif}(2,3)$.}
    \label{table:dc_array_sims_mod}
\end{table}

\begin{table}[H] \centering
    \begin{tabular}{crcccccrccccc}\toprule
        && \multicolumn{5}{c}{SgnT} && \multicolumn{5}{c}{SgnQ} \\
        \cmidrule{3-7} \cmidrule{9-13}
        $p$ && $0.6$ & $0.7$& $0.8$ & $0.9$ & $1.0$ && $0.6$ & $0.7$& $0.8$ & $0.9$ & $1.0$\\
        \midrule
        $\alpha=0.01$ && 0.37 & 0.03 & 0.01 & 0 & 0 && 0.95 & 0.19 & 0.02 & 0.02 & 0\\
        $\alpha=0.05$ && 0.64 & 0.15 & 0.08 & 0.05 & 0.06 && 0.96 & 0.35 & 0.06 & 0.05 & 0.04\\
        $\alpha=0.10$ && 0.80 & 0.29 & 0.14 & 0.09 & 0.12 && 0.98 & 0.43 & 0.13 & 0.11 & 0.08\\
        $\alpha=0.25$ && 0.95 & 0.52 & 0.29 & 0.26 & 0.22 && 1 & 0.61 & 0.23 & 0.22 & 0.25\\
        \bottomrule
    \end{tabular}
    \caption{The proportion of rejections of the proposed array test under severe degree heterogeneity level with fixed $\Tilde{\theta}_u = \sqrt{u}$.}
    \label{table:dc_array_sims_sev}
\end{table}

}

\subsection{Traffic collision data}
\begin{figure}[h]
\centering
	\includegraphics[width=13cm]{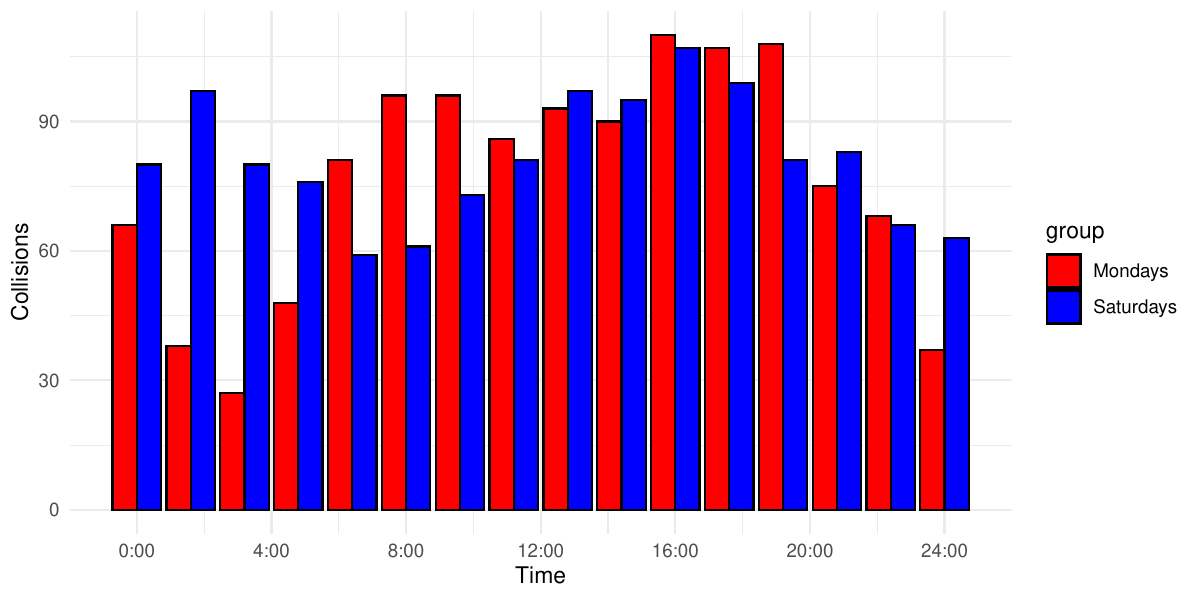}
	\caption{Number of NYC collisions on Mondays and Saturdays at different time of day.} 
\label{fig:collision_barplot}
\end{figure}

We conduct the two-sample test using data derived from motor vehicle collision events in New York City recorded by the New York Police Department during January 2022. We test whether there exists a significant difference between the occurrences of crash events on Mondays and Saturdays during this period. To achieve this, we treat the time-stamps of the collisions on Mondays and Saturdays as two realizations of Poisson Processes. Let $\bm{I}$ be the time interval between 00:00 and 24:00 and with the number of resolution level $R=4$, we evenly partition $\bm{I}$ by dyadic splitting such that $I_1^{(1)}=[00:00, 12:00)$, $I_1^{(2)}=[12:00, 24:00)$ and so on. we provide a barplot of the number of collision events within each discretized interval at the most granular resolution level, see Figure~\ref{fig:collision_barplot}. We can see that the distribution of collisions are very similar on Mondays and Saturdays from noon to midnight. However, in the morning, there are many more collisions happening on Monday on Saturday. On the other hand, in late night,  collisions are more frequent on Saturday than on Sunday.

{\renewcommand{\baselinestretch}{1}
\begin{figure}[!h]
\begin{center}
\begin{tikzpicture}[level distance=10mm, 
                    node/.style={circle,draw, minimum size=5mm},
                    level 1/.style={sibling distance=80mm},
                    level 2/.style={sibling distance=40mm},
                    level 3/.style={sibling distance=19mm},
                    level 4/.style={sibling distance=10mm}]
\node[label=above:$p_F^{(0,1)}$] {\bf 0.0} 
    child {node[label=above:$p_F^{(1,1)}$] {\bf 0.0} 
        child {node[label=above:$p_F^{(2,1)}$] {\bf 0.0} 
            child {node[label=above:$p_F^{(3,1)}$] {\bf 0.0}
                child {node {1}}
                child {node {\bf 0.0}}
            }
            child {node[label=above:$p_F^{(3,2)}$] {\bf 0.0}
                child {node {\bf 0.0}}
                child {node {0.144}}
            }
        }
        child {node[label=above:$p_F^{(2,2)}$] {\bf 0.0}
            child {node [label=above:$p_F^{(3,3)}$]{\bf 0.0}
                child {node {0.59}}
                child {node {\bf 0.03}}
            }
            child {node [label=above:$p_F^{(3,4)}$]{1}
                child {node {0.69}}
                child {node {1}}
            }
        }
    }
    child {node[label=above:$p_F^{(1,2)}$] {1}
        child {node[label=above:$p_F^{(2,3)}$] {1}
            child {node [label=above:$p_F^{(3,5)}$]{1}
                child {node {1}}
                child {node {1}}
            }
            child {node [label=above:$p_F^{(3,6)}$]{1}
                child {node {1}}
                child {node {1}}
            }
        }
        child {node[label=above:$p_F^{(2,4)}$] {0.352}
            child {node [label=above:$p_F^{(3,7)}$]{1}
                child {node {0.480}}
                child {node {1}}
            }
            child {node [label=above:$p_F^{(3,8)}$]{0.496}
                child {node {1}}
                child {node {0.064}}
            }
        }
    };
  \end{tikzpicture}
  \end{center}
  \vspace{-10pt}
\caption{Simultaneously valid p-values for testing NYC collision occurrences at Mondays and Saturdays. Bold indicates rejection at level $\alpha=0.05$}
\label{fig:collision_pvals}
\end{figure}
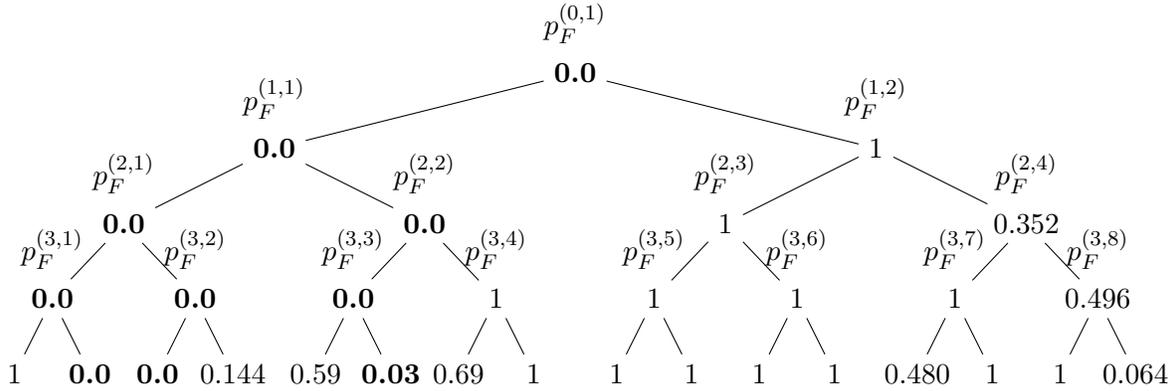
}

We perform the proposed testing procedure with 1000 bootstrap samples and use the Fisher combination method to combine p-values at the same resolution level. We provide the tree of valid p-values in Figure~\ref{fig:collision_pvals}. At level $\alpha=0.05$, we reject the global null $H_0^{(0,1)}$, which make sense as the two occurrences data do have different patterns. Notice that we also successfully rejected the local nulls $H_0^{(4,2)}$, $H_0^{(4,3)}$ and $H_0^{(4,6)}$, where the difference in number of occurrences are among the largest at the resolution level $r=4$. Although we see from Figure~\ref{fig:collision_barplot} that the number of collisions are different around midnight, we are not able to reject the corresponding local null $H_0^{(4,16)}$ due to the sequential testing adjustment.

\subsection{Primates interaction data}


In this study, we analyze a network of pairwise interactions within a group of 13 Guinea baboons residing in an enclosure at a Primate Center in France starting from June 13th, 2019 \citep{gelardi2020measuring}. The dataset was gathered using wearable proximity sensors attached to leather collars worn on the front side of the 13 baboons. These sensors utilized low-power radio communication, exchanging packets when the distance between two baboons was approximately less than 1.5 meters. The collected data was aggregated with a temporal resolution of 20 seconds, defining interaction between two individuals if their sensors exchanged at least one packet during a 20-second interval.
We consider here mainly three days of data collected between July 8th and July 10th 2019, capturing a total of 6458 interactions among the 13 baboons. To better analyze how the network evolves over the course of a typical day,  we aggregate the interactions across all the days so that each interaction has only the hour/minute/second information. We notice that the degrees are quite heterogeneous and thus believe that a degree corrected model is best suited for the analysis.

{\renewcommand{\baselinestretch}{1}
\begin{figure}[htp]
\centering
	\includegraphics[width=14cm, trim={0 0 0 10cm}, clip]{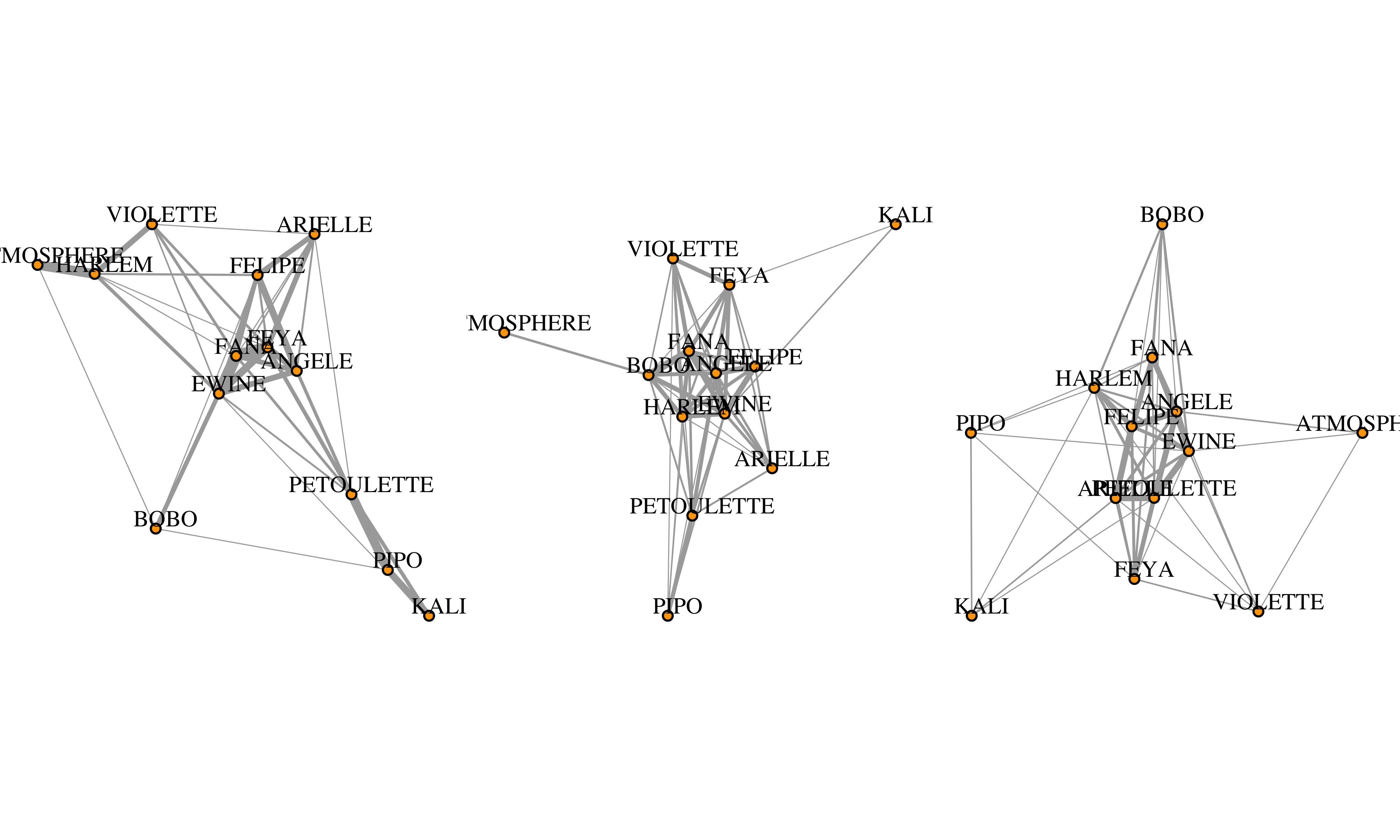}
	\vspace{-50pt}
	\caption{Subgraphs of baboon interaction network at different times of day. \textbf{Left:} Interactions in the morning between 6:48AM and 7:48AM. \textbf{Center:} In the afternoon between 3:53PM and 4:53PM. \textbf{Right:} At night between 7:55PM and 8:55PM.}
\label{fig:baboon_network}
\end{figure}
}

{\renewcommand{\baselinestretch}{1}
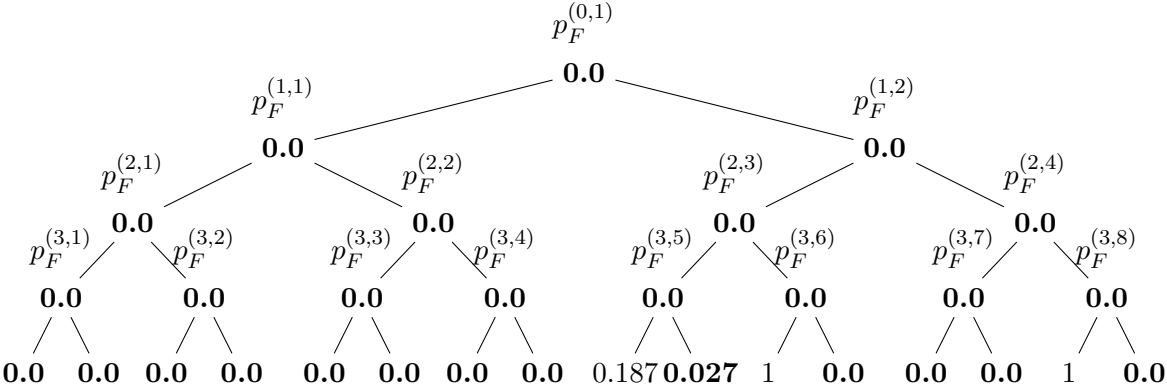
\begin{figure}[!h]
\begin{center}
\begin{tikzpicture}[level distance=10mm, 
                    node/.style={circle,draw, minimum size=5mm},
                    level 1/.style={sibling distance=80mm},
                    level 2/.style={sibling distance=40mm},
                    level 3/.style={sibling distance=19mm},
                    level 4/.style={sibling distance=10mm}]
\node[label=above:$p_F^{(0,1)}$] {\bf 0.0} 
    child {node[label=above:$p_F^{(1,1)}$] {\bf 0.0} 
        child {node[label=above:$p_F^{(2,1)}$] {\bf 0.0} 
            child {node[label=above:$p_F^{(3,1)}$] {\bf 0.0}
                child {node {\bf 0.0}}
                child {node {\bf 0.0}}
            }
            child {node[label=above:$p_F^{(3,2)}$] {\bf 0.0}
                child {node {\bf 0.0}}
                child {node {\bf 0.0}}
            }
        }
        child {node[label=above:$p_F^{(2,2)}$] {\bf 0.0}
            child {node [label=above:$p_F^{(3,3)}$]{\bf 0.0}
                child {node {\bf 0.0}}
                child {node {\bf 0.0}}
            }
            child {node [label=above:$p_F^{(3,4)}$]{\bf 0.0}
                child {node {\bf 0.0}}
                child {node {\bf 0.0}}
            }
        }
    }
    child {node[label=above:$p_F^{(1,2)}$] {\bf 0.0}
        child {node[label=above:$p_F^{(2,3)}$] {\bf 0.0}
            child {node [label=above:$p_F^{(3,5)}$]{\bf 0.0}
                child {node {0.187}}
                child {node {\bf 0.027}}
            }
            child {node [label=above:$p_F^{(3,6)}$]{\bf 0.0}
                child {node {1}}
                child {node {\bf 0.0}}
            }
        }
        child {node[label=above:$p_F^{(2,4)}$] {\bf 0.0}
            child {node [label=above:$p_F^{(3,7)}$]{\bf 0.0}
                child {node {\bf 0.0}}
                child {node {\bf 0.0}}
            }
            child {node [label=above:$p_F^{(3,8)}$]{\bf 0.0}
                child {node {1}}
                child {node {\bf 0.0}}
            }
        }
    };
  \end{tikzpicture}
  \end{center}
  \vspace{-10pt}
\caption{Simultaneously valid p-values for testing baboons interaction network. Bold indicates rejection at level $\alpha=0.01$}
\label{fig:baboon_pvals}
\end{figure}
}

We perform the multi-bining test using SgnQ statistics with boot strap sample size $B=600$ and number of resolution number $R=4$ and the intervals are evenly discretized between 5:30AM and 10:00PM. At the most granular level, each bin corresponds to approximately 1 hour. We provide the the tree of simultaneous valid p-values in Figure~\ref{fig:baboon_pvals}. At level $\alpha=0.05$, our testing procedure reject the global null and all the local nulls at resolution level $r=1,2,3$. Notice that there are three local nulls $H_0^{(4,10)}, H_0^{(4,11)}$, and $H_0^{(4,15)}$ at resolution level $r=4$ that our test failed to reject. We display the discretized networks that correspond to these two of these intervals in Figure~\ref{fig:baboon_network}. We also display another instance where we do reject the null as a comparison. We see that the results make sense based on the visualization since there appears to be 3 distinct clusters {\color{black}($\{$Violette, Mosphere, Harlem$\}$, $\{$Petoulette, Pipo, Kali$\}$, and others)} for the morning interaction graph.

\bibliographystyle{apalike}
\bibliography{ref}

\clearpage

\setcounter{section}{0}
\setcounter{equation}{0}
\setcounter{theorem}{0}
\def\theequation{S\arabic{section}.\arabic{equation}}
\def\thesection{S\arabic{section}}
\def\thetheorem{S\arabic{theorem}}

\begin{center}
\Large{Supplementary material to ``Multiscale Tests for Point Processes and Longitudinal Networks''} \\ \vspace{0.2in}
\end{center}

\section{Supplementary material for Sections~\ref{sec:process_tests} and~\ref{sec:array_tests}}

\subsection{Concise description of algorithms for longitudinal networks}

\begin{algorithm}[htp]
	\caption{Computing simultaneously valid p-values for $H_0^{(s,j)}$ in the symmetric array case.}
	\label{alg:symm}
	\begin{flushleft}
		\textbf{INPUT:} Poisson process realizations $\{N_{uv}(\cdot), u<v \in [n] \}$  and a hierarchical partitioning $\bm{I} = \{ I^{(r)}_\ell \}_{r \in [R], \ell \in [2^r]}$ of the domain. \\
		\textbf{OUTPUT:} p-values $\{ p_F^{(s,j)} \}$.
		\begin{algorithmic}[1]
			\For{ each $r \in [R]$ }
			\For{ each $\ell \in [2^r]$} 
            \State Define integer matrix $A^{(r,\ell)}$ as~\eqref{eq:ad_matrix_local}.
			\State Set $\bar{p}^{(r, \ell)} = 2 \min \biggl(F_{\text{TW1}}\Big(n^{2/3}\big(\lambda_1(A^{(r,\ell)})) - 2\big)\Big), 1-F_{\text{TW1}} \Big(n^{2/3}\big(\lambda_1(A^{(r,\ell)})) - 2\big)\Big)\bigg).$
			\EndFor
			\EndFor
            \State Apply Algorithm~\ref{alg:all_local_null} on $\{\bar{p}^{(r, \ell)} \}$ to obtain $\{ \{ p_F^{(s,j,r)} \}_{r=s}^R  \}_{s \in [R], j \in [2^s]}$.
			\State Compute $\tilde{p}_F^{(s,j)} = \min \{ p_F^{(s,j,r)} \,:\, r \in \{s, s+1, \ldots, R\} \}$.
			\State  Run Metropolis--Hastings described in Section~\ref{sec:symmetric_resample_method} to generate $M_1^{(b^*)}, \ldots, M_N^{(b^*)}$ for $b^* \in [B]$ and let $\{N_{uv}^{(b^*)}(\cdot)\}$ be the corresponding point process realizations.
			\For { $b^* \in \{1,2,\ldots, B\}$ }:
			\State For each $r \in [R]$ and $\ell \in [2^r]$, construct $A_{uv}^{(r,\ell) \, (b^*)}$ from $\{ N_{uv}^{(b^*)}(\cdot) \}$.
			\State Repeat lines 1 to 8 to obtain $\tilde{p}_{F, b^*}^{(s,j)}$.
			\EndFor
			\State Compute $\check{p}_F^{(s,j)} := \frac{1}{B} \sum_{b^* = 1}^B \mathbbm{1}\{ \tilde{p}^{(s,j)}_{F, b^*} \leq \tilde{p}^{(s,j)}_F \} $.
            \State Compute $p_F^{(s,j)} = \check{p}_F^{(s,j)} 2^s$.
		\end{algorithmic}
	\end{flushleft}
\end{algorithm}

\begin{algorithm}[htp]
	\caption{Computing simultaneously valid p-values for $H_0^{(s,j)}$ in the degree-corrected setting.}
	\label{alg:dc_graph}
	\begin{flushleft}
		\textbf{INPUT:} Poisson process realizations $\{N_{uv}(\cdot), u<v \in [n] \}$  and a hierarchical partitioning $\bm{I} = \{ I^{(r)}_\ell \}_{r \in [R], \ell \in [2^r]}$ of the domain. \\
		\textbf{OUTPUT:} p-values $\{ p_F^{(s,j)} \}$
		\begin{algorithmic}[1]
			\For{ each $r \in [R]$ }
			\For{ each $\ell \in [2^r]$} 
			\State Set $p^{(r, \ell)} =  2\biggl[ 1-\Phi\biggl( \biggl| \frac{T^{(r, \ell)}}{\sqrt{6} (\| \hat{\eta}^{(r, \ell)}\|^2 - 1)^{3/2} }
    \biggr| \biggr) \biggr].$		
			\EndFor
			\EndFor
            \State Apply Algorithm~\ref{alg:all_local_null} on $\{ \bar{p}^{(r, \ell)} \}$ to obtain $\{ \{ p_F^{(s,j,r)} \}_{r=s}^R \}_{s \in [R], j\in [2^s]}$.
			\State Compute $\tilde{p}^{(s,j)}_F := \min\{ p_F^{(s,j,r)} \,:\, r \in \{s, s+1, \ldots, R \} \}$. 
			\State  Run Metropolis--Hastings algorithm described in Section~\ref{sec:sampling_dc_network} to generate $\bm{m}^{(b^*)}$ for $b^* \in [B]$ and let $\{N_{uv}^{(b^*)}(\cdot)\}$ be the corresponding point process realizations.
			\For { $b^* \in \{1,2,\ldots, B\}$ }:
			\State For each $r \in [R]$ and $\ell \in [2^r]$, construct $A_{uv}^{(r,\ell) \, (b^*)}$ from $\{ N_{uv}^{(b^*)}(\cdot) \}$.
			\State Repeat lines 1 to 8 to obtain $\tilde{p}_F^{(b^*)}$.
			\EndFor
			\State Compute the adjusted p-value $\check{p}^{(s,j)}_F := \frac{1}{B} \sum_{b^* = 1}^B \mathbbm{1}\{ \tilde{p}^{(s,j)}_{F, (b^*)} \leq \tilde{p}^{(s,j)}_F \} $.
            \State Compute $p^{(s,j)} = \check{p}^{(s,j)}_F \cdot 2^{-s}$.
		\end{algorithmic}
	\end{flushleft}
\end{algorithm}

\subsection{Asymmetric arrays of interactions}
\label{sec:asymmetric}

So far we've considered testing symmetric interaction processes among a single group of individuals, we can also extend this to the problem of testing the asymmetric interactions between two possibly different groups of individuals. Let $V_1$ and $V_2$ be two sets of individuals and suppose $|V_1| = m$ and $|V_2|= n$. Now we let $N_{uv}(\cdot)$ represents the temporal interactions events between individual $u\in V_1$ and individual $v \in V_2$, resulting in a collection of asymmetric point process realizations $\{N_{uv}(\cdot)\,:\, u\in V_1,\ v\in V_2\}$ where
\[
N_{uv}(\cdot) \sim PP(\Lambda_{uv}), \quad \text{ for intensity measure $\Lambda_{uv}$}.
\]
Now we have $m\times n$ realizations in total and similarly we can reduce the the dimensionality of this problem by assuming community structures in both groups. Suppose there are $K_1$ and $K_2$ communities respectively in groups $V_1$ and $V_2$, we again assume the intensity function of the realization between two individuals only depends on their community memberships. More precisely, let $\bm{\sigma}_1\,:\, [m] \rightarrow [K_1]$ and $\bm{\sigma}_2\,:\, [n] \rightarrow [K_2]$ be two clustering function on groups $V_1$ and $V_2$, then we assume
\begin{align*}
\Lambda_{uv} = \Gamma_{\bm{\sigma}_1(u) \bm{\sigma}_2(v)}, \, \text{for any $u\in V_1$ and $v\in V_2$ }
\end{align*}
where $\{\Gamma_{st}\}_{s,t \in [K_1] \times [K_2]}$ is a collection of $K_1 K_2$ intensity measures. We can again consider the goodness-of-fit test of the community structure with null hypothesis
\begin{equation}
 H_0\,:\, K_1 = K_2 = 1 \quad vs. \quad  H_0\,:\, K_1\cdot K_2 > 1
\end{equation}
and with a partition of $\bm{I}$ of the support $\mathcal{X}$, we can define each discretized local null as 
\begin{align}
\label{eq:local_asymmetric_test}
\bar{H}^{(r,\ell)}_0 : \Lambda_{uv}(I_{\ell}^{(r)}) = \gamma^{(r,\ell)}, \text{ for some common $\gamma^{(r,\ell)} \geq 0$ and for all  $u \in V_1, v \in V_2$}.
\end{align}
Similar to the local adjacency matrix $A^{(r,\ell)}$ defined in previous section, we let $B^{(r,\ell)}$ be a $m\times n$ matrix with entries being counts of interactions between any two individuals from groups $V_1$ and $V_2$ respectively, within interval $I^{(r)}_{\ell}$
\begin{align*}
B^{(r,\ell)}_{uv} = N_{uv}(I^{(r)}_{\ell}), \ \text{for any $u \in V_1$ and $v\in V_2$}
\end{align*}
To test each discretized local null $\bar{H}^{(r,\ell)}_0$ given observed matrix $B^{(r,\ell)}$, we again remove the mean effect and check whether the residual matrix looks like random noise. We define
\begin{align}
\label{eq:asymmetric_mean_estimate}
\hat{\gamma}^{(r,\ell)} = \sum_{u\in V_1}\sum_{v\in V_2}\frac{B^{(r,\ell)}_{uv}}{mn}.
\end{align}
Moreover, we define
\begin{equation}
\label{eq:asymmetric_empirical_ad}
\tilde{B}^{(r,l)}= \frac{B^{(r,\ell)}-\hat{\gamma}^{(r,\ell)}}{\sqrt{m\cdot\hat{\gamma}^{(r,\ell)}}} \in \mathbb{R}^{n \times m}
\end{equation}
as the empirically scaled and centered counterpart of $B^{(r,\ell)}$, with $\hat{\gamma}^{(r,\ell)}$ defined as in (\ref{eq:asymmetric_mean_estimate}) and let $\tilde{W}^{(r,\ell)}= (\tilde{B}^{(r,\ell)})^\mathsf{T}\tilde{B}^{(r,\ell)}$. Then we have the following limiting distribution of the largest eigenvalues of $\tilde{W}^{(r,\ell)}$.

\begin{theorem}
\label{thm:marchenko-pastur}
Let $\lambda_{1}(\tilde{W}^{(r,\ell)})$ be the largest eigenvalue of matrix $\tilde{W}^{(r,\ell)}$ and suppose $\lim_{n\rightarrow \infty}n/m\in(0,\infty)$. Then for each $r\in[R] $ and $\ell\in[2^r]$, under the discretized local null hypothesis $\bar{H}^{(r,\ell)}_0$ given in \eqref{eq:local_asymmetric_test}, we have, as $n,m \rightarrow \infty$,
\begin{equation}
\frac{m\cdot\lambda_{1}(\tilde{W}^{(r,\ell)})-(\sqrt{n}+\sqrt{m})^2}{(\sqrt{n}+\sqrt{m})(\frac{1}{\sqrt{n}}+\frac{1}{\sqrt{m}})^{1/3}}\stackrel{d}{\longrightarrow} \text{TW}_1.
\end{equation}
\end{theorem}

We relegate the proof of Theorem~\ref{thm:marchenko-pastur} to Section~\ref{sec:marchenko-pastur-proof} of the Appendix.

Using Theorem~\ref{thm:marchenko-pastur}, we can let $\lambda_{1}(\tilde{W}^{(r,\ell)})$ be the test statistics for the local test (\ref{eq:local_asymmetric_test}), and derive the p-value for the discretized local null as
\begin{align*}
\bar{p}^{(r,\ell)} &\equiv p^{(r,\ell)}(\tilde{W}^{(r,\ell)}) \\
&:= 2\text{min}\bigg(F_{\scalebox{1}{$\scriptscriptstyle \text{TW1}$}}\Big(\frac{m\cdot\lambda_{1}(\tilde{W}^{(r,\ell)})-(\sqrt{n}+\sqrt{m})^2}{(\sqrt{n}+\sqrt{m})(\frac{1}{\sqrt{n}}+\frac{1}{\sqrt{m}})^{1/3}}\Big), 1-F_{\scalebox{1}{$\scriptscriptstyle \text{TW1}$}}\Big(\frac{m\cdot\lambda_{1}(\tilde{W}^{(r,\ell)})-(\sqrt{n}+\sqrt{m})^2}{(\sqrt{n}+\sqrt{m})(\frac{1}{\sqrt{n}}+\frac{1}{\sqrt{m}})^{1/3}}\Big)\bigg)
\end{align*}

Steps 2, 3, and 4 proceed in the same way as the symmetric case, except that the resampling method changes slightly. To generate samples under the null in this scenario, we can just change the distribution of the random marks to be $\mathbb{P}\bigl(M_i = (u,v)\bigr)=\frac{1}{mn}$, $\forall i\in[N] $, $u \in V_1$ and $v \in V_2$. Then we generate sequences of random marks $\{M_1^{(b^*)}, \ldots, M_N^{(b^*)}\}$ under the aforementioned distribution and let the collection $\{N_{uv}^{(b^*)}(\cdot): u\in V_1, v\in V_2\}$ be a resample of the observed asymmetric array.

\subsection{Randomizing p-value}
\label{sec:pval_transform}

Let $X$ be a discrete random variable taking value on $\{x_1,\ldots, x_m\} \subset \mathbb{R}$ where we have the ordering $x_1 \leq x_2 \leq \ldots x_m$. Define $S(x) = \mathbb{P}(X \geq x)$ and 
\[
q_1 := S(x_1) = 1,\,\, q_2 := S(x_2),\ldots, q_m := S(x_m),\,\, q_{m+1} := 0,
\]
so that $S(X)$ takes value on $\{q_1, q_2,\ldots, q_m \}$. We define random variable $\tilde{S}$ such that if $S(X) = q_i$, then $\tilde{S} = q_{i+1}$. 

\begin{proposition}
\label{prop:pval_transform}
Let $U \sim \text{Unif}[0,1]$ be independent of $X$. Define
\begin{align}
Z := U \cdot S(X) + (1 - U)\cdot \tilde{S}.
\label{eq:pval_transform}
\end{align}
Then, we have that $Z \leq S(X)$ and that $Z \sim \text{Unif}[0,1]$.
\end{proposition}

\begin{proof}
Since $\tilde{S} < S(X)$ by definition, it is clear that $Z \leq S(X)$ as well. To show that $Z$ has the $\text{Unif}[0,1]$ distribution, fix $t \in (0, 1)$. Then there exists $i \in [m]$ such that $q_i \geq t > q_{i+1}$. We then have that
\begin{align*}
\mathbb{P}( Z \leq t) 
&= \mathbb{P}( S(X) \leq q_{i+1} ) + \mathbb{P}( Z \leq t, S(X) = q_i, \tilde{S} = q_{i+1}) \\
&= q_{i+1} + \mathbb{P}\biggl( U \leq \frac{t-q_{i+1}}{q_i - q_{i+1}} \biggr) \mathbb{P}( S(X) = q_i) = t,
\end{align*}
where the last inequality follows because $\mathbb{P}( S(X) = q_i) = \mathbb{P}(X = x_i) = S(x_i) - S(x_{i+1}) = q_i - q_{i+1}$. The Proposition follows as desired.

\end{proof}

\subsection{Proof of Proposition~\ref{prop:mcmc_ergodicity}}
\label{sec:mcmc_ergodicity_proof}

\begin{proof}

Let $T(\cdot|\cdot)$ be the transition probability of the Markov Chain specified via \eqref{eq:MH_proposal}, we first verify that the Metropolis-Hastings ratio is 1 by showing that $T(\cdot|\cdot)$ is a symmetric distribution, i.e., for any two sample vector $\bm{m}^{(1)}\ne\bm{m}^{(2)}\in \mathcal{M}_{\bm{d}}$, we have $T(\bm{m}^{(1)}|\bm{m}^{(2)}) = T(\bm{m}^{(2)}|\bm{m}^{(1)})$. It is obvious that the necessary condition for $T(\bm{m}^{(1)}|\bm{m}^{(2)})$ to be positive, is that there must exist exactly two indices $i\ne j\in[N]$ such that $m^{(1)}_i\ne m^{(2)}_i$, $m^{(1)}_j\ne m^{(2)}_j$ while the other elements are all the same for the two vectors. We can see that $T(\bm{m}^{(1)}|\bm{m}^{(2)}) = T(\bm{m}^{(2)}|\bm{m}^{(1)})=\frac{1}{5\binom{[N]}{2}}$ regardless of the values of $m^{(1)}_i, m^{(1)}_j, m^{(2)}_i, m^{(2)}_j$.

Next, we show that the Markov Chain is irreducible on the support $\mathcal{M}_{\bm{d}}$. By definition it suffice to show that for any two sample $\bm{m}^{(1)}\ne\bm{m}^{(2)}\in \mathcal{M}_{\bm{d}}$, there exist a finite steps path $\bm{m}^{(1)}\longrightarrow\bm{m}^{(2)}$. Since the vector of all degrees $\bm{D(m)}$ are identical for all $\bm{m}\in\mathcal{M}_{\bm{d}}$, we can show there exist a path $\bm{m}^{(1)}\longrightarrow\bm{m}^{(2)}$ which sequentially matching each element of $\bm{m^{(1)}}$ to be the same as in $\bm{m^{(2)}}$. To be specific, we denote the $j_{th}$ elements of vectors $\bm{m}^{(1)}$ and $\bm{m}^{(2)}$ as $\bm{m}^{(1)}_j = (u_j, v_j)$ and $\bm{m}^{(2)}_j = (u^\prime_j, v^\prime_j)$ respectively. Let $i = \min\left\{j\in[N]:\bm{m}^{(1)}_j\ne\bm{m}^{(2)}_j\right\}$, we first show that we can go from $\bm{m}^{(1)}$ to an intermediate state $\bm{m}^{(1,i)}\in \mathcal{M}_{\bm{d}}$ in finite steps, such that $\bm{m}^{(1,i)}_j = \bm{m}^{(2)}_j$ for all $j\leq i$. We can easily see for $\bm{m}^{(1)}_i \ne \bm{m}^{(2)}_i$, there could only be two cases 
\begin{enumerate}
    \item $u_i = u_i^\prime,  v_i\ne v_i^\prime$ or $u_i \ne u_i^\prime,  v_i=v_i^\prime$ \label{case:1}
    \item $u_i\ne v_i\ne u_i^\prime\ne v_i^\prime$ \label{case:2}
\end{enumerate}
For the first case, suppose $u_i = u_i^\prime,  v_i\ne v_i^\prime$, then by the fact $\bm{D}(\bm{m^{(1)}}) = \bm{D}(\bm{m^{(2)}})$ there must exist $s>i$ such that $u_s = v_i^\prime$ or $v_s = v_i^\prime$. Then by \eqref{eq:MH_proposal} we can easily check that we can go from $\bm{m}^{(1)}$ to $\bm{m}^{(1,i)}$ in one step with $T(\bm{m}^{(1,i)}|\bm{m}^{(1)})=\frac{1}{5\binom{[N]}{2}}$. For the second case, we can go from $\bm{m}^{(1)}$ to $\bm{m}^{(1,i)}$ in two steps where in the first step we move to a state $\bm{m}^{(1,i)\prime}$ such that $\bm{m}^{(1,i)\prime}_i = (u_i^\prime, v_j)$ which is just the state in the first case, so by the same reason we can move to $\bm{m}^{(1,i)}$ in the second step. Notice that the above paths does not depend on the index i, thus there exist an integer $t\leq2N$ such that we can sequentially move from $\bm{m}^{(1)}\longrightarrow\bm{m}^{(1,i)}\longrightarrow\cdots\longrightarrow\bm{m}^{(1,N)}=\bm{m}^{(2)}$ in $t$ steps.

Given any state $\bm{m}$, if not all edges are the same, i.e., $m_i = {(u,v)}$ for some $u,v\in[n]$ and all $i\in[N]$, then we can always find $m_i=(u_i,v_i)$ and $m_j=(u_j,v_j)$ with $i<j\in[N]$ such that the nodes $u_i,v_i,u_j,v_j$ satisfy one of the two cases listed above. For case \ref{case:1}, we can see that the five outcomes contains multi-edges, so the Markov Chain can stay at current state with positive probability and thus the period for this state is 1. For \ref{case:2}, we can easily check the state can return in $t$ steps for any $t\geq2$, thus the period for this state is also 1. Then by irreducibility, we can conclude that the Markov Chain is aperiodic.

With the Markov Chain being irreducible and aperiodic, it converges to its unique stationary distribution. Then by the construction of the Metropolis-Hastings algorithm below \eqref{eq:MH_proposal}, it is guaranteed that the stationary distribution is the target distribution, i.e., the uniform distribution on $\mathcal{M}_{\bm{d}}$.

\end{proof}

\subsection{Supplementary material for Section~\ref{sec:test_array_homogeneous}}

\label{sec:sup_test_array_homogeneous}
Recall that we let $A^{(r,\ell)}$, defined in \eqref{eq:ad_matrix_local}, be the adjacency matrix of a undirected Poisson Stochastic Block Model with $K$ communities. We denote $\bm{\sigma}$ as the membership vector and $\bm{\gamma}$ as the connection intensity between different communities, as discussed in Section~\ref{sec:array_homo_symm_step1}. Without loss of generality, we omit all the superscripts of $A^{(r,\ell)}$ that represents the partition of the support $\mathcal{X}$ and just use $A$ to denote adjacency matrix generated from Poisson Stochastic Block Model in all subsequent analysis in Section \ref{sec:sup_test_array_homogeneous} for simplicity. For the same reason, we also omit the subscripts of $B^{(r, \ell)}$, the matrix with entries being counts of interactions between two groups of individuals, in the proof of theorem \ref{thm:marchenko-pastur}.  

\subsubsection{Proof of Theorem~\ref{thm:tracy-widom}}

\label{sec:tracy-widom-proof}

\begin{proof}
Under the null hypothesis, we have that $\bm{P} =\gamma \mathbf{1}_n \mathbf{1}_n^\mathsf{T}$,  for some constant $\gamma > 0$.

Let $\tilde{A^\prime}$ be a $n \times n$ matrix such that
\begin{align*}
\tilde{A^\prime}_{uv} = \begin{cases}
(A_{uv} - \hat{\gamma})/\sqrt{(n-1)\gamma}, & u \ne v \\
(\gamma-\hat{\gamma})/\sqrt{(n-1)\gamma}, & u=v
\end{cases}
\end{align*}
Where $\hat{\gamma}=\frac{2}{n^2-n}\sum_{u<v}A_{uv}$ is an estimator of $\gamma$. Let $C_n = n(\gamma-\hat{\gamma})/\sqrt{(n-1)\gamma}$ and matrix $\tilde{A}^*$ be as defined in \eqref{eq:ad_matrix_star}. Then by definition we have that $\tilde{A^\prime} = \tilde{A}^* + \Delta^\prime$, where $\Delta^\prime=(\gamma-\hat{\gamma})\mathbf{1}_n \mathbf{1}_n^\mathsf{T}/\sqrt{(n-1)\gamma} = C_n \mathbf{1}_n \mathbf{1}_n^\mathsf{T}/n$.
Note that $\hat{\gamma}$ is the sample mean of $n(n-1)/2$ \emph{i.i.d} Poisson random variables with mean $\gamma$, we can apply the Poisson tail bound \eqref{eq:poisson_tail_bound} again and get
\begin{align*}
\mathbb{P}\left(\left|\gamma-\hat{\gamma}\right|>s\right) \leq 2\text{exp}\big\{-\frac{n(n-1) s^2}{4(\gamma + s)}\big\}
\end{align*} 
And thus we have $\left|\gamma-\hat{\gamma}\right| = o_p(\log n/n)$ and that $C_n = o_p(\log{n}/\sqrt{n})$.

Let $\mu_i^*$ be the eigenvector of $\tilde{A}^*$ corresponding to its $i_{th}$ largest eigenvalue. Then by Lemma~\ref{lem:ev_delocalization}, we have a lower bound on the largest eigenvalue of $\tilde{A^\prime}$:
\begin{align*}
\lambda_{1}(\tilde{A^\prime}) &\geq (\mu_1^*)^\mathsf{T}\tilde{A^\prime}\mu_1^*\\ 
&= \lambda_{1}(\tilde{A^*}) + (\mu_1^*)^\mathsf{T}\Delta^\prime\mu_1^*\\
&= \lambda_{1}(\tilde{A^*}) +C_n(\mu_1^*)^\mathsf{T}\mathbf{1}_n \mathbf{1}_n^\mathsf{T}\mu_1^*/n \\
&= \lambda_{1}(\tilde{A^*}) + \tilde{O}_p(1/n)\cdot o_p(\log{n}/\sqrt{n})\\
&\geq \lambda_{1}(\tilde{A^*}) - o_p(n^{-2/3}) 
\end{align*}

To derive the upper bound of $\lambda_{1}(\tilde{A^\prime})$, we denote $\mu_1^\prime$ as the eigenvector corresponding to the largest eigenvalue of $\tilde{A^\prime}$. Let $\{a_1,\ldots, a_n\}$ be the coordinates of the vector $\mu_1^\prime$ with respect to the basis $\{\mu_1^*, \ldots, \mu_n^*\}$, i.e., $\mu_1^\prime=\sum_{i=1}^{n}a_i \mu_i^*$.  Define $\mathcal{S}_{C_n}\subset[n]:=\{i\in[n]:\lambda_{i}(\tilde{A^*}) > \big(\lambda_{1}(\tilde{A^*}) - |C_n|\big)\}$ as the set of indices of those eigenvalues of $\tilde{A^*}$ that lies in the interval $(\lambda_{1}(\tilde{A^*})-|C_n|, \lambda_{1}(\tilde{A^*})]$. Then By Lemma \ref{lem:ev_delocalization} and the fact that $|C_n|$ is the largest eigenvalue of $\Delta^\prime$, we have
\begin{align*}
\lambda_{1}(\tilde{A^\prime})&=(\mu_1^\prime)^\mathsf{T}\tilde{A^\prime}\mu_1^\prime\\ 
&= (\mu_1^\prime)^\mathsf{T}\tilde{A^*}\mu_1^\prime + (\mu_1^\prime)^\mathsf{T}\Delta^\prime\mu_1^\prime \\
&\leq \sum_{i=1}^{n}a^2_i\lambda_{i}(\tilde{A^*}) + \bigl(\sum_{i\in \mathcal{S}_{Cn}}a_{i}(\mu_i^*)^\mathsf{T}\bigr)|\Delta^\prime|\bigl(\sum_{i\in \mathcal{S}_{Cn}}a_{i}(\mu_i^*)\bigr)\\
& \quad + \bigl(\sum_{i\in ([n]/\mathcal{S}_{C_n})}a_{i}(\mu_i^*)^\mathsf{T}\bigr)|\Delta^\prime| \bigl(\sum_{i\in ([n]/\mathcal{S}_{C_n})}a_{i}\mu_i^*\bigr)\\
&\leq \lambda_{1}(\tilde{A^*})\sum_{i\in \mathcal{S}_{C_n}}^{m}a^2_i + \big(\lambda_{1}(\tilde{A^*}) - |C_n|\big)\sum_{i\in ([n]/\mathcal{S}_{C_n})}a^2_i \\
&+ |\mathcal{S}_{C_n}|\cdot\sum_{i\in ([n]/\mathcal{S}_{C_n})}^{m}a^2_i(\mu_i^*)^{\mathsf{T}}|\Delta^\prime|\mu^*_i\quad + |C_n|\sum_{i\in ([n]/\mathcal{S}_{C_n})}^{m}a^2_i\\
&\leq \lambda_{1}(\tilde{A^*}) + |\mathcal{S}_{C_n}||C_n|\cdot\bigl(\sum_{i\in ([n]/\mathcal{S}_{C_n})}^{m}a^2_i\cdot\tilde{O}_p(1/n)\bigr)\\
&= \lambda_{1}(\tilde{A^*}) + |\mathcal{S}_{C_n}|\cdot\tilde{O}_p(1/n)\cdot o_p(\log{n}/\sqrt{n})
\end{align*}
Then we could bound the size of $\mathcal{S}_{C_n}$ by using the results from \cite{erdHos2012rigidity} and \cite{bickel2016hypothesis}, where the main idea is that the empirical counting of the eigenvalues is close to the semicircle counting functions.

Let $N(a,b)$ be the number of eigenvalues of $\tilde{A^*}$ lying in interval $(a,b]$, and define $N_{sc}(a,b):=n\int_a^b\rho_{sc}(x)dx$, where $\rho_{sc}=(1/2\pi)((4-x^2)_+)^{1/2}$ denote the the density of the semicircle law discussed in \cite{erdHos2012rigidity}. Following Theorem 2.2 in \cite{erdHos2012rigidity} and the discussion in \cite{bickel2016hypothesis}, there exist constant $A_0>1$, $C$, $c$ and $d<1$, such that for any $L$ satisfying the following:
\begin{align*}
A_0\text{loglog}n\leq L\leq \text{log}(10n)/\text{loglog}n    
\end{align*}
and for $|a|$, $|b|<5$, we have :
\begin{align*}
&\mathbb{P}\left(|N(a,b)-N_{sc}(a,b)|\geq 2(\text{log}n)^L\right)\\
\leq&\mathbb{P}\left(|N(-\infty,b)-N_{sc}(\infty,b)|\geq(\text{log}n)^L \right)+ \mathbb{P}\left(|N(-\infty,a)-N_{sc}(\infty,a)|\geq(\text{log}n)^L \right)\\
\leq&2C\text{exp}\{-c(\text{log}n)^{(-dL)}\}
\end{align*}
Notice that $\mathcal{S}_{C_n}=N\big(\lambda_{1}(\tilde{A^*}) - |C_n|,\lambda_{1}(\tilde{A^*})\big)$, and from the above inequality we have that: 
\begin{equation}
\mathcal{S}_{C_n}=N\big(\lambda_{1}(\tilde{A^*}) - |C_n|,\lambda_{1}(\tilde{A^*})\big)=N_{sc}\big(\lambda_{1}(\tilde{A^*}) - |C_n|,\lambda_{1}(\tilde{A^*})\big))+O_p(\text{log}n)^L
\end{equation}
And
\begin{align*}
N_{sc}\big(\lambda_{1}(\tilde{A^*}) - |C_n|,\lambda_{1}(\tilde{A^*})\big))&=n\int_{\lambda_{1}(\tilde{A^*})-|C_n|}^{\lambda_{1}(\tilde{A^*})}(\frac{1}{2\pi}((4-x^2)_+)^{1/2}dx \\
&\leq n\int_{\lambda_{1}(\tilde{A^*})-|C_n|}^{2}(\frac{1}{2\pi}((4-x^2)_+)^{1/2}dx\\
&=O(n|C_n|^{3/2})\\
&=o_p\big(n^{1/4}(\log n)^{3/2})\big)
\end{align*}
Where the second to last equality holds by using the area of a rectangle with side length ($2-|C_n|$) and $\sqrt{4-(2-|C_n|)^2}$ to cover the actual size of the integral.

Now we can see that
\begin{align*}
\lambda_{1}(\tilde{A^\prime}) &\leq \lambda_{1}(\tilde{A^*}) + |\mathcal{S}_{C_n}|\cdot\tilde{O}_p(1/n)\cdot o_p(\log{n}/\sqrt{n})\\
&= \lambda_{1}(\tilde{A^*}) + o_p\big(n^{1/4}(\log n)^{3/2})\big)\cdot \tilde{O}_p(1/n)\cdot o_p(\log{n}/\sqrt{n})\\
&= \lambda_{1}(\tilde{A^*}) + \tilde{O}_p((\log n)^{5/2}n^{-5/4})\\
&\leq \lambda_{1}(\tilde{A^*}) + o_p(n^{-2/3})
\end{align*}
And combining the lower and upper bound we have that
\begin{equation}
\lambda_{1}(\tilde{A^\prime}) = \lambda_{1}(\tilde{A^*}) + o_p(n^{-2/3})
\end{equation}

Now let's get back to the target matrix $\tilde{A}=\sqrt{\frac{\hat{\gamma}}{\gamma}}\Big(\tilde{A^\prime}-\frac{C_n}{n}\bm{I}_n\Big)$. By triangle inequality of matrix norm we have
\begin{align*}
\big\|\tilde{A^\prime}\big\| - \big\|\frac{C_n}{n}\bm{I}_n\big\| \leq \big\|\tilde{A^\prime}-\frac{C_n}{n}\bm{I}_n\big\| \leq \big\|\tilde{A^\prime}\big\| + \big\|\frac{C_n}{n}\bm{I}_n\big\|
\end{align*}
And since $\|\frac{C_n}{n}\bm{I}_n\| = |C_n/n| = o_p(\log n\cdot n^{-3/2}) $, we could easily see that
\begin{align*}
\lambda_{1}(\tilde{A}) &= \sqrt{\hat{\gamma}/{\gamma}}\cdot\big\|\tilde{A^\prime}-\frac{C_n}{n}\bm{I}_n\big\|\\
&= \big(1+o_p(\log n/n)\big)\big(\lambda_{1}(\tilde{A^\prime})+o_p(\log n\cdot n^{-3/2})\big)\\
&= \lambda_{1}(\tilde{A^\prime}) + o_p(n^{-2/3}) \\
&= \lambda_{1}(\tilde{A^*}) + o_p(n^{-2/3})
\end{align*}
Finally by Lemma \ref{lem:nonrandom_tracy_widom} and Slutsky's lemma, we have
\begin{align*}
n^{2/3}\big(\lambda_{1}(\tilde{A}) - 2\big) \stackrel{d}{\longrightarrow} \text{TW}_1.
\end{align*}
\end{proof} 

\begin{lemma}
\label{lem:nonrandom_tracy_widom}
Let $\bm{P}$ be defined in \eqref{eq:nodes_prob_matrix} and $\tilde{A}^*$ be a matrix such that
\begin{align}
\label{eq:ad_matrix_star}
\tilde{A}^*_{uv} = \begin{cases}
(A_{uv} - \bm{P}_{uv})/\sqrt{(n-1)\bm{P}_{uv}}, & u \ne v \\
0, & u=v
\end{cases}
\end{align}
Then we have 
\[
n^{2/3}(\lambda_{1}(\tilde{A}^*)-2) \stackrel{d}{\longrightarrow} \text{TW}_1.
\]
\end{lemma}
\begin{proof}
Consider a $ n\times n$ real symmetric Wigner matrix
\[
G^*_{uv} = \frac{1}{\sqrt{n-1}}x_{uv},\quad  1\leq u,v \leq n
\]
Where the off-diagonal elements are \emph{i.i.d.} standard normal distributed random variables and the diagonal elements are zeros. Theorem 1.2. in \cite{lee2014necessary} implies that the largest eigenvalue of $G^*$ weakly converges to the Tracy-Widom distribution.

Next, by tail bound of Poisson random variables, for any $s>0$ and $1 \leq u,v \leq n$ we have
\begin{align}
\label{eq:poisson_tail_bound}
\mathbb{P}\left(\left|\tilde{A}^*_{uv}\right|>\frac{s}{\sqrt{n-1}}\right) \leq 2\text{exp}(-\frac{\bm{P}_{uv} s^2}{2(\bm{P}_{uv}+\sqrt{\bm{P}_{uv}}s)})
\end{align}
And thus there exist a constant $\mathit{v}$ independent of $n$, such that for any $s \geq 1$ we have
\begin{align*}
\mathbb{P}\left(\left|\tilde{A}^*_{uv}\right|>\frac{s}{\sqrt{n-1}}\right) \leq \mathit{v}^{-1}\text{exp}(-s^{\mathit{v}})
\end{align*}
The above inequality shows that the entries of $\tilde{A}^{*}$ have a uniformly subexponential decay, and thus by Theorem 2.4 in \cite{erdHos2012rigidity}, we have that $n^{2/3}(\lambda_{1}(\tilde{A}^*)-2)$ converges to $n^{2/3}(\lambda_{1}(G^*)-2)$ in distribution.
\end{proof}

\begin{lemma}
\label{lem:ev_delocalization}
	For each $1\leq i\leq n$, let $\mu_i^*$ be the eigenvector of $\tilde{A}^*$ corresponding to the $i_{th}$ largest eigenvalue $\lambda_i(\tilde{A}^*)$. Then for any deterministic unit vector $\mathbf{v}$, we have 
\begin{align}
((\mu_i^*)^\mathsf{T} \mathbf{v})^2=\tilde{O}_p(1/n), \ \ \text{uniformly for all}\ i\in[n]
\end{align}
Where we define $a_{n}=\tilde{O}_p(b_n)$, if for any $\epsilon>0$ and $D>0$, there exists $n_0=n_0(\epsilon,D)$ such that
\begin{align*}
\mathbb{P}(a_n\geq n^{\epsilon}b_n)\leq n^{-D} \ \  \text{for all} \ n\geq n_0.
\end{align*}
\end{lemma}
Lemma \ref{lem:ev_delocalization} is a direct application of the eigenvector delocalization theorem proposed in \cite{alex2014isotropic}. Note that the conditions of Theorem 2.16 in \cite{alex2014isotropic} does not apply to our configuration of $\tilde{A}^*$ since the diagonal entries are made to be all zeros while the original condition requires that all elements of the matrix should have positive variance. However, \cite{erdHos2013local} provides a local semicircle law(Theorem 2.3) which holds even when some entries of a generalized Wigner matrix have zero variance, and as a result of the local semicircle law, the eigenvector delocalization theorem still holds in our setting. See also discussions in \cite{bickel2016hypothesis} and \cite{lei2016goodness}.

\subsubsection{Maximum eigenvalue test statistic under an alternative}
\label{sec:eigen_dist_alt}

We consider the limiting distribution of $\lambda_1(\tilde{A}^{(r,\ell)})$ under some alternative cases. When the adjacency matrix is generated from a Stochastic Block Model with $K > 1$ communities and Bernoulli entries, \cite{bickel2013hypothesis} shows that the largest eigenvalue of the scaled and centered adjacency matrix is $O(\sqrt{n})$, given that the community probability matrix $\bm{\psi}$ is diagonally dominant. \cite{lei2016goodness} provided a more general result  which requires that each community has size at least proportional to $n/K$. The following proposition is a direct extension of Theorem 3.3 in \cite{lei2016goodness} to the Poisson network.
\begin{proposition}
\label{prop:eigen_dist_alt}
Let $A^{(r, \ell)}$ be an adjacency matrix from Poisson stochastic model with K communities and let $\mathcal{G}_{k}=\{u\in[n]: \sigma(u)=k\}$ be the set of vertices that belong to group $k$ for $k \in [K]$.

Assume there exist a constant $C_K > 0$ such that for all $n$ we have 
\begin{equation}
\label{eq:community_size}
\underset{k\in[K]}{\mathrm{min}}|\mathcal{G}_k| \geq C_K \cdot n
\end{equation}
Then for any $r\in [R]$ and $\ell \in [2^r]$, if $K>1$ we have
\begin{align}
\lambda_{1}(\tilde{A}^{(r,\ell)}) \geq \frac{\sqrt{n}\delta C_K - O_p(1)}{\big(\|\bm{\gamma}\|_{\text{max}}+o_p(\log{n}/n)\big)^{1/2}}
\end{align}
where $\delta$ is the minimum $\ell_\infty$  distance between any two distinct rows of $\bm{\gamma}$.
\end{proposition}

\begin{proof}
	
Let $\hat{\bm{P}} = \hat{\gamma}\mathbf{1}_n \mathbf{1}_n^\mathsf{T} $, we have
\begin{align*}
\big\|\tilde{A}\big\| &= \big((n-1)\hat{\gamma}\big)^{-1/2}\big\|A-\big(\hat{\bm{P}} - \text{diag}(\hat{\bm{P}})\big)\big\| \\
&\geq \big((n-1)\hat{\gamma}\big)^{-1/2}\Big(\big\|\bm{P}-\hat{\bm{P}}-\text{diag}(\bm{P}-\hat{\bm{P}})\big\|-\big\|A-\big(\bm{P}-\text{diag}(\bm{P})\big)\big\|\Big)
\end{align*}
We can see that the matrix $A-\big(\bm{P}-\text{diag}(\bm{P})\big)$ has off-diagonal entries being independent, centered Poisson random variables and diagonal entries being all zeros.  By Theorem 2 in \cite{latala2005some}, we have that there exist some $C^\prime>0$ such that
\begin{equation}
\mathbb{E}\big\|A-\big(\bm{P}-\text{diag}(\bm{P})\big)\big\| \leq C^\prime \sqrt{n}
\end{equation}
and thus $\big\|A-\big(\bm{P}-\text{diag}(\bm{P})\big)\big\| \leq O_p(\sqrt{n})$.

To derive an upper bound of $\big\|\bm{P}-\hat{\bm{P}}-\text{diag}(\bm{P}-\hat{\bm{P}})\big\|$, we notice that since $K>1$, there exist two community $k_1 \ne k_2$. Let $\bm{g}_{k_1} = \{u\in[n]: \sigma(u)=k_1\}$ and $\bm{g}_{k_2} = \{u\in[n]: \sigma(u)=k_2\}$ be the set of vertices that belong to $k_1$ and $k_2$ respectively. Since we assume the matrix $\bm{\gamma}$ have pairwise distinct rows, there must exist a group $k_3\in[K]$ such that $\bm{\gamma}_{k_1 k_3} \ne \bm{\gamma}_{k_2 k_3}$, and we can choose
\[
 k_3 = \underset{k^\prime\in[K]}{\mathrm{argmin}}|\bm{\gamma}_{k_1 k^\prime} - \bm{\gamma}_{k_2 k^\prime}|.
\]
Note that $k_3$ can be equal to $k_1$ or $k_2$. Now let $\bm{D}$ be a submatrix of $\bm{P}-\hat{\bm{P}}-\text{diag}(\bm{P}-\hat{\bm{P}})$, which only consist the rows in $k_1 \cup k_2$ and columns in $k_3$. We can see that when $k_1\ne k_2 \ne k_3$, after some row permutaions D could be seen as:
\[ \bm{D} = \begin{pmatrix}
D_1\\
D_2
\end{pmatrix}
\]
where $D_1$ is a $|k_1|\times |k_3|$ matrix with all entries equal to $\bm{\gamma}_{k_1 k_3}-\hat{\gamma}$ and $D_2$ is a $|k_2|\times |k_3|$ matrix with all entries equal to $\bm{\gamma}_{k_2 k_3}-\hat{\gamma}$. Then we have
\begin{align*}
\big\|\bm{D}\big\| &\geq \text{max}\Big((\bm{\gamma}_{k_1 k_3}-\hat{\gamma})\sqrt{|k_1|\cdot|k_3|}, (\bm{\gamma}_{k_2 k_3}-\hat{\gamma})\sqrt{|k_2|\cdot|k_3|}\Big)\\
& \geq n\delta C_K
\end{align*}

When $k_3=k_1$ or $k_3=k_2$, we can see $\bm{D}$ could still be permuted into a block matrix with blocks $D_1$ and $D_2$. However in this case, one of the blocks have all diagonal entries being zeros, since we do not allow self-loops. Without loss of generality, we assume that $k_3 = k_1$, and we still have the same lower bound of $\big\|\bm{D}\big\|$ by
\begin{align*}
\big\|\bm{D}\big\| &\geq \text{max}\Big((\bm{\gamma}_{k_1 k_1}-\hat{\gamma})(|k_1|-1), (\bm{\gamma}_{k_2 k_1}-\hat{\gamma})\sqrt{|k_2|\cdot|k_1|}\Big)\\
& \geq n\delta C_K - O_p(1)
\end{align*}

Finally we have
\begin{align*}
\big\|\tilde{A}\big\| &\geq  \big((n-1)\hat{\gamma}\big)^{-1/2}\big( n\delta C_K - O_p(\sqrt{n})\big)\\
&\geq \frac{\sqrt{n}\delta C_K - O_p(1)}{\big(\|\bm{\gamma}\|_{\text{max}}+o_p(\log{n}/n)\big)^{1/2}}
\end{align*}
\end{proof} 

\subsubsection{Proof of Theorem~\ref{thm:marchenko-pastur}}
\label{sec:marchenko-pastur-proof}

\begin{proof}
Under the null hypothesis, we have that $B_{uv} \sim \text{Poisson}(\gamma)$ for some $\gamma>0$ and for any $u\in V_1, v \in V_2$.
Recall that we denote $\hat{\gamma}$ \eqref{eq:asymmetric_mean_estimate} as an estimator of $\gamma$, matrix $\Tilde{B}$ as the empirically centered and scaled counterpart of $B$ and $\Tilde{W} = \Tilde{B}^\mathsf{T}\Tilde{B}$.

Let $\Tilde{B}^*$ be as defined in \eqref{eq:asymmetric_ad_star} and  $\Tilde{B}^\prime$ be a $m \times n$ matrix with entries $\Tilde{B}_{uv}^\prime = (B_{uv} - \hat{\gamma}) / \sqrt{m \gamma}$. Then we have $\Tilde{B}^\prime=\Tilde{B}^{*}+\alpha \Delta$, where $\Delta=\mathbf{1}_m\mathbf{1}_n^\mathsf{T}$ and $\alpha=\frac{\gamma-\hat{\gamma}}{\sqrt{m\gamma}}$. Denote $\Tilde{W}^* = \Tilde{B}^{*\mathsf{T}}\Tilde{B}^*$ and $\Tilde{W}^\prime = \Tilde{B}^{\prime\mathsf{T}}\Tilde{B}^\prime$.
Then we let $\bigl(\lambda_i(\tilde{W}^*),\mu_i^*\bigr)_{i=1}^{n}$ be the pairs of eigenvalue and eigenvector of matrix $\tilde{W}^*$ with the eigenvalues in a non-increasing order, namely $\lambda_1(\tilde{W}^*)\geq \lambda_2(\tilde{W}^*) \cdots \geq \lambda_n(\tilde{W}^*)$. Similarly, we let $(\lambda_i(\Tilde{W}^\prime), \mu_i^\prime)_{i=1}^{n}$ be the pairs of eigenvalue and eigenvectors of $\Tilde{W}^\prime$, where the eigenvalues are in non-increasing order as well.

First let us derive the a lower bound on $\lambda_1(\tilde{W}^\prime)$, the largest eigenvalue of $\Tilde{W}^\prime = \Tilde{B}^{\prime\mathsf{T}}\Tilde{B}^\prime$:
\begin{align*}
\lambda_1(\tilde{W}^\prime)&\geq(\mu_1^*)^\mathsf{T}\tilde{B}^{\prime\mathsf{T}}\Tilde{B}^\prime\mu_1^*\\
&=(\mu_1^*)^\mathsf{T}\tilde{B}^{*\mathsf{T}}\tilde{B}^*\mu_1^*+\alpha(\mu_1^*)^\mathsf{T}(\tilde{B}^{*\mathsf{T}}\Delta+\Delta^\mathsf{T}\tilde{B}^*+\alpha\Delta^\mathsf{T}\Delta)\mu_1^*\\
&=\lambda_1(\Tilde{W}^*)+\alpha(\mu_1^*)^\mathsf{T}(\tilde{B}^{*\mathsf{T}}\Delta+\Delta^\mathsf{T}\tilde{B}^*+\alpha\Delta^\mathsf{T}\Delta)\mu_1^*\\
&\geq \lambda_1(\Tilde{W}^*) -|\alpha(\mu_1^*)^\mathsf{T}(\tilde{B}^{*\mathsf{T}}\Delta+\Delta^\mathsf{T}\tilde{B}^*+\alpha\Delta^\mathsf{T}\Delta)\mu_1^*|
\end{align*}

Let $\tilde{B}^*=\sum_{i=1}^{n}\sqrt{\lambda_i(\Tilde{W}^*)}s_i^*\mu_i^{*\mathsf{T}}$ be the singular value decomposition of $\tilde{B}^*$. Then we have:
\begin{align*}
\mu_i^{*\mathsf{T}}\tilde{B}^{*\mathsf{T}}=\sum_{i=1}^{n}\sqrt{\lambda_i(\Tilde{W}^*)}\mu_1^{*\mathsf{T}}\mu_i^*s_i^{*\mathsf{T}}=\sqrt{\lambda_i(\Tilde{W}^*)}s_i^{*\mathsf{T}}  
\end{align*}
Notice that $s_i^*, \mu^*_i$ are the eigenvectors of the matrix $\tilde{B}^*\tilde{B}^{*\mathsf{T}}, \tilde{B}^{*\mathsf{T}}\tilde{B}^*$ respectively, and we can easily check the conditions in Lemma \ref{lem:isotropic_delocalization} hold for both matrix $\Tilde{B}^*$ and its transpose, thus we have 
\begin{align*}
\alpha\mu_1^{*\mathsf{T}}\tilde{B}^{*\mathsf{T}}\Delta\mu_1^*&=\alpha\sqrt{\lambda_1(\Tilde{W}^*)}s_i^{*\mathsf{T}}\mathbf{1}_m\mathbf{1}_n^{\mathsf{T}}\mu_1^*\\
&=o_{p}(\log n/n^{3/2})\cdot\tilde{O}_p(1)\cdot\lambda_1(\Tilde{W}^*)\\
&=\tilde{O}_p(\log n/n^{3/2})\cdot O_p(n^{-1/3})
 \end{align*}
Where the $o_{p}(\log n/n^{3/2})$ term is derived by noticing that $\hat{\gamma}$ is the sample mean of $m\times n$ independent Poisson random variables, and again by the Poisson tail bound \eqref{eq:poisson_tail_bound} we get $\alpha=\frac{\gamma-\hat{\gamma}}{\sqrt{m\gamma}}=o_{p}(\log n/n^{3/2})$. On the other hand it is easily seen that $\alpha^{2}\mu_1^{*\mathsf{T}}\Delta^\mathsf{T}\Delta\mu_1^*=\tilde{O}_p(1)\cdot o_{p}((\log n)^2/n^2)$, which indicates that:
\begin{align}
\label{eq:lower_bound_covar_eigenvalue}
\lambda_1(\tilde{W}^\prime) &\geq \lambda_1(\Tilde{W}^*) -|\alpha(\mu_1^*)^\mathsf{T}(\tilde{B}^{*\mathsf{T}}\Delta+\Delta^\mathsf{T}\tilde{B}^*+\alpha\Delta^\mathsf{T}\Delta)\mu_1^*|\nonumber \\
&\geq \lambda_1(\Tilde{W}^*)  -\tilde{O}_p(\log n/n^{3/2})\cdot O_p(n^{-1/3}) -\tilde{O}_p((\log n)^2/n^2)\nonumber \\
&=\lambda_1(\Tilde{W}^*) -\tilde{O}_P(\log n\cdot n^{-11/6})
\end{align}

Next, we  derive an upper bound for the largest eigenvalue of $\Tilde{W}^\prime$. Let $\{a_1, \ldots,a_n\}$ be the coordinates of $\mu_1^\prime$, the eigenvector of $\Tilde{W}^\prime$ associated with its largest eigenvalue, with respect to the basis consisting of eigenvector of $\Tilde{W}^*$, i.e., $\mu_1^\prime = \sum_{i=1}^{n}a_i\mu_i^*$. Let $\mathcal{S} = \bigl\{i\in[n]:\lambda_i(\Tilde{W}^*) >\lambda_1(\Tilde{W}^*)-2\|\alpha(\tilde{B}^{*\mathsf{T}}\Delta+\Delta^\mathsf{T}\tilde{B}^*+\alpha\Delta^\mathsf{T}\Delta)\| \bigr\}$, such that $|\mathcal{S}|$ is the number of $\lambda_i(\Tilde{W}^*)^\prime$s in the interval $\bigl(\lambda_1(\Tilde{W}^*)-2\|\alpha(\tilde{B}^{*\mathsf{T}}\Delta+\Delta^\mathsf{T}\tilde{B}^*+\alpha\Delta^\mathsf{T}\Delta)\|,\lambda_1(\Tilde{W}^*)\bigr)$. Let $\mathbf{v}_1=\sum_{i=1}^{m}a_i\mu_i^*$ and $\mathbf{v}_2=\sum_{i=m+1}^{n}a_i\mu_i^*$ so that $\mu_1^\prime = \mathbf{v}_1 + \mathbf{v}_2$. we have:
\begin{align*}
\lambda_1(\Tilde{W}^\prime) &= (\mu_1^{\prime})^\mathsf{T}\Tilde{W}^\prime\mu_1^\prime\\
&=(\mu_1^\prime)^\mathsf{T}\tilde{B}^{*\mathsf{T}}\tilde{B}^*\mu_1^\prime+\alpha(\mu_1^\prime)^\mathsf{T}(\tilde{B}^{*\mathsf{T}}\Delta+\Delta^\mathsf{T}\tilde{B}^*+\alpha\Delta^\mathsf{T}\Delta)\mu_1^\prime\\
&\leq \lambda_1(\Tilde{W}^*)\cdot\sum_{j\in\mathcal{S}}a^2_j+\bigl(\lambda_1(\Tilde{W}^*)-2\|\alpha(\tilde{B}^{*\mathsf{T}}\Delta+\Delta^\mathsf{T}\tilde{B}^*+\alpha\Delta^\mathsf{T}\Delta)\|\bigr)\cdot\sum_{j\in([n]/\mathcal{S})}a^2_j\\
&\quad +2\mathbf{v}_1^\mathsf{T}|\alpha(\tilde{B}^{*\mathsf{T}}\Delta+\Delta^\mathsf{T}\tilde{B}^*+\alpha\Delta^\mathsf{T}\Delta)|\mathbf{v}_1+2\mathbf{v}^\mathsf{T}_2|\alpha(\tilde{B}^{*\mathsf{T}}\Delta+\Delta^\mathsf{T}\tilde{B}^*+\alpha\Delta^\mathsf{T}\Delta)|\mathbf{v}_2\\
&\leq\lambda_1(\Tilde{W}^*)-2\|\alpha(\tilde{B}^{*\mathsf{T}}\Delta+\Delta^\mathsf{T}\tilde{B}^*+\alpha\Delta^\mathsf{T}\Delta)\|\sum_{j\in([n]/\mathcal{S})}a^2_j\\
&\quad +2m\sum_{j\in\mathcal{S}}a^2_j(\mu_j^*)^\mathsf{T}|\alpha(\tilde{B}^{*\mathsf{T}}\Delta+\Delta^\mathsf{T}\tilde{B}^*+\alpha\Delta^\mathsf{T}\Delta)|\mu^*_j\\
&\quad +2\|\alpha(\tilde{B}^{*\mathsf{T}}\Delta+\Delta^\mathsf{T}\tilde{B}^*+\alpha\Delta^\mathsf{T}\Delta)\|\sum_{j\in([n]/\mathcal{S})}a^2_j\\
&\leq\lambda_1(\Tilde{W}^*)+2|\mathcal{S}|\sum_{j\in\mathcal{S}}a^2_j(\mu_j^*)^\mathsf{T}\bigl(|2\alpha(\tilde{B}^{*\mathsf{T}}\Delta|+|\alpha^2\Delta^\mathsf{T}\Delta)|\bigr)\mu^*_j\\
&\leq \lambda_1(\Tilde{W}^*) + 2|\mathcal{S}|\sum_{j\in\mathcal{S}}a^2_j\bigl(\lambda_j(\Tilde{W}^*)\cdot\Tilde{O}_p(\log n/n^{3/2}) + \Tilde{O}_p((\log n)^2/n^2)\bigr)\\
&\leq \lambda_1(\Tilde{W}^*) + 2|\mathcal{S}|\bigl(O_p(n^{-1/3})\cdot\Tilde{O}_p(\log n/n^{3/2}) + \Tilde{O}_p((\log n)^2/n^2)\bigr)\\
&\leq \lambda_1(\Tilde{W}^*) + 2|\mathcal{S}|\cdot \Tilde{O}_p(\log n \cdot n^{-11/6})
\end{align*}
Now let $a=\lambda_1(\Tilde{W}^*)-2\|\alpha(\tilde{B}'\Delta+\Delta'\tilde{B}+\alpha\Delta'\Delta)\|$ and $b=\lambda_1(\Tilde{W}^*)$. We can see that $|\mathcal{S}|=\mathcal{N}(a)-\mathcal{N}(b)$. Noticing that $\left|(\mathcal{N}(a)-\mathcal{N}(b))-(\mathcal{N}_{m}(a)-\mathcal{N}_{m}(b))\right|=|(\mathcal{N}(a)-\mathcal{N}_{m}(a))-(\mathcal{N}(b)-\mathcal{N}_{m}(b))|$ and together with Lemma \ref{lem:rigidity_covariance} we have that for any $\varepsilon > 0$
\begin{align*}
&\mathbb{P}(|(\mathcal{N}(a)-\mathcal{N}(b))-(\mathcal{N}_{m}(a)-\mathcal{N}_{m}(b))|\geq 2n^{-1}\log(n)^{C_\varepsilon \log\log(n)})\\
\leq&\mathbb{P}[(\mathcal{N}(a)-\mathcal{N}_{m}(a))\geq n^{-1}\log(n)^{C_\varepsilon \log\log(n)}]+\mathbb{P}[(\mathcal{N}(b)-\mathcal{N}_{m}(b))\geq n^{-1}\log(n)^{C_\varepsilon \log\log(n)}]\\
\leq&2n^{C_\varepsilon}\exp(-\log(n)^{\varepsilon\log\log(n)})
\end{align*}
and which indicates that $|\mathcal{S}|=|\mathcal{N}_{m}(a)-\mathcal{N}_{m}(b)|+O_p\left(n^{-1}\log(n)^{C_\varepsilon \log\log(N)}\right)$. Since $\mathcal{N}_{m}(a)-\mathcal{N}_{m}(b)=n\int_{a}^{b}\varrho_{m}(x)dx$, and it is easily seen by simple calculus that $\varrho_{m}(x)$ achieves it's local maximum at $x=\frac{(1-n/m)^2}{1+n/m}$, and $\varrho_{m}(\frac{(1-n/m)^2}{1+n/m})=\frac{1}{\pi|1-n/m|\sqrt{n/m}}$. Thus we could have the following bound on the size of $|\mathcal{N}_{m}(a)-\mathcal{N}_{m}(b)|$:
\begin{align*}
|\mathcal{N}_{m}(a)-\mathcal{N}_{m}(b)|\leq 2n\|\alpha(\tilde{B}^{*\mathsf{T}}\Delta+\Delta^\mathsf{T}\tilde{B}^*+\alpha\Delta^\mathsf{T}\Delta)\|\frac{1}{\pi|1-\frac{n}{m}|\sqrt{\frac{n}{m}}}
\end{align*}
And we could see that
\begin{align*}
\norm{\alpha(\tilde{B}^{*\mathsf{T}}\Delta+\Delta^\mathsf{T}\tilde{B}^*+\alpha\Delta^\mathsf{T}\Delta)}&\leq |\alpha|\big(\norm{\tilde{B}^{}*\mathsf{T}\Delta}+\norm{\Delta'\tilde{B}^*}+|\alpha|\cdot\norm{\Delta'\Delta}\big)\\
&\leq|\alpha|\big(2\norm{\tilde{B}^*}\cdot\norm{\Delta}_F+|\alpha|\cdot\norm{\Delta}_{F}^2\big)\\
&=o_p(\log n\cdot n^{-3/2})\bigl(2\sqrt{mn\lambda_1(\Tilde{W}^*)}+mn\cdot o_p(\log n\cdot n^{-3/2})\bigr)\\
&=o_p(\log n \cdot n^{-5/6})
\end{align*}
Thus we have that $|\mathcal{S}|=|\mathcal{N}_m(a)-\mathcal{N}_m(b)|+O_p\left(n^{-1}\log(n)^{C_\varepsilon \log\log(n)}\right)\leq o_p(\log n \cdot n^{1/6})$ and
\begin{align}
\label{eq:upper_bound_covar_eigenvalue}
\lambda_1(\Tilde{W}^\prime) 
&\leq \lambda_1(\Tilde{W}^*) + |\mathcal{S}|\cdot\tilde{O}_P(\log n \cdot n^{-11/6})\nonumber\\
&\leq \lambda_1(\Tilde{W}^*) + \Tilde{O}_p\bigl((\log n)^2\cdot n^{-17/6}\bigr)
\end{align}    
Combine \eqref{eq:lower_bound_covar_eigenvalue} and \eqref{eq:upper_bound_covar_eigenvalue} we conclude that
\begin{equation}
\lambda_1(\Tilde{W}^\prime) = \lambda_1(\Tilde{W}^*) + o_p(n^{-2/3})
\end{equation}
Finally we can look at the target matrix $\tilde{B}$. We see that $\tilde{B}=\tilde{B}^\prime\times \sqrt{\frac{\gamma}{\hat{\gamma}}}$ and we can derive from the Poisson tail bond \eqref{eq:poisson_tail_bound} that $\sqrt{\frac{\gamma}{\hat{\gamma}}} = o_p(\log n \cdot n^{-1/2})$. Thus we can also have
\begin{align*}
\lambda_1(\Tilde{W}^\prime)&=\sqrt{\frac{\gamma}{\hat{\gamma}}}\lambda_1(\Tilde{W}^\prime)\\
&=\bigl(1+o_p(\log n \cdot n^{-1/2})\bigr)\bigl(\lambda_1(\Tilde{W}^*) + o_p(n^{-2/3})\bigr)\\
&= \lambda_1(\Tilde{W}^*) + o_p(n^{-2/3})
 \end{align*}
Then by Lemma \ref{lem:nonrandom_tw_covar} and Slutsky's theorem, we get the result of Theorem \ref{thm:marchenko-pastur}.
\end{proof}

\begin{lemma}
\label{lem:nonrandom_tw_covar}
Let $\Tilde{B}^*_{uv}$ be a matrix with entries:
\begin{equation}
\label{eq:asymmetric_ad_star}
\Tilde{B}^*_{uv}:=\frac{B_{uv}-\gamma}{\sqrt{m\gamma}}, \,\, \forall u\in[m], v\in[n]
\end{equation}
and let $\Tilde{W}^* = \Tilde{B}^{*\mathsf{T}}\Tilde{B}^*$ with $\lambda_1(\Tilde{W}^*)$ being its largest eigenvalue. Suppose that $\lim_{n\rightarrow \infty}n/m\in(0,\infty)$, then we have, as m,n $\rightarrow \infty$:
\begin{equation}
    \frac{m\lambda_1(\Tilde{W}^*)-(\sqrt{n}+\sqrt{m})^2}{(\sqrt{m}+\sqrt{n})(\frac{1}{\sqrt{n}}+\frac{1}{\sqrt{m}})^{1/3}} \stackrel{d}{\longrightarrow} \text{TW}_1, 
\end{equation}
\end{lemma}

\begin{lemma}[Theorem 2.8 in \cite{alex2014isotropic}]
\label{lem:isotropic_delocalization}
Let $G$ be an $m \times n$ random matrix with independent entries satisfying  
\begin{align*}
    \mathbb{E} G_{uv} = 0, \quad \mathbb{E}|G_{uv}|^2 = \frac{1}{\sqrt{nm}}.
\end{align*}
Assume that m and n satisfy the bounds $n^{1/C} \leq m \leq n^C$ for some $C>0$. Suppose for all $p\in \mathbb{N}$, there exist $C_p$ such that
\begin{align*}
    \mathbb{E}|(mn)^{1/4}G_{uv}|^p \leq C_p
\end{align*}
Let $\bm{\mu}_i$ be the eigenvalue of $G^\mathsf{T}G$ associated with its $i_{th}$ largest eigenvalue. Then for any $\varepsilon>0$ we have
\begin{align*}
|\bm{\mu}_i^\mathsf{T}\mathbf{v}|^2=\tilde{O}_P(1/n)    
\end{align*}
uniformly for all $i\leq(1-\varepsilon)\min{(m,n)}$ and any deterministic unit vector $\mathbf{v}\in \mathbb{R}^{n}$.
\end{lemma}

\begin{lemma}[Theorem 3.3 in \cite{pillai2014universality}]
\label{lem:rigidity_covariance}
Let $\xi_\pm=(1\pm \sqrt{\frac{n}{m}})^2$, and denote the Marchenko-Pastur law by $\varrho_m$, which is given by
\begin{align*}
\varrho_m(x)=\frac{m}{2\pi n}\sqrt{\frac{[(\xi_+-x)(x-\xi_-)]_+}{x^2}}
\end{align*}
Let $\beta \in \mathbb{R}$, define the empirical spectral distribution of $(\tilde{B}^*)^\mathsf{T}\tilde{B}^*$ by: 
\begin{equation*}
\mathcal{N}(\beta)\,:=\frac{1}{n}\sum_{i=1}^{n}\mathbbm{1}_{[\beta,\infty)}\bigl(\lambda_i(\Tilde{W}^*)\bigr)
\end{equation*}
And the distribution given by the Marchenko-Pastur law:
\begin{equation*}
\mathcal{N}_{m}(\beta)\,: =\int_{\beta}^{\infty}\varrho_m(x)dx
\end{equation*}
If $\lim_{n\rightarrow\infty}\frac{n}{m}\in(0,\infty)\setminus\{1\}$, then for any $\varepsilon>0$, there exists a constant $C_\varepsilon$ such that:
\begin{align*}
\mathbb{P}(|\mathcal{N}(\beta)-\mathcal{N}_{m}(\beta)|\geq n^{-1}\log(n)^{C_\varepsilon \log\log(n)})\leq n^{C_\varepsilon}\exp(-\log(n)^{\varepsilon\log\log(n)})
\end{align*}
\end{lemma}

\section{Supplementary material for Section~\ref{sec:power_analysis}}

For $r \in [R]$ and $\ell \in [2^r]$, define
\begin{align}
\delta^{(r,\ell)} := \frac{1}{2} \frac{\int_{I^{(r)}_\ell} \lambda_a - \lambda_b\, d\nu}
{ \int_{I^{(r)}_\ell} \lambda_a + \lambda_b\, d\nu}, \label{eq:delta_defn}
\end{align}
so that $\frac{\int_{I^{(r)}_{\ell}} \lambda_a \, d\nu}{ \int_{I^{(r)}_{\ell}} \lambda \, d\nu } = \frac{1}{2} + \delta^{(r,\ell)}$ and $\frac{\int_{I^{(r)}_{\ell}} \lambda_b \, d\nu}{ \int_{I^{(r)}_{\ell}} \lambda \, d\nu } = \frac{1}{2} - \delta^{(r,\ell)}$.

\begin{proposition}
\label{prop:bin_binomial}
For every $r \in [R]$ and $l \in [2^r]$, $N_a^{(r,\ell)} \sim \text{Bin}\bigl( \frac{1}{2} + \delta^{(r, \ell)}, N^{(r,\ell)} \bigr)$ conditional on $N^{(r,\ell)}$ and if $\delta^{(r,\ell)} = 0$, then $p^{(r,\ell)} \sim \text{Unif}[0,1]$. 

Moreover, for every $r \in [R]$, conditional on $\{ N^{(r,\ell)} \}_{\ell \in [2^r]}$, the collection of random variables $\{ N_a^{(r,\ell)} \}_{l \in [2^r]}$ are mutually independent.
\end{proposition}

\begin{proof}
Since $N_a^{(r,\ell)} = N_a( I^{(r)}_\ell )$, we have that $N_a^{(r,\ell)}$ has the Poisson distribution with mean $\int_{I^{(r)}_l} \lambda_a d\nu$. Since $N^{(r,\ell)} - N_a^{(r,\ell)} = N_b^{(r,\ell)}$ has the Poisson distribution with mean $\int_{I^{(r)}_l} \lambda_b d\nu$, and is independent of $N_a^{(r,\ell)}$, we have that, for any $s, t \in \mathbb{N}$ where $s \leq t$,
\begin{align*}
\mathbb{P}( N_a^{(r,\ell)} = s \,|\, N^{(r, \ell)}  = t) 
&= \frac{\mathbb{P}( N_a^{(r,\ell)} = s, N_b^{(r,\ell)} = t-s)}{ 
\mathbb{P}( N^{(r,\ell)} = t)} \\
&= \frac{ 
\frac{1}{s!} e^{ - \int_{I^{(r)}_l} \lambda_a d\nu} \biggl\{ \int_{I^{(r)}_l} \lambda_a d\nu \biggr\}^{s} 
\frac{1}{(t-s)!} e^{ - \int_{I^{(r)}_l} \lambda_b d\nu} \biggl\{ \int_{I^{(r)}_l} \lambda_b d\nu \biggr\}^{t-s}
}{ \frac{1}{t!} e^{ - \int_{I^{(r)}_l} \lambda_a + \lambda_b d\nu} \biggl\{ \int_{I^{(r)}_l} \lambda_a + \lambda_b d\nu \biggr\}^{t} } \\
&= \binom{t}{s} \biggl( \frac{1}{2} + \delta^{(r, \ell)} \biggr)^{s} \biggl( \frac{1}{2} - \delta^{(r, \ell)} \biggr)^{t-s},
\end{align*}
and the first claim follows directly. If $\delta^{(k)}_l = 0$, then $\hat{p}_l^{(k)}$ is uniform by Proposition~\ref{prop:pval_transform}.

The second claim follows from the independent increment property of a Poisson process.
\end{proof}

\subsection{Proof of Theorem~\ref{thm:fisher_power}}
\label{sec:fisher_power_proof}

\begin{proof} (of Theorem~\ref{thm:fisher_power})\\
Let $r^* \in [R]$ denote the resolution level that satisfies~\eqref{eq:special_r}.
Recalling that $\delta_l^{(r^*)}$ is defined as~\eqref{eq:delta_defn}, we define the event 
\begin{align*}
\mathcal{E}_{r^*} := \biggl\{ \sum_{l = 1}^{2^{r^*}} (N^{(r^*, l)} - 1) \delta_l^{(r^*)2} \geq 2^{r^*/2} \biggl( \frac{C^{1/2}}{\beta^{1/2}} + 2\log^{1/2} \frac{R}{\alpha} \biggr) + 2\log \frac{R}{\alpha} \biggr\},
\end{align*}
where $C$ is the universal constant specified in Theorem~\ref{thm:binomial_fisher_power} which we may assume to be greater than 1. Then, by Theorem~\ref{thm:binomial_fisher_power},
\begin{align*}
    \mathbb{P} &\bigl( p_F \geq \alpha\bigr) 
    \leq \mathbb{P}\bigl( p_F^{(r^*)} \geq \frac{\alpha}{R} \bigr) \\
    &\leq \mathbb{P}\biggl( \biggl\{ p_F^{(r^*)} \geq \frac{\alpha}{R} \biggr\}  \cap \mathcal{E}_{r^*} \biggr) + \mathbb{P}(\mathcal{E}_{r^*}^c) \leq \beta + \mathbb{P}( \mathcal{E}_{r^*}^c).
\end{align*}  

In order to upper bound the probability of $\mathcal{E}_{r^*}^c$, we observe, by our assumption that $\mathbb{E} N^{(r^*, l)} = \int_{I^{(r^*)}_l} \lambda d\nu \geq 2$ for all $l \in [2^{r^*}]$ and the fact that $|\delta_l^{(r^*)2}| \leq \frac{1}{2}$, that
\begin{align}
\mathbb{E} \sum_{l=1}^{2^{r^*}} (N^{(r^*, l)}-1) \delta_l^{(r^*)2} 
&\geq \frac{1}{2} \sum_{l=1}^{2^{r^*}} \biggl(\int_{I^{(r^*)}_l} \lambda d\nu \biggr) \delta_l^{(r^*)2} \quad \text{and} \label{eq:Ebound1}\\
\text{Var} \sum_{l=1}^{2^{r^*}} (N^{(r^*,l)}-1) \delta_l^{(r^*)2} &\leq \frac{1}{4} \sum_{l=1}^{2^{r^*}} 
  \biggl( \int_{I^{(r^*)}_l} \lambda d\nu \biggr) \delta_l^{(r^*)2}.  \label{eq:Vbound1}
\end{align}

As a short hand, we write
\begin{align*}
W &\equiv \sum_{l=1}^{2^{r^*}} (N^{(r^*,l)} - 1) \delta_l^{(r^*)2}, \quad \text{ and } \\
T_{r^*, R, \alpha, \beta} &\equiv 2^{r^*/2} \bigl( \frac{C^{1/2}}{\beta^{1/2}} + 2\log^{1/2} \frac{R}{\alpha} \bigr) + 2\log \frac{R}{\alpha}.
\end{align*}

We note that by~\eqref{eq:special_r}, we have
$\mathbb{E} W \geq \frac{1}{2} \sum_{l=1}^{2^{r^*}} \biggl( \int_{I_l^{(r^*)}} \lambda d\nu\biggr) \geq T_{r^*, R, \alpha, \beta}$. By this, Chebyshev' inequality, and~\eqref{eq:Ebound1} and~\eqref{eq:Vbound1}, we have
\begin{align*}
\mathbb{P}( \mathcal{E}_{r^*}^c) 
&= \mathbb{P} \biggl\{ \sum_{l =1}^{2^{r^*}} (N^{(r^*,l)}-1) \delta_l^{(r^*)2} \leq T_{r^*,R,\alpha,\beta} \biggr\} \\
&= \mathbb{P} \bigl\{ W \leq T_{r^*,R,\alpha,\beta} \bigr\} \\
&= \mathbb{P} \bigl\{ W - \mathbb{E} W \leq T_{r^*,R,\alpha,\beta} - \mathbb{E} W \bigr\} \\
&\leq \text{Var}(W) \cdot \{ \mathbb{E} W - T_{r^*,R,\alpha,\beta} \}^{-2} \\
&\leq \text{Var} \biggl\{ \sum_{l=1}^{2^{r^*}}  (N^{(r^*,l)}-1) \delta_l^{(r^*)2}  \biggr\}
 \biggl\{ \mathbb{E} \biggl(\sum_{l=1}^{2^{r^*}}  (N^{(r^*,l)}-1) \delta_l^{(r^*)2}  \biggr) -T_{r^*,R,\alpha,\beta} \biggr\}^{-2} \\
&\leq \biggl\{ \frac{1}{4} \sum_{l=1}^{2^{r^*}} \biggl( \int_{I^{(r^*)}_l} \lambda d\nu \biggr) \delta_l^{(r^*)2}  \biggr\}
 \biggl\{\frac{1}{2} \sum_{l=1}^{2^{r^*}}\biggl( \int_{I^{(r^*)}_l} \lambda d\nu \biggr)  \delta_l^{(r^*)2}   -T_{r^*,R,\alpha,\beta} \biggr\}^{-2} \\
&\leq \biggl\{ \frac{1}{4} \sum_{l=1}^{2^{r^*}} \biggl( \int_{I^{(r^*)}_l} \lambda d\nu \biggr) \delta_l^{(r^*)2}  \biggr\}
 \biggl\{\frac{1}{4} \sum_{l=1}^{2^{r^*}}\biggl( \int_{I^{(r^*)}_l} \lambda d\nu \biggr)  \delta_l^{(r^*)2} \biggr\}^{-2} \\
 &\leq \biggl\{ \frac{1}{4} \sum_{l=1}^{2^{r^*}}\biggl( \int_{I^{(r^*)}_l} \lambda d\nu \biggr)  \delta_l^{(r^*)2} \biggr\}^{-1} \leq \beta,
\end{align*}
where the penultimate inequality follows from~\eqref{eq:special_r} and the fact that $C \geq 1$. The Theorem follows as desired. 

\end{proof}

\subsection{Proof of Theorem~\ref{thm:fisher_power_smooth}}
\label{sec:fisher_power_smooth_proof}

\begin{proof} (of Theorem~\ref{thm:fisher_power_smooth}) \\
We first claim that, writing $C$ as the universal constant specified in Theorem~\ref{thm:fisher_power}, 
\begin{align}
\frac{1}{4 n} \int_I \biggl( \frac{\lambda_a - \lambda_b}{\lambda} \biggr)^2 \lambda\, d \nu \geq 
\min_{r \in [R]} \frac{2^{r/2}}{n} \biggl( \frac{C^{1/2}}{\beta} + 2 \log^{1/2} \frac{R}{\alpha} \biggr) + \frac{2}{n} \log \frac{R}{\alpha} + \frac{C_H C_d}{2} 2^{ - \frac{r\gamma}{q} }.
\label{eq:overall_hellinger_condition}
\end{align}
To see that this claim is true, define
\[
\tilde{r} = \min\biggl( R, \biggl\lfloor \frac{\log_2 \frac{n}{2}}{\frac{1}{2} + \frac{2 \gamma}{q}} - \log_2 \frac{c_{\max}}{c_{\min}}  \biggr \rfloor \biggr),
\]
or equivalently, 
\[
\tilde{r} = \begin{cases} 
\bigl\lfloor \frac{\log_2 \frac{n}{2} }{\frac{1}{2} + \frac{2 \gamma}{q} }  - \log_2 \frac{c_{\max}}{c_{\min}} \bigr \rfloor & \text{ if $\gamma/q \geq 1/4$} \\
R & \text{ if $\gamma/q \leq 1/4$} \end{cases}.
\]
Then, using the fact that $\log_2 \frac{n}{2} - \log_2 \frac{c_{\max}}{c_{\min}} - 1 \leq R \leq \log_2 \frac{n}{2} - \log_2 \frac{c_{\max}}{c_{\min}}$, we have
\begin{align*}
& \frac{2^{\tilde{r}/2}}{n} \biggl( \frac{C^{1/2}}{\beta} + 2 \log^{1/2} \frac{R}{\alpha} \biggr) + \frac{2}{n} \log \frac{R}{\alpha} + \frac{C_H C_d}{2} 2^{- \frac{2 \tilde{r} \gamma}{q}} \\
&\leq \begin{cases}
\frac{1}{4} C_1 n^{-\frac{4\gamma}{q + 4\gamma}} \bigl(\beta^{-1} + \log \frac{\log n}{\alpha} \bigr) & \text{ if $\gamma/q \geq 1/4$} \\
\frac{1}{4} C_1 n^{-\frac{2\gamma}{q}} \bigl( \beta^{-1} + \log \frac{\log n}{\alpha} \bigr) & \text{ if $\gamma/q \leq 1/4$}
\end{cases}
\end{align*}
for some $C_1 > 0$ whose value depends only on $\frac{c_{\max}}{c_{\min}}$, $C_H$, and $C_d$. Therefore, we have from assumption~\eqref{eq:strong_signal} that claim~\eqref{eq:overall_hellinger_condition} holds. Then, by Lemma~\ref{lem:hellinger_approximation}, we have that for every $r \in [R]$,
\begin{align*}
&\frac{1}{n} \sum_{l=1}^{2^r} \biggl( 
\frac{ \int_{I_l^{(r)}} \lambda_a - \lambda_b d \nu }{\int_{I_l^{(r)}} \lambda d \nu } \biggr)^2 \int_{I_l^{(r)}} \lambda d \nu \\
&\geq \frac{1}{n} \int_{I} \biggl( \frac{\lambda_a - \lambda_b}{\lambda} \biggr)^2 \lambda \, d\nu - \frac{C_H C_d}{2} 2^{- \frac{2\gamma r}{q}}.
\end{align*}

Thus, using~\eqref{eq:overall_hellinger_condition}, we may conclude that there exists a $r \in [R]$ such that
\[
\frac{1}{4 n} \sum_{l=1}^{2^r} \biggl( 
\frac{ \int_{I_l^{(r)}} \lambda_a - \lambda_b d \nu }{2 \int_{I_l^{(r)}} \lambda d \nu } \biggr)^2 \int_{I_l^{(r)}} \lambda d \nu 
\geq 
 \frac{2^{r/2} }{n}  \biggl( \frac{C^{1/2}}{\beta} + 2 \log^{1/2} \frac{R}{\alpha} \biggr) + \frac{2}{n} \log \frac{R}{\alpha}.
\]

From the hypothesis of the theorem, we also have that for all $l \in [2^R]$,
\begin{align*}
\int_{I_l^{(R)}} \lambda d \nu = 
n \frac{ \int_{I_l^{(R)}} \lambda\, \nu}{ \int_I \lambda \, d\nu} 
\geq n \frac{c_{\min}}{c_{\max}} \frac{\nu(I^{(R)}_l)}{\nu(I)}  \geq n \frac{c_{\min}}{c_{\max}} 2^{-R} \geq 2,
\end{align*}
where, in the final inequality, we use the assumption that $R \leq \log_2 \frac{n}{2} - \log_2 \frac{c_{\max}}{c_{\min}}$.

Then, from Theorem~\ref{thm:fisher_power}, it holds that $\mathbb{P}(p_F \leq \alpha) \geq 1 - 2\beta$ and the conclusion of the Theorem follows as desired.

\end{proof}

\subsection{Proof of Theorem~\ref{thm:min_power}}
\label{sec:min_power_proof}

\begin{proof} (of Theorem~\ref{thm:min_power}) \\


The proof is similar to that of Theorem~\ref{thm:fisher_power}. Let $r^* \in [R]$ and $l^* \in [2^{r^*}]$ denote the resolution level and bin such that
\[
\frac{1}{4} \biggl( \frac{ \int_{I_{l^*}^{(r^*)}} \lambda_a - \lambda_b d\nu }{ 2 \int_{I_{l^*}^{(r^*)}} \lambda d \nu} \biggr)^2 \int_{I_{l^*}^{(r^*)}} \lambda d \nu \geq 2 r^* + \frac{C^{1/2}}{\beta} + 2 \log \frac{K}{\alpha}.
\]

Define the event
\[
\mathcal{E}_{r^* l^*} := \biggl\{ (m_{l^*}^{(r^*)} - 1) \delta_{l^*}^{(r^*) 2} \geq 2r^* + \frac{C^{1/2}}{\beta^{1/2}} + 2 \log \frac{K}{\alpha} \biggr\}.
\]

By Theorem~\ref{thm:binomial_min_power}, we have that
\begin{align*}
\mathbb{P}( \hat{p}_{\min} \geq \frac{\alpha}{K} ) 
&\leq \mathbb{P}( \hat{p}^{(k)} \geq \frac{\alpha}{K} ) \\
&\leq \mathbb{P}\biggl( \biggl\{ \hat{p}^{(k)} \geq \frac{\alpha}{K} \biggr\} \cap \mathcal{E}_{r^* l^*} \biggr) + \mathbb{P}(\mathcal{E}_{r^* l^*}^c ) \\
&\leq \beta +  \mathbb{P}(\mathcal{E}_{r^* l^*}^c ).
\end{align*}

To bound $\mathbb{P}( \mathcal{E}_{r^* l^*}^c)$, we use our assumption that $\mathbb{E} m_{l^*}^{(r^*)} = \int_{I_{l^*}^{(r^*)}} \lambda d \nu \geq 2$ and the fact that $\delta_{l^*}^{(r^*)} \leq 1$ to obtain
\begin{align*}
\mathbb{E} (m_{l^*}^{(r^*)} - 1) \delta_{l^*}^{(r^*) 2} &\geq \frac{1}{2} \biggl( \int_{I_{l^*}^{(r^*)}} \lambda d\nu \biggr) \delta_{l^*}^{(r^*)2} \qquad \text{and} \\
\text{Var} (m_{l^*}^{(r^*)} - 1) \delta_{l^*}^{(r^*) 2} &\leq \biggl( \int_{I_{l^*}^{(r^*)}} \lambda d\nu \biggr) \delta_{l^*}^{(r^*)2}.
\end{align*}

We have then 
\begin{align*}
\mathbb{P}( \mathcal{E}_{l^* r^*}^c) 
&= \mathbb{P} \biggl( (m_{l^*}^{(r^*)} - 1) \delta_{l^*}^{(r^*)2} \leq \frac{C^{1/2}}{\beta^{1/2}} + 2r^* + 2 \log \frac{K}{\alpha} \biggr) \\
&\leq \bigl\{ \text{Var} (m_{l^*}^{(r^*)} - 1) \delta_{l^*}^{(r^*)2} \bigr\}
\biggl\{ \mathbb{E}  (m_{l^*}^{(r^*)} - 1) \delta_{l^*}^{(r^*)2} - \biggl(2r^* + \frac{C^{1/2}}{\beta^{1/2}} + 2 \log \frac{K}{\alpha}\biggr) \biggr\}^{-2} \\
&\leq \biggl\{ \frac{1}{4} \biggl( \int_{I_{l^*}^{(r^*)}} \lambda d \nu \biggr) \delta_{l^*}^{(r^*)2}  \biggr\}^{-1} \leq \beta. 
\end{align*}

\end{proof}

\subsection{Proof of Theorem~\ref{thm:min_power_smooth}}
\label{sec:min_power_smooth_proof}

\begin{proof}

Let $r := \bigl \lceil \log_2 \frac{\nu(I)}{\nu(S)} \bigr \rceil$ so that 
\[
\frac{\nu(I)}{2^{r - 1}} \geq \nu(S) \geq \frac{\nu(I)}{2^r}.
\]
We observe that since $\frac{\nu(S)}{\nu(I)} \geq \frac{c_{\max}}{c_{\min}} \frac{8}{n}$ by assumption, 
\[
r \leq \bigl \lceil \log_2 \frac{\nu(I)}{\nu(S)} \bigr \rceil < \bigl \lfloor \log_2 \frac{n}{2} - \log_2 \frac{ c_{\max}}{c_{\min}} \bigr \rfloor = R.
\]
Hence, $\{I_l^{(r+1)} \}$ exists in our dyadic partitioning and there exists $l^* \in [2^{r+1}]$ such that the interval $I_{l^*}^{(r+1)} \subset S$. Let $C$ be the universal constant specified in Theorem~\ref{thm:min_power} and let $C_2 := \frac{32 c_{\max}}{c_{\min}} C^{1/2}$. From~\eqref{eq:delta_S_condition}, we have that
\begin{align*}
&\frac{1}{4} \max_{l \in [2^{r+1}]} \biggl( \frac{ \int_{I_l^{(r+1)}} \lambda_a - \lambda_b d \nu}{  \int_{I_l^{(r+1)}} \lambda d \nu} \biggr)^2 \int_{I_l^{(r+1)}} \lambda d\nu  \\
&\geq  \frac{1}{4} \biggl( \frac{ \int_{I_{l^*}^{(r+1)}} \frac{\lambda_a - \lambda_b}{ \lambda}  \lambda d \nu}{  \int_{I_{l^*}^{(r+1)}} \lambda d \nu} \biggr)^2 \int_{I_{l^*}^{(r+1)}} \lambda d\nu \\
&\geq \frac{1}{4} \delta_S^2 \int_{I_{l^*}^{(r+1)}} \lambda d\nu 
\stackrel{(a)}{\geq} \frac{1}{4} \delta_S^2 n \frac{c_{\min}}{c_{\max}} 2^{-(r+1)} \\
&\geq \delta^2 n 2^{-(r-1)} \frac{c_{\min}}{16 c_{\max}} \geq \delta_S^2 n \frac{\nu(S)}{\nu(I)} \frac{c_{\min}}{16 c_{\max}} \geq n \delta_S^2 \frac{\nu(S)}{\nu(I)} \frac{2 C^{1/2}}{C_2} \\
&\geq 2\log n + \frac{C^{1/2}}{\beta} + 2\log \frac{1}{\alpha} \geq 2r + \frac{C^{1/2}}{\beta} + 2 \log \frac{R}{\alpha},
\end{align*}
where inequality $(a)$ follows from the fact that 
\[
\int_{I_{l^*}^{(r+1)}} \lambda d \nu = \frac{ \int_{I_{l^*}^{(r+1)}} \lambda d \nu }{ \int_{I} \lambda d \nu} \geq n \frac{c_{min}}{c_{max}} 2^{-(r+1)}.
\]

The conclusion of the theorem follows from Theorem~\ref{thm:min_power}.

\end{proof}

\subsection{Auxiliary results}
\label{sec:aux_result}

Recall that, for a positive integer $m$, we define 
\begin{align}
S_{\text{Bin}(\frac{1}{2},m)}(t) &:= \mathbb{P}( |\text{Bin}(\frac{1}{2}, m) - \frac{m}{2} | \geq t)  \label{eq:Sbin_defn} \\
S_{\chi^2_m}(t) &:= \mathbb{P}( \chi^2_m \geq t). \label{eq:Schi_defn}
\end{align}

Moreover, define 
\begin{align}
\mathcal{M}_m &:= 
\biggl\{ - \frac{m}{2}, - \frac{m}{2} + 1, \ldots, \frac{m}{2}-1, \frac{m}{2} \biggr \} \label{eq:Mm_defn} \\
\mathcal{M}^+_m &:=
\begin{cases}
\bigl\{ 0, 1, \ldots, \frac{m}{2} \bigr\}
& \text{ if $m$ is even,} \\
\bigl\{ \frac{1}{2}, \frac{3}{2}, \ldots, \frac{m}{2} \bigr\} & \text{ if $m$ is odd.}
\end{cases} \label{eq:Mmplus_defn}
\end{align}

\begin{theorem}
\label{thm:binomial_fisher_power}
Let $d$ be a positive integer. For each $l \in [d]$, let $m_l \in \mathbb{N}$, $\delta_l \in [0, \frac{1}{2}]$, and let $A_1, \ldots, A_d$ be independent random variables where $A_l \sim \text{Bin}\bigl(\frac{1}{2} + \delta_l, m_l \bigr)$,

Let $U_1, \ldots, U_d$ be independent random variables distributed uniform on $[0,1]$ and independent of $A_1, \ldots, A_d$. Define $p_l := U_l \cdot  S_{\text{Bin}(\frac{1}{2}, m_l)}(|A_l - \frac{m_l}{2}|) + (1 - U_l) S_{\text{Bin}(\frac{1}{2}, m_l)}(|A_l - \frac{m_l}{2}| + 1)$, define the set $L := \{ l \in [d] \,:\, m_l \geq 2 \}$, and define $p := S_{\chi^2_{2|L|}}\bigl( \sum_{l \in L} -2 \log p_l \bigr)$.

Then, there exists a universal constant $C > 0$ such that, for any $\alpha, \beta \in (0,1)$, if 
\begin{align}
 \sum_{l \in L} (m_l-1) \delta_l^{2} \geq |L|^{1/2} \biggl( \frac{C^{1/2}}{\beta^{1/2}} + 2\log^{1/2} \frac{1}{\alpha} \biggr) + 2\log \frac{1}{\alpha}, \label{eq:strong_signal_assumption1}
\end{align}
then $\mathbb{P}(p \leq \alpha) \geq 1 - \beta$. 

\end{theorem}

\begin{proof}

Define
\begin{align*}
L_1 := \{ l \in L \,:\, (m_l-1) \delta_l^{2} \geq 2 \} \quad \text{and} \quad L_2 := \{ l \in L \,:\, (m_l-1) \delta_l^2 < 2 \}.
\end{align*}

For simplicity of presentation, we write $Z_l :=  -2 \log p_l$ and $\tilde{Z}_l := \frac{4}{m_l} \bigl( A_l - \frac{m_l}{2} \bigr)^2$ for $l \in L$. By Hoeffding's inequality, it holds that $S_{\text{Bin}(\frac{1}{2}, m)}(t) \leq 2 \exp\bigl\{ - 2 \frac{t^2}{m} \bigr\}$. Therefore, we have that
\begin{align*}
Z_l &= - 2 \log \biggl\{ U_l \cdot S_{\text{Bin}(\frac{1}{2}, m)}( |A_l - \frac{m_l}{2} |) 
+ (1 - U_l) S_{\text{Bin}(\frac{1}{2}, m)}\biggl( |A_l - \frac{m_l}{2} |+1 \biggr) \biggr\} \\
&\qquad \geq -2 \log \biggl\{ S_{\text{Bin}(\frac{1}{2}, m)}( |A_l - \frac{m_l}{2} |)
\biggr\}
\geq \frac{4}{m_l} \biggl( A_l - \frac{m_l}{2} \biggr)^2 - 2\log 2 = \tilde{Z}_l - 2 \log 2.
\end{align*}

By Lemma~\ref{lem:chi_squared_concentration}, we have
\begin{align}
S_{\chi^2_{2|L|}}\biggl( 2|L| + \sqrt{8} |L|^{1/2} \log^{1/2} \frac{1}{\alpha}  + 2 \log \frac{1}{\alpha} \biggr)
\leq \alpha, \label{eq:S_chi_bound}
\end{align}
By~\eqref{eq:S_chi_bound}, the fact that $S_{\chi^2_{2|L|}}(\cdot)$ is monotone decreasing, and the fact that $2 \log2 \leq 2$, we have
\begin{align*}
\mathbb{P}( p \geq \alpha) 
&= \mathbb{P}\biggl\{ 
S_{\chi_{2|L|}^2}\biggl( \sum_{l \in L} - 2 \log p_l \biggr) \geq \alpha \biggr\} \\
&\leq \mathbb{P} \biggl\{ 
     \sum_{l \in L} -2 \log p_l \leq 2 |L| + \sqrt{8} |L|^{\frac{1}{2}} \log^{\frac{1}{2}} \frac{1}{\alpha} + 2\log \frac{1}{\alpha} \biggr\}\\
& \leq \mathbb{P} \biggl\{ 
        \sum_{l \in L_1} (\tilde{Z}_l - 2\log 2) + \sum_{l \in L_2} Z_l \leq 2|L| + \sqrt{8} |L|^{\frac{1}{2}} \log^{\frac{1}{2}} \frac{1}{\alpha} + 2\log \frac{1}{\alpha} \biggr\}\\
&\leq \mathbb{P}
    \biggl\{ \sum_{l \in L_1} (\tilde{Z}_l - \mathbb{E} \tilde{Z}_l) + \sum_{l \in L_2} (Z_l - \mathbb{E} Z_l) \leq \sum_{l \in L_1} (4 - \mathbb{E} \tilde{Z}_l)\\
&\qquad \qquad \qquad   + \sum_{l \in L_2} (2 - \mathbb{E} Z_l) + 2 |L|^{\frac{1}{2}} \log^{\frac{1}{2}} \frac{1}{\alpha} + 2\log \frac{1}{\alpha} \biggr\}. \quad (\star)
\end{align*}
We now observe that by Lemma~\ref{lem:expectation_bound}, 
\begin{align*}
&\sum_{l \in L_1} (4 - \mathbb{E} \tilde{Z}_l) + \sum_{l \in L_2} (2 - \mathbb{E}Z_l) + 2 |L|^{1/2} \log^{1/2} \frac{1}{\alpha} + 2 \log \frac{1}{\alpha} \\
&\leq - \sum_{l \in L} 2(m_l-1) \delta_l^2 + 2 |L|^{1/2} \log^{1/2} \frac{1}{\alpha} + 2 \log \frac{1}{\alpha} \leq 0,
\end{align*}
where the final inequality follows by our assumption~\eqref{eq:strong_signal_assumption1}. Therefore, returning to $(\star)$, we may apply Chebyshev's inequality to obtain
\begin{align*}
(\star) &\leq  \frac{ \sum_{l \in L_1} \text{Var}\tilde{Z}_l + \sum_{l \in L_2} \text{Var} Z_l}{ \biggl\{ 
-\sum_{l \in L} 2 (m_l - 1) \delta_l^2 + 2 |L|^{\frac{1}{2}} \log^{\frac{1}{2}} \frac{1}{\alpha} + 2\log \frac{1}{\alpha} \biggr\}^2 } \\
  &\leq  C d \biggl\{ - 2\sum_{l \in L} (m_l-1) \delta_l^2 + 2 |L|^{\frac{1}{2}} \log^{\frac{1}{2}} \frac{1}{\alpha} + 2\log \frac{1}{\alpha} \biggr\}^{-2}   \leq\beta.
\end{align*}
where, in the penultimate inequality, we used Lemmas~\ref{lem:variance_bound} and where $C > 0$ is the universal constant specified in Lemma~\ref{lem:variance_bound}. The conclusion of the Theorem follows as desired.

\end{proof}

\begin{theorem}
\label{thm:binomial_min_power}
Let $d$ be a positive integer. For each $l \in [d]$, let $m_l \in \mathbb{N}$, $\delta_l \in [0, 1/2]$, and let $A_1, \ldots, A_d$ be independent random variables where $A_l \sim \text{Bin}\bigl( \frac{1}{2} + \delta_l, m_l\bigr)$. 

Let $U_1, \ldots, U_d$ be independent random variables distributed uniform on $[0,1]$ and independent of $A_1, \ldots, A_d$. Define $p_l := U_l \cdot S_{\text{Bin}(\frac{1}{2}, m_l)}(|A_l - \frac{m_l}{2}|) + (1 - U_l) S_{\text{Bin}(\frac{1}{2}, m_l)}(|A_l - \frac{m_l}{2}| + 1) $, define the set $L := \{ l \in [d] \,:\, m_l \geq 2\}$, and define $p_{\text{min}} := F_{\text{Beta},|L|}\bigl( \min_{l \in L} p_l \bigr)$ where $F_{\text{Beta},|L|}(x) := \mathbb{P}( \text{Beta}(1, |L|+1) \leq x)$ for any $x \in \mathbb{R}$.

There exists universal constants $C > 0$ such that for any $\alpha, \beta \in (0,1)$, if 
\[
 \max_{l \in [d]} \, (m_l-1) \delta_l^{2} \geq 
 \frac{C^{1/2}}{\beta^{1/2}} + 2 \log \frac{ |L|}{\alpha},
\]
then $\mathbb{P}(p_{\text{min}} \leq \alpha) \geq 1 - \beta$. 
\end{theorem}

\begin{proof}
Let $C$ be the maximum of $4$ and the universal constant specified in Lemma~\ref{lem:variance_bound}. By assumption, there exists $l^* \in [d]$ be such that 
\begin{align}
(m_{l^*} - 1) \delta_{l^*}^2 \geq \frac{C^{1/2}}{\beta^{1/2}} + 2 \log \frac{|L|}{\alpha} \geq 2, \label{eq:lstar_cond}
\end{align}
where the last inequality follows since $\beta, \alpha \in (0, 1)$. 

By Hoeffding's inequality, we have that
\begin{align*}
-2 \log p_{l^*} &\geq - 2 \log S_{\text{Bin}(\frac{1}{2}, m_{l^*})}( | A_{l^*} - \frac{m_{l^*}}{2}|) \\
&\geq \frac{4}{m_{l^*}} \biggl( A_{l^*} - \frac{m_{l^*}}{2} \biggr)^2 - 2\log 2.
\end{align*}
We write $\tilde{Z}_{l^*} := \frac{4}{m_{l^*}} \bigl( A_{l^*} - \frac{m_{l^*}}{2} \bigr)^2$ so that $- 2 \log p_{l^*} \geq \tilde{Z}_{l^*} - 2 \log 2 \geq \tilde{Z}_{l^*} - 4$.

For any $\alpha, \beta \in (0,1)$, we may use the fact that $F_{\text{Beta},|L|}(x) \leq |L| x$ to show that
\begin{align*}
\mathbb{P}( p_{\text{min}} \geq \alpha) 
&\leq \mathbb{P}\bigl( \min_{l \in L} p_l \geq \frac{\alpha}{|L|} \bigr) 
\leq \mathbb{P} \bigl( p_{l^*} \geq \frac{\alpha}{|L|} \bigr) \\
&= \mathbb{P} \bigl( - 2 \log p_{l^*} \leq  2 \log \frac{|L|}{\alpha} \bigr) \\
&\leq \mathbb{P} \biggl( \tilde{Z}_{l^*} - \mathbb{E} \tilde{Z}_{l^*} \leq (4 - \mathbb{E} \tilde{Z}_{l^*}) + 2 \log \frac{|L|}{\alpha} \biggr). \quad (\star) 
\end{align*}

By Lemma~\ref{lem:expectation_bound} and~\eqref{eq:lstar_cond}, we have that
\begin{align*}
(4 - \mathbb{E} \tilde{Z}_{l^*}) + 2 \log \frac{|L|}{\alpha} 
&\leq -2 (m_{l^*} - 1)\delta^2_{l^*} + 2 \log \frac{|L|}{\alpha} \leq 0.
\end{align*}

Therefore, continuing on from $(\star)$, we have by Chebyshev inequality and Lemma~\ref{lem:variance_bound} that

\begin{align*}
(\star) &\leq \frac{\text{Var} (\tilde{Z}_{l^*}) }{ \bigl\{ - (m_{l^*}-1) \delta^2_{l^*} + 2 \log \frac{|L|}{\alpha} \bigr\}^2} \leq C \biggl\{ -  (m_{l^*} - 1) \delta_{l^*}^2 + 2 \log \frac{|L|}{\alpha} \biggr\}^{-2} \leq \beta.
\end{align*} 
The conclusion of the Theorem follows as desired. 

\end{proof}

\begin{lemma}
\label{lem:variance_bound}
Let $m$ be a positive integer and let $\delta \in [0, 1/2]$. Let $A \sim \text{Bin}(\frac{1}{2} + \delta, m)$ and let $U \sim \text{Unif}[0,1]$ be independent of $A$. Define $Z := -2 \log \bigl\{ U \cdot S_{\text{Bin}(\frac{1}{2},m)}( | A - \frac{m}{2} | ) + (1-U) S_{\text{Bin}(\frac{1}{2},m)}( | A - \frac{m}{2} | +1 ) \bigr\}$ and $\tilde{Z} := \frac{4}{m} ( A - \frac{m}{2} )^2$.

There exists a universal constant $C > 0$ such that 
\begin{enumerate}
\item if $(m-1) \delta^2 < 2$, then $\text{Var}(Z) \leq C$,
\item and $\text{Var} (\tilde{Z}) \leq C$.
\end{enumerate} 

\end{lemma}

\begin{proof}
First assume that $(m-1)\delta^2 < 2$. By increasing the value of the universal constant $C$ if necessary, we may assume without the loss of generality that $m \geq 17$. 

Define $\mathcal{M}_m$ as~\eqref{eq:Mm_defn}. Let $P, Q$ be probability measures on $\mathcal{M}_m$ such that $P$ is the distribution of $|A - \frac{m}{2}|$ and $Q$ is the distribution of $|\text{Bin}(\frac{1}{2}, m) - \frac{m}{2} |$. 

Then, let $\tilde{S}_0(\cdot)$ be defined as in Lemma~\ref{lem:Stilde0_cdf}, we have by the same lemma that
\begin{align}
\text{Var} Z
&\leq \mathbb{E} Z^2 \nonumber \\
&= \int_0^1 \sum_{s \in \mathcal{M}_m}  4 \log^2 \tilde{S}_0(|s| + u)  \cdot \frac{P(s)}{Q(s)} Q(s) \, du  \nonumber \\
&\leq  \underbrace{\biggl\{ \sum_{s \in \mathcal{M}_m} \biggl( \frac{P(s)}{Q(s)} \biggr)^2 Q(s) \biggr\}^{1/2}}_{\text{Term 1}}\\
&\qquad + \underbrace{ \biggl\{ \int_0^1 \sum_{s \in \mathcal{M}_m}  \biggl( 4 \log^2 \bigl\{ \tilde{S}_0(|s|+u) \bigr \} \biggr)^2  Q(s) \, du \biggr\}^{1/2}  }_{\text{Term 2}}. \label{eqn:tricky_variance_decomp}
\end{align}

Term 2 of~\eqref{eqn:tricky_variance_decomp} is equal to $16 \cdot \mathbb{E} \log^4 \tilde{S}_0( | \text{Bin}(\frac{1}{2}, m) - \frac{m}{2}| + U)$. Since the random variable $\tilde{S}_0( | \text{Bin}(\frac{1}{2}, m) - \frac{m}{2}| + U)$ is uniformly distributed on $[0,1]$, we have that Term 2 is upper bounded by a universal constant. 

For Term 1, we define $r := \frac{1 + 2\delta}{1 - 2\delta}$ and observe that for any $s \in \mathcal{M}_m$,
\begin{align*}
\frac{P(s)}{Q(s)} &= \frac{1}{2} 2^m \biggl\{ \biggl( \frac{1}{2} + \delta \biggr)^{\frac{m}{2}+s} \biggl( \frac{1}{2} - \delta \biggr)^{\frac{m}{2}-s} +\biggl( \frac{1}{2} + \delta \biggr)^{\frac{m}{2}-s} \biggl( \frac{1}{2} - \delta \biggr)^{\frac{m}{2}+s} \biggr\} \\
&= \frac{1}{2} (1 - 4 \delta^2)^{\frac{m}{2}} ( r^s + r^{-s} ) \leq r^s.
\end{align*}

Since we assume $(m-1) \delta^2 \leq 2$ and since we assume that $m \geq 17$, we have that $\delta^2 \leq \frac{1}{8}$ and thus $0 \leq \log r \leq 8 \delta$. Let $W$ be a random variable distributed as $| \text{Bin}(\frac{1}{2}, m ) - \frac{m}{2} |$. Then
\begin{align*}
\sum_{s \in \mathcal{M}} r^{2s} Q(s) 
&= \mathbb{E} r^{2W}
= \int_1^\infty \mathbb{P}( r^{2W} \geq t ) \,dt \\
&= \int_1^\infty \mathbb{P}\biggl( W \geq \frac{\log t}{2\log r} \biggr) \, dt \\
&\leq \int_1^\infty \exp\biggl( - \frac{\log^2 t}{4 m \log^2 r} \biggr) \,dt \\
&\leq \int_1^\infty t^{ - \frac{\log t}{2^{12}} } \, dt \leq C,
\end{align*}
where $C > 0$ is a universal constant.

Now we turn to the second claim. Write $A = \sum_{i=1}^{m} \epsilon_i$ where $\epsilon_1, \ldots, \epsilon_m$ are independent and identically distributed $\text{Ber}(\frac{1}{2} + \delta)$ random variables. 

For any $i \in [m]$, we have
\begin{align*}
\text{Var}_{\cdot \,|\, \{\epsilon_{-i}\} } \biggl[ \biggl(A - \frac{m}{2} \biggr)^2 \biggr]
&\leq \sup_{z \in [- \frac{m}{2}, \frac{m}{2}] } \text{Var} \bigl[ (z + \epsilon_i)^2 \bigr] \leq m.
\end{align*}

Thus, by the Efron--Stein inequality, 
\begin{align*}
\text{Var} \tilde{Z} 
&= \frac{16}{m^2} \text{Var} \biggl[\biggl( A - \frac{m}{2} \biggr)^2 \biggr]\\
&\leq  \frac{16}{m^2} \mathbb{E} \sum_{i=1}^{m} \text{Var}_{\cdot \,|\, \{\epsilon_{-i}\} } \biggl[ \biggl(A - \frac{m}{2} \biggr)^2\biggr] \leq 16.
\end{align*}

The conclusion of the lemma follows as desired. 

\end{proof}

\begin{lemma}
\label{lem:expectation_bound}
Let $m$ be a positive integer and let $\delta \in [0, 1/2]$. Let $A \sim \text{Bin}(m, \frac{1}{2} + \delta)$ and let $U \sim \text{Unif}[0,1]$ be independent of $A$. Define $Z := -2 \log \bigl\{ U \cdot S_{\text{Bin}(\frac{1}{2},m)}( | A - \frac{m}{2} | ) + (1-U) S_{\text{Bin}(\frac{1}{2},m)}( | A - \frac{m}{2} | +1 ) \bigr\}$ and $\tilde{Z} := \frac{4}{m} ( A - \frac{m}{2} )^2$. 

We have that
\begin{enumerate}
\item $\mathbb{E} Z - 2 \geq 8(m-1) \delta^{2}$,
\item and if $(m-1)\delta^2 \geq 2$, then $\mathbb{E} \tilde{Z} - 4 \geq 2 (m-1) \delta^2$.
\end{enumerate}
\end{lemma}

\begin{proof}
Define $\mathcal{M}_m$ as~\eqref{eq:Mm_defn} and note that $\mathcal{M}_m = - \mathcal{M}_m$. For $s \in \mathcal{M}_m$, write $P_m(s, \delta) = \binom{m}{\frac{m}{2} + s} (\frac{1}{2} + \delta)^{\frac{m}{2} + s} (\frac{1}{2} - \delta)^{\frac{m}{2}-s}$ as the probability that $\text{Bin}(m, \frac{1}{2}+\delta)$ random variable is equal to $\frac{m}{2} + s$ and $Q_m(s) :=  \binom{m}{\frac{m}{2} + s} (\frac{1}{2})^{m} = P_m(s, 0)$. 

Define $W = |A - \frac{m}{2}| + U$. We also define
\begin{align*}
F_m(\delta) &= \mathbb{E} Z = \mathbb{E}\bigl[ - 2 \log \tilde{S}_0(W) \bigr] \\
&= \sum_{s \in \mathcal{M}_m} P_m(s, \delta) \int_0^1 \bigl\{ - 2 \log \tilde{S}_0( |s| + u) \bigr\} \, du,
\end{align*}
where the definition of $\tilde{S}_0(\cdot)$ and the second equality follow from Lemma~\ref{lem:Stilde0_cdf}. We note then that 
\begin{align*}
\mathbb{E}Z - 2 \geq F_m(\delta) - F_m(0). 
\end{align*}

Moreover, since the function $\delta \mapsto P_m(s,\delta)$ is equal to its Taylor series expansion for all $\delta \in (-\frac{1}{2}, \frac{1}{2})$, the same holds for $F_m(\delta)$, that is,

\begin{align*}
F_m(\delta) = F_m(0) + \sum_{j=1}^\infty F_m^{(j)}(0) \frac{\delta^j }{j!}, \quad \text{ for all $\delta \in (-\frac{1}{2}, \frac{1}{2})$.}
\end{align*}
By symmetry, $P_m(s, \delta) = P_m(-s, -\delta)$ and thus, $F_m(\delta) = F_m(-\delta)$ and $F_m^{(j)}(0) = 0$ when $j$ is an odd integer. When $j$ is an even integer, we have that, by Lemma~\ref{lem:Pm_second_derivative}, 
\begin{align*}
F_m^{(j)}(0) &= \sum_{s \mathcal{M}_m} \biggl( \partial_{\delta}^{(j)} P_m(s, \delta) \bigl|_{\delta=0} \biggr) \int_0^1 \bigl\{ - 2 \log \tilde{S}_0( |s| + u) \bigr\} \, du \geq 0.
\end{align*} 

We now claim that $F_m^{(2)}(0) \geq 8(m-1)$. To see this, first observe that, by Hoeffding's inequality, it holds that $-2 \log \bigl\{ S_{\text{Bin}(\frac{1}{2},m)}(|s|) \bigr\} \geq \frac{4}{m} s^2 - 2 \log2$. Moreover, since $\sum_{s \in \mathcal{M}_m} P_m(s,\delta) = 1$,  writing $P_m^{(2)}(s,\delta)$ as second derivative of $P_m(s, \delta)$ with respect to $\delta$, we have $\sum_{s \in \mathcal{M}_m} P_m^{(2)}(s, \delta) = 0$. Thus, using the fact that $P^{(2)}_m(s,0) \geq 0$ for all $s \in \mathcal{M}_m$ (by Lemma~\ref{lem:Pm_second_derivative}), we have that, for any $\delta \in (-1/2, 1/2)$,
\begin{align*}
F_m^{(2)}(\delta) 
&= \sum_{s \in \mathcal{M}_m} P_m^{(2)}(s, \delta) \int_0^1 \bigl\{ - 2 \log \tilde{S}_0( |s| + u) \bigr\} \, du \\
&\geq  \sum_{s \in \mathcal{M}_m} P_m^{(2)}(s, \delta) \bigl( - 2 \log \bigl\{ S_{\text{Bin}(\frac{1}{2},m)}(|s|) \bigr\} \bigr) \\
&\geq \sum_{s \in \mathcal{M}_m} P_m^{(2)}(s, \delta) \frac{4}{m} s^2 \\
&\geq \frac{4}{m} \biggl(\frac{d}{d \delta} \biggr)^{2} \underbrace{ \biggl\{ \sum_{s \in \mathcal{M}_m} P_m(s, \delta) s^2  \biggr\} }_{ \mathbb{E}(A - m/2)^2 \text{ where } A \sim \text{Bin}(1/2+\delta, m)} \\
&= \frac{4}{m} \biggl(\frac{d}{d \delta} \biggr)^{2} \biggl\{ \frac{m}{4} (1 - 4\delta^2) + m^2 \delta^2 \biggr\} = 1 + 8(m-1).
\end{align*}

We may conclude then that $F^{(2)}_m(0) \geq 8(m-1)$ as desired. Therefore, we have that
\begin{align*}
\mathbb{E} Z - 2 = F_m(\delta) - F_m(0) \geq 8(m-1) \delta^2. 
\end{align*}

For the second claim of the Lemma, we note that $A \sim \text{Bin}(\frac{1}{2} + \delta, m)$. Therefore, assuming $(m-1) \delta^2 \geq 2$, we have that
\begin{align*}
\mathbb{E} \tilde{Z} - 4 
&= \frac{4}{m} \mathbb{E} \bigl( A - \frac{m}{2} \bigr)^2 - 4 \\
&= \frac{4}{m} \biggl\{ \mathbb{E} \bigl( A - \frac{m}{2} - m \delta \bigr)^2 + m^2 \delta^2 \biggr\} - 4 \\
&= (1 - 4\delta^2) + 4m\delta^2 - 4 \geq 2 m\delta^2
\end{align*}
as desired. The conclusion of the lemma thus follows. 
\end{proof}

\begin{lemma}
\label{lem:hellinger_approximation}
Let $I \subset \mathbb{R}^q$ and let $I_1, \ldots, I_L$ be a partition of $I$ such that $\text{diam}(I_l) \leq C_d L^{-1/q}$ for all $l \in [L]$ for some $C_d > 0$. Write $\delta := \frac{\lambda_a - \lambda_b}{\lambda}$ and suppose that $\delta$ is $\gamma$-Holder continuous for $\gamma \in (0, 1]$, i.e., $|\delta(x) - \delta(y)| \leq C_H \| x - y\|_2^\gamma$ for all $x, y \in I$, for some $C_H > 0$.

Then, we have that
\[
0 \leq \int_I \biggl( \frac{\lambda_a - \lambda_b}{\lambda} \biggr)^2 \lambda\, d\nu - \sum_{l=1}^L \bigg\{ \biggl( \frac{ \int_{I_l} \lambda_a - \lambda_b \, d\nu}{ \int_{I_l} \lambda \, d\nu} \biggr)^2 \int_{I_l} \lambda \, d\nu \biggr\} \leq 2 C_H C_d d^{- \frac{2\gamma}{q}} \biggl( \int_I \lambda\, d \nu \biggr).
\]
\end{lemma}

\begin{proof}
Fix an arbitrary $l \in [L]$ and define $\mathbb{E}^{(l)}[ \,\cdot\, ]$ as expectation with respect to the probability measure with density $\frac{\lambda}{\int_{I_l} \lambda d\nu}$. We then have that

\begin{align*}
\frac{ \int_{I_l} \bigl( \frac{\lambda_a - \lambda_b}{\lambda} \bigr)^2 \lambda \, d\nu}{\int_{I_l} \lambda \, d\nu} = \mathbb{E}^{(l)}[ \delta^2 ] 
\geq \{ \mathbb{E}^{(l)} \delta \}^2 = \biggl\{ \frac{ \int_{I_l} \lambda_a - \lambda_b d\nu }{\int_{I_l} \lambda d \nu } \biggr\}^2.
\end{align*}

For the other direction, we observe that
\begin{align*}
& \frac{ \int_{I_l} \bigl( \frac{\lambda_a - \lambda_b}{\lambda} \bigr)^2 \lambda \, d\nu}{\int_{I_l} \lambda \, d\nu} - \biggl\{ \frac{ \int_{I_l} \lambda_a - \lambda_b d\nu }{\int_{I_l} \lambda d \nu } \biggr\}^2 \\
&\qquad = \mathbb{E}^{(l)} [\delta^2] - \{ \mathbb{E}^{(l)} \delta \}^2 = \text{Var}^{(l)} (\delta) \\
&\qquad \stackrel{(a)}{=} \frac{1}{2} \mathbb{E}^{(l)}\bigl[ (\delta(X) - \delta(Y))^2 \bigr] \\
&\qquad \leq \frac{1}{2} C_H \mathbb{E}^{(l)} \| X - Y \|_2^{2\gamma}
\leq \frac{1}{2} C_H \sup_{x, y \in I_l} \|x - y\|_2^{2\gamma} \stackrel{(b)}{\leq} \frac{C_H C_d}{2} L^{-2 \gamma / q}.
\end{align*}
where in inequality $(a)$, the random variables $X, Y$ are independent and distributed with density $\frac{\lambda}{\int_{I_l} \lambda \, d\nu}$ and where in inequality $(b)$, we use the assumption that $\text{diam}(I_l) \leq C_d L^{-1/q}$.

In summary, we have that, for each $l \in [L]$, 
\begin{align*}
0 \leq \int_{I_l} \biggl( \frac{\lambda_a - \lambda_b}{\lambda} \biggr)^2 \lambda \, d\nu - \biggl( \frac{\int_{I_l} \lambda_a - \lambda_b d\nu }{\int_{I_l} \lambda \, d\nu } \biggr)^2 \int_{I_l} \lambda \, d\nu \leq \frac{C_H C_d}{2} L^{-2\gamma/q} \int_{I_l} \lambda \, d\nu.
\end{align*}
By summing over $l \in [L]$, the conclusion of the theorem follows as desired. 

\end{proof}

\subsection{Technical lemmas}

\begin{lemma}
\label{lem:chi_squared_concentration}
Let $X \sim \chi^2_{2k}$. Then, we have that for all $t > 0$,
\begin{align*}
\mathbb{P}( X \geq 2k + 2 \sqrt{2 k t} + 2t) \leq e^{-t}.
\end{align*}
\end{lemma}

\begin{proof}
If $X \sim \chi^2_{2k}$, then $\mathbb{E}X = 2k$ and $X$ is also a $\text{Gamma}(k, 2)$ random variable and hence its moment generation function is bounded by $\mathbb{E} e^{\lambda (X - \mathbb{E}X)} \leq \frac{4k \lambda^2 }{2(1 - 2\lambda)}$ for all $\lambda \in (0, \frac{1}{c})$ by \citet[][Section 2.4]{boucheron2013concentration}. Then, it holds by \citet[][Theorem 2.3]{boucheron2013concentration} that $\mathbb{P}(X - \mathbb{E}X \geq \sqrt{8kt} + ct) \leq e^{-t}$ for all $t > 0$. The Lemma immediately follows.
\end{proof}

\begin{lemma}
\label{lem:Stilde0_cdf}
Let $m \in \mathbb{N}$ and let $\tilde{A} \sim \text{Bin}(\frac{1}{2}, m)$. Define $\tilde{W} = | \tilde{A} - \frac{m}{2}| + U$ where $U \sim \text{Unif}[0,1]$ is independent of $\tilde{A}$. We let $\mathcal{M}^+_m$ be as defined in~\eqref{eq:Mmplus_defn}.

Write $\tilde{S}_0(z) := \mathbb{P}( \tilde{W} \geq z )$. We have that, for any $z \in \mathbb{R}$,
\begin{align}
\tilde{S}_0(z) = 
\begin{cases}
(1 - (z - k_1)) S_{\text{Bin}(\frac{1}{2}, m)}(k_1) + (z - k_1) S_{\text{Bin}(\frac{1}{2}, m)}(k_1 + 1) 
& \text{ if $z \in \bigl[ \min \mathcal{M}^+_m, 1 + \max \mathcal{M}^+_m \bigr)$} \\
1 & \text{ if $z < \min \mathcal{M}^+_m$} \\
0 & \text{ if $z \geq 1 + \max \mathcal{M}^+_m$},
\end{cases} \label{eq:Stilde0_defn}
\end{align}
where in the first case, $k_1$ is defined as $k_1 := \max \{k \in \mathcal{M}^+_m \,:\, k \leq z \}$.

Moreover, we have that
\begin{align*}
\tilde{S}_0'(z) = 
\begin{cases}
- \mathbb{P}\bigl( | \tilde{A} - \frac{m}{2}| = k_1\bigr) 
& \text{ if $z \in \bigl[ \min \mathcal{M}^+_m, 1 + \max \mathcal{M}^+_m \bigr)$} \\
0 & \text{ else }.
\end{cases}
\end{align*}

Finally, let $A \sim \text{Bin}(\frac{1}{2} + \delta, m)$ and let $W = |A - \frac{m}{2}| + U$ where $U \sim \text{Unif}[0,1]$ is independent of $A$, we have that
\[
\tilde{S}_0(W) \stackrel{d}{=} 
(1 - U) S_{\text{Bin}(\frac{1}{2},m)}\biggl( |A - \frac{m}{2}|\biggr) + U\cdot S_{\text{Bin}(\frac{1}{2},m)}\biggl( |A - \frac{m}{2}| + 1\biggr).
\]

\end{lemma}

\begin{proof}
To establish the first claim, let $z \in \bigl[ \min \mathcal{M}^+_m, 1 + \max \mathcal{M}^+_m \bigr)$ and let $k_1 := \max \{ k \in \mathcal{M}^+_m \,:\, k \leq z \}$. Define the event
\[
\mathcal{E}_{k_1} = \biggl\{ \bigl| \tilde{A} - \frac{m}{2} \bigr| = k_1 \biggr\}.
\]
Then, we have that
\begin{align*}
\tilde{S}_0(z) &= \mathbb{P}( \tilde{W} \geq z) \\
&= \mathbb{P}( \mathcal{E}_{k_1}) \mathbb{P}( \tilde{W} \geq z \,|\, \mathcal{E}_{k_1}) +
\mathbb{P}\biggl( \bigl| \tilde{A} - \frac{m}{2} \bigr| > k_1 \biggr). \\
&= \bigl\{ S_{\text{Bin}(\frac{1}{2}, m)}(k_1) - S_{\text{Bin}(\frac{1}{2}, m)}(k_1 +1) \bigr\} (1 - (z - k_1)) + S_{\text{Bin}(\frac{1}{2}, m)}(k_1 +1).
\end{align*}
The first claim~\eqref{eq:Stilde0_defn} follows immediately.

The second claim follows by direct differentiation, and the third claim follows directly from the first claim. The whole Lemma thus follows as desired. 
\end{proof}

\begin{lemma}
\label{lem:Pm_second_derivative}
For $m \in \mathbb{N}$, $s \in \mathcal{M}_m$ (defined as~\eqref{eq:Mm_defn}), and $\delta \in (-\frac{1}{2}, \frac{1}{2})$, define $P_m(s, \delta) = \binom{m}{\frac{m}{2}+s} (\frac{1}{2} + \delta)^{\frac{m}{2}+s} (\frac{1}{2} - \delta)^{\frac{m}{2}-s} $. We then have that, for any integer $j \geq 1$,
\[
\frac{\partial^{(2j)}}{\partial \delta^{(2j)}} P_m(s, \delta) \bigg|_{\delta = 0} \geq 0.
\]
\end{lemma}

\begin{proof}
First suppose $s \geq 0$. 
Since $|2 \delta| < 1$, we have that,
\begin{align*}
P_m(s,\delta) 
&= \binom{m}{\frac{m}{2}+s} 2^{-m} \biggl( \frac{1+2\delta}{1-2\delta} \biggr)^s \\
&= \binom{m}{\frac{m}{2}+s} 2^{-m} (1 + 2\delta)^s \biggl( 1 + \sum_{k=1}^\infty (2\delta)^k \biggr)^s.
\end{align*}
It is thus clear that in Taylor series expansion of $\delta \mapsto P_m(s, \delta)$, all the coefficients are non-negative and and thus, $\frac{\partial^{(2j)}}{\partial \delta^{(2j)}} P_m(s,\delta) \geq 0$. 

If $s \leq 0$ on the other hand, the same claim follows by the fact that 
\[
P_m(-s, \delta) = P_m(s, -\delta). 
\]
The lemma thus immediately follows.
\end{proof}

\section{Supplementary material for Section~\ref{sec:experiments}}

\subsection{Two sample test Type I error}
\label{sec:type1exp}

We consider the following three intensities
\begin{center}
	\begin{minipage}{3in}
		\begin{enumerate}
			
			\item $\lambda_a(x) = \lambda_b(x) =  40\cdot\bm{1}_{[0,1]}(x)$
			\item $\lambda_a(x) = \lambda_b(x) =  40\cdot\bigl(\text{sin}(2\pi x)+1\bigr)$
			\item $\lambda_a(x) = \lambda_b(x) =  40\cdot\frac{x(1-x)^4}{\int_{0}^{1}x(1-x)^4 dx}\bm{1}_{[0,1]}(x)$
		\end{enumerate}
	\end{minipage}
\end{center}
The first function is uniform, while the other two are not, indicating the intensities changes on the support. Note that the third function is the scaled beta density function with parameters (2,5). For each of the three cases under the null hypothesis, we conduct 2000 simulations of two independent Poisson processes with the intensities functions given in the corresponding case and present the proportions of rejections out of all simulations based on the adjusted p-value of each test. We generate 500 bootstrap resamples of each of the 2000 pairs of Poisson processes conditional on the total number of observations of the pooled process $N = N_a + N_b$, and use the same resamples to derive adjusted p-values for all tests. We provide the percentage of rejections at level $\alpha = 0.05, 0.1$ and $0.25$ for the five test procedures under 3 different intensities, the results are given in Table \ref{table:two_sample_sims_null}. We can see from the results that these five tests all attains the corresponding nominal levels, which is not a surprise due to the Monte Carlo Approximation of the exact rejection threshold.
\begin{table}[H]\centering
    \caption{The empirical level (\% of rejections) of different tests under the null}
		\begin{tabular}{@{}rcrrrcrrrcrrr@{}}\toprule
		\multirow{2}{*}{Test} && \multicolumn{3}{c}{case 1} & \phantom{abc} & \multicolumn{3}{c}{case 2} & \phantom{abc} & \multicolumn{3}{c}{case 3}\\
			\cmidrule{3-5} \cmidrule{7-9} \cmidrule{11-13}
			&& 5\% & 10\% & 25\% && 5\% & 10\% & 25\% && 5\% & 10\% & 25\%  \\ \midrule
		$MF$   && 4.9 & 9.8 & 25.7 && 5.5 & 8.9 & 23.7 && 5.1 & 10.2 & 24.5         \\	
        $MM$   && 4.6 & 8.9 & 22.4 && 5.4 & 10.4& 23.9 && 4.9 & 10.4 & 26.0         \\
        $KN_1$ && 4.8 & 9.6 & 23.5 && 5.1 & 9.5 & 24.2 && 6.1 & 11.3 & 27.2         \\
        $KN_2$ && 4.7 & 9.7 & 25.9 && 4.4 & 9.3 & 24.1 && 5.9 & 10.9 & 25.8         \\
        $KS$   && 5.1 & 9.9 & 24.9 && 4.5 & 9.1 & 24.6 && 6.2 & 11.4 & 25.9         \\
			\bottomrule
		\end{tabular} %
\label{table:two_sample_sims_null}  
\end{table}

\subsection{Testing homogeneous array}
\label{sec:homogeneous_simulation}

As an empirical verification of Theorem~\ref{thm:tracy-widom}, in Figure~\ref{figure:empirical_tracy_widom}, we plot the finite sample distributions of the largest eigenvalue of the adjacency matrix $A^{(r, \ell)}$ under the null hypothesis. We give the details of the experimental set-up in Section~\ref{sec:tracywidom} of the appendix; in that section, we also discuss the bootstrap correction method proposed by~\cite{bickel2016hypothesis} to improve the Tracy-Widom approximation.

Next, we consider two alternative Poisson SBM models with $K=2$ and $K=3$ equally sized communities respectively. We let the probability distribution of the interactions between two nodes $u,v$ only depends on whether they are in the same community and we denote the intensity function of realizations between individuals within the same community as $\lambda_{\text{same}}(\cdot)$ and from different communities as $\lambda_{\text{diff}}(\cdot)$. We then define
\begin{align*}
   \lambda_{\text{same}}(x) = s\cdot\bm{1}_{[0,1]}(x), \quad \lambda_{\text{diff}}(x) = s\cdot\frac{x(1-x)^4}{\int_{0}^{1}x(1-x)^4 dx}\bm{1}_{[0,1]}(x)  
\end{align*}
for both the two alternative Poisson SBM models, where $s$ is a parameter that controls the sparsity levels of the networks in this experiments. We again have $n=200$ and either $K=2$ or $K=3$ equally sized communities. 
We then generate 200 sample collections of realizations on the same support for each of the two models and for each value of $s\in\{0.1, 0.175, 0.25, 0.5, 1\}$ and conduct our proposed test on these samples where the bootstrap sample size and partitioning of the support are exactly the same as in the preceding experiment. The proportions of rejections for the two SBM models under different sparsity levels are recorded in Table \ref{table:array_test_sims_sparsity}. 

{
\renewcommand{\baselinestretch}{1}
\begin{table}[H] \centering
    \begin{tabular}{crcccccrccccc}\toprule
        && \multicolumn{5}{c}{$K=2$} && \multicolumn{5}{c}{$K=3$} \\
        \cmidrule{3-7} \cmidrule{9-13}
        $s$ && $1.0$ & $0.5$& $0.25$ & $0.175$ & $0.1$ && $1.0$ & $0.5$& $0.25$ & $0.175$ & $0.1$\\
        \midrule
        $\alpha=0.01$ && 1 & 1 & 0.98 & 0.41 & 0.055 && 1 & 1 & 1 & 0.785 & 0.05 \\
        $\alpha=0.05$ && 1 & 1 & 0.99 & 0.575 & 0.115 && 1 & 1 & 1 & 0.905 & 0.145 \\
        $\alpha=0.10$ && 1 & 1 & 0.995 & 0.63 & 0.165 && 1 & 1 & 1 & 0.97 & 0.24   \\
        $\alpha=0.25$ && 1 & 1 & 1 & 0.74 & 0.36 && 1 & 1 & 1 & 1 & 0.38  \\
        \bottomrule
    \end{tabular}
    \caption{The proportion of rejections of the proposed array test out of 200 simulated samples of networks of 200 nodes at different sparsity and confidence levels $\alpha\in\{0.01, 0.05, 0.10, 0.25\}$.}
    \label{table:array_test_sims_sparsity}
\end{table}
}

\subsection{Empirical verification of Tracy-Widom approximation and bootstrap correction}
\label{sec:tracywidom}

To see how fast the largest eigenvalues converge to the limiting distribution, we consider two cases with the numbers of nodes $n=300$ and $n=1600$ respectively. For each case we simulate 1000 adjacency matrix $A$ whose entries $\{A_{ij}:i\ne j\leq n\}$ are independent and identically distributed Poisson random variables with mean equals to 20. Then we plot the sample distribution of the test statistics, i.e., $n^{2/3}\big(\lambda_{1}(\tilde{A})-2\big)$ against the Tracy-Widom distribution, where $\tilde{A}$ is the empirically centered and scaled version of $A$. 

\begin{figure}[h]
     \includegraphics[width=15cm]{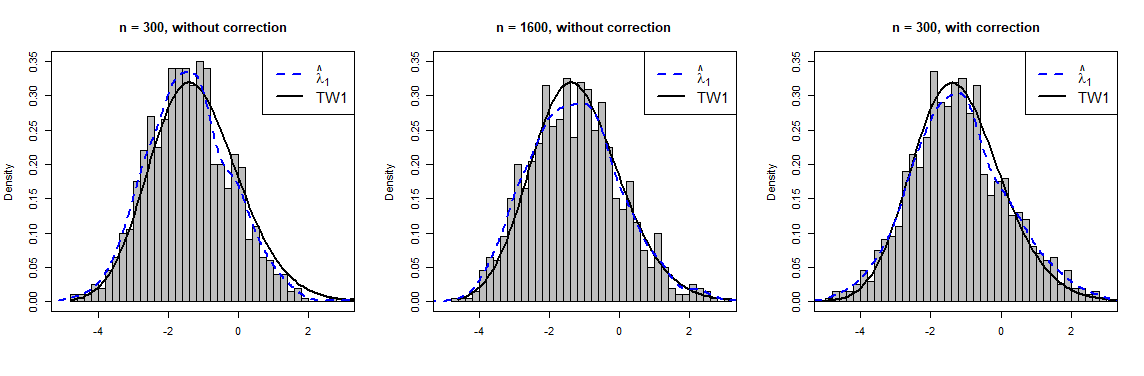}
    \caption{The empirical distribution of 1000 simulated samples of centered and scaled largest eigenvalues of $\Tilde{A}$, compared with the Tracy-Widom distribution.}
 \label{figure:empirical_tracy_widom}
\end{figure}

We can see from the first two graphs in Figure \ref{figure:empirical_tracy_widom} that when $n=300$ the sample distribution deviates in location compared with the target distribution and when $n=1600$ the location is corrected but there still is some difference in scale. Though there are some differences in location and scale, we can see the sample distribution does have similar shape with the Tracy-Widom distribution even when the number of nodes is as small as 300. In similar experiments where adjacency matrices have Bernoulli distributed entries, \cite{bickel2016hypothesis} proposed to apply bootstrap correction to the largest eigenvalue, where they generate parametric bootstrap samples of the adjacency matrices and use the bootstrapped mean and variance of the largest eigenvalues to shift and scale the test statistics to have a better match with the Tracy-Widom distribution. Here we adapted the same bootstrap correction technique to the eigenvalues of adjacency matrix with $n=300$ nodes, where we generate 50 bootstrap samples for each sample adjacency matrix. We plot the empirical distribution of the test statistics after bootstrap correction as the third graph in Figure \ref{figure:empirical_tracy_widom}. We can see that even with just 50 bootstrap samples, the sample distributions of the test statistics looks much closer to the target distribution.

Remember that here we are using the Tracy-widom distribution to compute p-values for every local null hypotheses $\bar{H}_0^{(r, \ell)}$ and we generate bootstrap samples to estimate the exact critical threshold for the global null hypothesis $\bar{H}_0$, thus we could simply use the same bootstrap samples generated for testing the global null to correct the location and scale of the largest eigenvalue of each local adjacency matrix $\tilde{A}^{(r, \ell)}$. Given observed collection of Poisson process realizations $\{N_{uv}(\cdot): u<v\in[n\}$, we describe the procedure to derive the local p-values with bootstrap correction of the location and scale of the largest eigenvalue in the following steps:
\begin{enumerate}
	\item For $b^* = 1,2,\dots,B$, generate bootstrap sample collections $\{N^{b^*}_{uv}(\cdot): u<v\in[n]\}$ as described in Section \ref{sec:symmetric_resample_method}.
	
    \item For the observed realization, estimate the Poisson mean $\hat{\lambda}^{(r,\ell)}$ of each discretized interval as $\hat{\lambda}^{(r,\ell)} = \frac{1}{n^2-n}\sum_{u \ne v}N_{uv}^{(r,\ell)}$ and let $\tilde{A}^{(r, \ell)}$ be the centered and re-scaled adjacency matrix for interval $I^{(r,\ell)}$
    \begin{align*}
    \tilde{A}^{(r,\ell)}_{uv}:=\begin{cases} 
    \frac{N^{(r,\ell)}_{uv}-\hat{\lambda}^{(r,\ell)}}{\sqrt{(n-1)\hat{\lambda}^{(r,\ell)}}} , & u\neq v, \\
    0, & u=v.
    \end{cases}
    \end{align*}
    and let $\lambda_1(\tilde{A}^{(r,\ell)})$ be the largest eigenvalue of adjacency matrix $\tilde{A}^{(r,\ell)}$.
    
    \item Do step 2 for every bootstrap resamples to derive their largest eigenvalues $\lambda_1(\tilde{A}^{(r,\ell,b^*)})$ at every discretized interval. Then we calculate the sample mean and standard deviation of $\big\{\lambda_{1}(\tilde{A}^{(r,\ell,b^*)}): b^*\in\{1,2,\dots,B\}\big\}$ for each $r\in[R], \ell\in[2^r]$ and denote them as  $\hat{\mu}^{(r, \ell)}_1, \hat{s}^{(r, \ell)}_1$ respectively.
    
    \item Denote $\mu_{\text{tw}}$ and $s_\text{tw}$ as the mean and standard deviation of Tracy-Widom distribution with $\beta=1$ and let $$\lambda_{bc}^{(r,\ell)} = \mu_{\text{tw}} + s_\text{tw}\frac{\lambda_1(\Tilde{A}^{(r,\ell)})-\hat{\mu}^{(r, \ell)}_1}{\hat{s}^{(r, \ell)}_1}$$ be the test statistic after bootstrap correction.
    
    \item Finally we compute the p-value for the discretized local null $\bar{H}_)^{(r,\ell)}$ as 
    \[
      p^{(r, \ell)} \equiv p^{(r,\ell)}\bigl(\lambda_{bc}^{(r,\ell)}\bigr) := 2\text{min}\bigg(F_{\scalebox{1}{$\scriptscriptstyle \text{TW1}$}}\Big(\lambda_{bc}^{(r,\ell)}\Big), 1-F_{\scalebox{1}{$\scriptscriptstyle \text{TW1}$}}\Big(\lambda_{bc}^{(r,\ell)}\Big)\bigg)
    \]
\end{enumerate}

\end{document}